%% file: cpl_44.tex
\documentclass[11pt,oldtoc,a4paper]{cernart}
\usepackage{times}
\usepackage{amsmath} 
\usepackage{amsfonts}
\usepackage{amssymb}
\usepackage{pifont}
\usepackage{graphicx}
\usepackage{cite}
\usepackage{multicol} 
\begin{document}
\setcounter{tocdepth}{2}
\include{preamble}

\include{page_title}

\thispagestyle{empty}
\cleardoublepage
%=====================
\pagenumbering{arabic}
\section{Introduction}\label{sec:intro}

A neutral kaon oscillates forth and back between itself and its antiparticle, because the
physical quantity, strangeness, which distinguishes antikaons from kaons, is not conserved
by the interaction which governs the time evolution \cite{gell,pais}.\\
The observed behaviour is the interference
pattern of the matter waves of the \ks\ and of the \kl\ , which act as highly stable damped
oscillators with a relative frequency difference of $\dm /\m_\kn \approx 7 \times 10^{-15}$.
The corresponding beat frequency of \
$\omega =\dm /\hbar \approx 5.3\times 10^9 \ \mathrm{s}^{-1}$, and the scale of the
resulting spatial interference pattern of \ $2\pi c/\omega \approx 0.36$ m\ , fit well
the present technical capabilities of detectors in high-energy physics.\\
Some fraction of the size of $\dm \approx 3.5 \times 10^{-15}\geve$, depending on the
measurement's precision, indicates the magnitude of effects, which may be present inside of the
equation of motion, to which the measurements are sensitive: $10^{-18}$ to $10^{-21}\geve$.

This article describes the relations of the parameters, which specify the properties of the
laws of the time evolution and of the decays of the neutral kaons, to the measured quantities,
with emphasis to \CPTz\ invariance and to \Tz\ violations.

Experimental tests of \CPTz\ invariance are of great interest. This invariance plays a fundamental
r\^ole for a causal description of physical phenomena. We quote Res Jost \cite{jost}, pioneer
of the\ \CPTz\ theorem \cite{schw, bellcpt, lued, paulicpt, jostcpt}, \ \ '... : eine kausale Beschreibung
ist nur m\"oglich, wenn man den Naturgesetzen eine Symmetrie zugesteht, welche den Zeitpfeil
umkehrt'.
We shall describe experimental results that could have contradicted \CPTz\ invariance, but
did not.

\Tz\ violation accompanies \CPz\ violation \cite{schub}. Recently, an experiment \cite{pen2}
at CERN has measured that the evolution \ \knb \ra\ \kn\ is faster than \ \kn \ra\ \knb\ , which,
as explained below, formally contradicts \Tz\ invariance in the neutral kaon's weak
interaction.

\Tz\ and \ \CPTz\ violation have \CPz\ violation as a necessary condition. For this one, a simple criterion on the
quark level has been given in the form of one single relation \cite{jarl}.

We also consider \Tz\ violation versus \Tz\ oddness.

We shall discuss the general time evolution as well of a single neutral kaon as  of an entangled
pair of neutral kaons by applying the density matrix formalism. This allows one to design
experiments which test, whether isolated kaons in a pure state do not evolve into ones in a
mixed state. Such transitions would be forbidden by quantum mechanics and would violate
\CPTz\ invariance.\\
First experimental results on this subject have been obtained at CERN \cite{Ad95}.

In our description we shall, as most authors do, assume, that the kaon's time evolution,
and the decay, are distinct processes. The time evolution is derived from a Schr\"odinger
equation, and it becomes parametrized via perturbation theory (\ra\PLij),
while specific decays are described
by time-independent amplitudes. The conservation of probability, inherent in the perturbation result, links
the two processes.

We present the results
which concern \CPTz\ invariance and \Tz\ violation derived mostly from neutral kaon decays
into \pipa\ and \pen\  final states.

The history of the unveilings of the characteristics of the neutral kaons is a sparkling
succession of brilliant ideas and achievments, theoretical and  experimental
\cite{femn, kbir, jrlg, wo, cago, bloi, klei}.\\
We have reasons to assume that neutral kaons will enable one to make more basic discoveries.
Some of the kaons' properties (e. g. entanglement) have up to now only scarecely been exploited.

\section{The neutral-kaon system}\label{sec:motiv}
\subsection{Time evolution}\label{sub:evol}

The time evolution of a neutral kaon and of its decay products may be 
represented by the state vector
\begin{equation}\label{eq:2.01}
\ket{\psi} = \psi_{\kn }(t)\ket{\kn } + \psi_{\knb }(t)\ket{\knb }
           +  \sum_m c_m (t)\ket{m}
\end{equation}
which satisfies the Schr\"odinger equation
\begin{equation}
\label{eq:2.02}
\i \frac{\dd \ket{\psi }}{\dd t} = \Hz \ket{\psi }.
\end{equation}
In the Hamiltonian, $\Hz = \Hz _0 + \Hwk $, $\Hz _0$ governs the strong and 
electromagnetic interactions. It is invariant with respect to the 
transformations \Cz , \Pz , \Tz , and it conserves the strangeness $S$. 
The states $\ket{\kn }$ and $\ket{\knb }$ are common stationary eigenstates 
of $\Hz _0$ and $S$, with the mass $m_0$ and with opposite strangeness: 
$\Hz _0\ket{\kn } = m_0 \ket{\kn }$, $\Hz _0\ket{\knb } = m_0 \ket{\knb }$,
$S\ket{\kn } = \ket {\kn } $, $S\ket{\knb } = -\ket{\knb} $. The states 
$\ket{m}$ are continuum eigenstates of $\Hz _0$ and represent the decay 
products. They are absent at the instant of production $(t=0)$ of the neutral 
kaon. The initial condition is thus
\begin{equation}\label{eq:2.03}
\ket{\psi_0} = \psi_{\kn }(0)\ket{\kn } + \psi_{\knb }(0)\ket{\knb } \ .
\end{equation}
\Hwk\  governs the weak interactions. Since these do not conserve strangeness, 
a neutral kaon will, in general,  change its strangeness as time evolves.

Equation (\ref{eq:2.02}) may be solved for the unknown functions $\psi _{\kn}
(t)$ and $\psi_{\knb}(t)$ , by using a perturbation
approximation~\cite{wigner} which yields~\cite{trei, lee2}
\begin{equation}\label{eq:2.04}
\psi = \e^{-\i\Lz t} \psi_0 \ .
\end{equation}
$\psi$ is the column vector  with components $\psi_{\kn}(t)$ and 
$\psi_{\knb}(t)$, $\psi_0$ equals $\psi$ at $t=0$, and \Lz\ is the 
time-independent $2 \times 2$ matrix $\left(\PLik \right)$, whose components
refer to the two-dimensional basis $\ket{\kn },\ket{\knb}$ and may be written
as \ \PLik\ = \ $\bra{\alpha}\Lambda \ket{\alpha^{\prime}}$ with
$\alpha , \alpha^{\prime} = \kn , \knb$.
\\\\
Since the kaons decay, we have 
\begin{equation}\label{eq:2.05}
0 > \frac{\dd |\psi|^2}{\dd t}  = -\i\psi^{\dagger}\left(\Lz-\Lz^{\dagger}\right)\psi \,.
\end{equation}
\Lz \ is thus not hermitian, $\e^{-\i\Lz t}$ is not unitary, in general.
\\\\
This motivates the definition of $\Mz$ and $\Gz$ as
\begin{subequations}\label{eq:2.06}
\begin{eqnarray}
\Lz &=& \Mz - \frac{\i}{2} \Gz \;,\label{eq:2.06a}\\ 
\Mz &=& \Mz ^{\dagger},\quad \Gz = \Gz ^{\dagger}\;.\label{eq:2.06b}
\end{eqnarray}
\end{subequations}
We find
\begin{equation}\label{eq:2.07}
0 > \frac{\dd |\psi|^2}{\dd t}  = - \psi^{\dagger}\Gz\psi \;.
\end{equation}
\\
This expresses that \Gz\ has to be a positive matrix.
\\\\
The perturbation approximation also establishes the relation
from \Hwk\ to \Lz \ (including second order in \Hwk ) by
\begin{subequations}\label{eq:2.08}
\begin{align}
\Mz _{\alpha\alpha^{\prime}} & =   
   m_0 \del_{\alpha\alpha^{\prime}}
 + \sbraket{\alpha|\Hwk }{\alpha^{\prime}} +
 \Pz  \sum_{\beta} \left( \frac{\sbraket{\alpha|\Hwk }{\beta}
                                \sbraket{\beta|\Hwk }{\alpha^{\prime}}}
                               {m_0 - E_{\beta}}\right)\;, \label{eq:2.08a} \\
\Gz _{\alpha\alpha^{\prime}} & = 2\pi\sum_{\beta}
        \sbraket{\alpha|\Hwk }{\beta}\sbraket{\beta|\Hwk }{\alpha^{\prime}}
                              \del (m_0 - E_{\beta})\;,      \label{eq:2.08b}\\
(\alpha , \alpha^{\prime} &= \kn , \knb )\,.              \notag 
\end{align}
\end{subequations}
\\
Equations (\ref{eq:2.08a}, \ref{eq:2.08b}) enable one now to state directly 
the symmetry properties of \Hwk\  in terms of experimentally observable 
relations among the elements of \Lz , see Table \ref{tab:01}. We remark
that \CPTz\ invariance imposes no restrictions on the off-diagonal elements,
and that \Tz\ invariance imposes no restrictions on the diagonal elements
of \Lz\ . \ \CPz\ invariance is violated, whenever one, at least, of these invariances is
violated.\\

%\newpage
\begin{table}
\begin{center}
\caption{The symmetry properties of \Hwk\ induce symmetries in observable quantities.}
The last column indicates asymmetries of quantities which have been measured\\
by the CPLEAR experiment \cite{pr} at CERN. More explanations are given in the text.\\
\medskip
{\small
\begin{tabular}{llll}
\hline\hline \\[-0.2cm]
If \Hwk\ has the property   & called        &   then    & or \\[0.2cm]
\hline \\[0.2cm]
$\Tz ^{-1} \ \Hwk\ \ \Tz = \Hwk $ &\Tz\ invariance &
                                  $|\PLab | = |\PLba |$ & $\AT   = 0$ \\[0.2cm]
$(\CPTz )^{-1} \ \Hwk\ \ (\CPTz ) = \Hwk $ & \CPTz\ invariance & 
                                  $ \ \ \PLaa   = \ \PLbb  $ & $\ACPT = 0$ \\[0.2cm]
$(\CPz  )^{-1} \ \Hwk\ \ (\CPz  ) = \Hwk $ & \CPz\  invariance &
                                  $ \ \ \PLaa   = \ \PLbb   $  \ and &      \\[0.2cm]
                           & &    $|\PLab | = |\PLba |$ & $\ACP  = 0$ \\[0.2cm]
\hline\hline \\[-0.2cm] 
\end{tabular}}
\label{tab:01}
\end{center}
\end{table}

The definitions of $\ket{\kn }$ and $\ket{\knb }$ leave a real phase
$\vartheta $ undetermined: \\\\ Since $ S ^{-1} \ \Hz _0 \ S = \ \Hz _0 $, the states

\begin{subequations}\label{eq:2.09}
\begin{eqnarray}
\ket{\kn '} & = & \e^{\i\vartheta S}  \ket{\kn } \ = \ \e^{\i\vartheta} \ \ \ket{\kn } \ , \label{eq:2.09a}\\
\ket{{\knb} ^{'}}  & = & \e^{\i\vartheta S} \ket{\knb } \ = \ \e^{-\i\vartheta} \ket{\knb } \ . \label{eq:2.09b}
\end{eqnarray}
\end{subequations}
fulfil the definitions of $\ket{\kn}$ and $\ket{\knb}$ as well, and 
constitute thus an equivalent basis which is related to the original basis by
a unitary transformation. As the observables are always invariant with respect
to unitary base transformations, the parameter $\vartheta$ cannot be measured, and
remains undetermined. This has the effect that expressions which depend on
$\vartheta$ are not suited to represent experimental results, unless $\vartheta$ has
beforehand been fixed to a definite value by convention. Although such a
convention may simplify sometimes the arithmetic, it risks to obscure the
insight as to whether a certain result is genuine or whether it is just an
artifact of the convention.\\As an example we consider the elements of $\Lz $,
$\PLik $ = \ $\bra{\alpha}\Lambda \ket{\alpha^{\prime}}$ which refer to the basis
$\alpha , \alpha^{\prime} = \kn , \knb $. With respect to the basis
$\e^{\i \vartheta}\kn $, $\e^{-\i \vartheta}\knb $ we obtain the same diagonal elements,
whereas the off-diagonal elements change into
\begin{subequations}\label{eq:2.10}
\begin{eqnarray}
\PLab  & \longmapsto  &  \Lambda'_{\kn\knb }  \ = \ \e^{-2\i\vartheta} \ \ \PLab \ , \label{eq:2.10a}\\
\PLba  & \longmapsto  &  \Lambda'_{\knb\kn }  \ = \ \e^{2\i\vartheta} \ \ \ \ \PLba \  , \label{eq:2.10b}
\end{eqnarray}
\end{subequations}
and are thus convention dependent. However, their product, their absolute values,
the trace tr\{\Lz\}, its determinant, and its eigenvalues (not so its eigenvectors),
but also the partition into $M$ and $\Gz$, are convention independent ~\cite{lavoura}. (We will
introduce a phase convention later in view of comparing experimental
results).

The definition of the operations \CPz\ and \CPTz\ allows one to define two additional phase
angles. We select them such that \ $\cal{O}\kn =$ \knb\ and \ $\cal{O}$\knb = \kn, where
$\cal{O}$ stands for \CPz\ or \CPTz. See e. g. \cite{tya}.\\

In order to describe the time evolution of a neutral kaon, the matrix exponential
$\e^{-\i\Lz t}$ has to be calculated.\\
If the exponent matrix has $two$ dimensions, a generalized Euler formula gives a
straightforward answer. \\ Be $\Lz$ represented as a superposition of Pauli matrices
\begin{equation}\label{eq:2.11}
\Lz = \Lz^\mu \sigma^\mu
\end{equation}
with $\sigma^0$ = \ unit matrix, $\sigma^k$ = \ Pauli matrices, $\Lz^\mu$ complex.
(Summation over multiple indices: Greek 0 to 3, Roman 1 to 3). \\\\Then
\begin{equation}\label{eq:2.12}
\e^{-\i\Lz t} = \e^{-\i\Lz^\mu \sigma^\mu t} = \e^{-\i\Lz^0 t}( \sigma^0\cos(\Omega t)
- \i\Lz^m \sigma^m t \sin(\Omega t)/(\Omega t)),
\end{equation}\\
where \ $\Omega = \ (\Lz^m \Lz^m)^{1/2}$. Noting that $\Lz^\mu=\frac{1}{2}$tr$\{\sigma^\mu\Lz\}$,
we see that Eq. (\ref{eq:2.12}) expresses $\e^{-\i\Lz t}$ entirely in terms of the elements of $\Lz$. Since the
(complex) eigenvalues $\lzl$, $\lzs$ of $\Lz$ turn out to be observable (and are thus doubtlessly
phase transformation invariant) we introduce them into (\ref{eq:2.12}).\\
They fulfil $\lzl\lzs =det(\Lz)=\Lz^0\Lz^0 -\Lz^m\Lz^m$, and $\lzl+\lzs=\ $tr$\{\Lz\}=2\Lz^0$,
and thus, with $\dlz \equiv \lzl - \lzs$,
\begin{equation}\label{eq:2.13}
\Omega=\dlz/2 \;.
\end{equation}
We note here (with relief) that the calculation of the general time evolution, expressed
in Eq. (\ref{eq:2.04}), does not need the knowledge of the $eigenstates$ of $\Lz$, whose
physical interpretation needs special attention ~\cite{enz,gau1}.\\
The following corollary will be of interest:\\ The off-diagonal elements of a $2\times 2$ exponent matrix
factorize the off-diagonal elements of its matrix exponential, with equal factors:
\begin{equation}\label{eq:2.14}
(\e^{-\i \Lz t})_{j \neq k} = 
(-i \Lz t)_{j \neq k} \ \e^{-\frac{1}{2}\i (\PLaa + \PLbb) t} \ \sin (\Omega t)/(\Omega t) \ \ .
\end{equation}
This is valid for two dimensions.\\Independent of the dimension $n$ of the exponent matrix,
diagonalization allows one to calculate the matrix exponential. Find the two vectors
$\ket{\kls}$ which $\Lz$ transforms into a multiple of themselves
\begin{equation}\label{eq:2.15}
\Lz \ket{\kls} = \lzls \ket{\kls}\;.
\end{equation}
The eigenvalues $\lzls$ need to be
\begin{equation}\label{eq:2.16}
\lzls = \frac{1}{2}\ \mathrm{tr}\{\Lz\}\pm \sqrt{(\mathrm{tr}\{\Lz\})^2/4 - det(\Lz)} \ \;.
\end{equation}\\ We may express the solutions of (\ref{eq:2.15}) in the basis
$\ket{\kn},\ket{\knb}$ as
\begin{eqnarray}
\ket{\ks}&=&V^{11}\ket{\kn}+V^{21} \ket{\knb} \ \hat{=} \
\left ( \begin{array}{c} V^{11} \\ V^{21} \end{array} \right )\label{eq:2.17}\\
\ket{\kl}&=&V^{12}\ket{\kn}+V^{22} \ket{\knb} \ \hat{=} \
\left ( \begin{array}{c} V^{12} \\ V^{22} \end{array} \right )\label{eq:2.18}
\end{eqnarray}
and form the matrix $V=(V^{ij})$ whose columns are the components of the eigenvectors,
and also $W=V^{-1}$. The matrix $\Lz$ can now be represented as
\begin{equation}\label{eq:2.19}
\Lz = VDW
\end{equation}
where $D$ is diagonal,
\[ D= \left ( \begin{array}{cc} \lzs & 0 \\ 0 & \lzl \end{array} \right ). \] \\
Since we need to extract $V$ and $W$ from the exponent to obtain
\begin{equation}\label{eq:2.20}
\e^{-\i \Lz t}=\e^{-\i VDW t}=V\e^{-\i D t}W,
\end{equation}
it is important that \ $W=V^{-1}$ (and $not \ W=V^{\dagger} \ne V^{-1}$). Since $WV=1$
or \ $W^{ij}V^{jk}=\delta^{ik}$, the rows of $W$ are orthogonal to the columns of
$V$. A convenient solution is
\begin{equation}\label{eq:2.21}
W=\left(W^{ij}\right)=\frac{1}{\modulus{V}}\left ( \begin{array}{cc}
V^{22} & -V^{12} \\ -V^{21} & V^{11} \end{array} \right ).
\end{equation}\\
Inserting $V$, $D$, $W$ into (\ref{eq:2.20}) allows one to express $\e^{-\i \Lz t}$
in terms of the eigenelements $\ket{\kls}$, and $\lzls$. (Eq. (\ref{eq:2.19})
also shows how to construct a matrix with prescribed (non-parallel) eigenvectors and
independently prescribed eigenvalues).\\
If we define the vectors
\begin{eqnarray}
\bra{\widetilde{\ks}}=W^{11}\bra{\kn}+W^{12}\bra{\knb}\label{eq:2.22}\\
\bra{\widetilde{\kl}}=W^{21}\bra{\kn}+W^{22}\bra{\knb}\label{eq:2.23}
\end{eqnarray}
then we have 
\begin{equation}\label{eq:2.24}
\sbraket{\widetilde{\mathrm{K}_\kappa}}{\mathrm{K}_{\kappa'}}=\delta_{\kappa \kappa'}\;,\ \ \ \kappa,\kappa'=\mathrm{L,S}
\end{equation}
in contrast to
\begin{equation}\label{eq:2.25}
\sbraket{{\mathrm{K}_\kappa}}{\mathrm{K}_{\kappa'\ne \kappa}}\ne 0.
\end{equation}\\
The difficulty to interprete these vectors as states is discussed in \cite{enz}.
Eq. (\ref{eq:2.25}) shows that there is no clear state $\kl$, because the vector
$\ket{\kl}$ always has a component of $\ket{\ks}$, with the probability $\modulus{\sbraket{\ks}{\kl}}^2$.\\\\
We now solve (\ref{eq:2.15}): $\Lz^{ij}V^{j\kappa}=\lambda_{\kappa}V^{i\kappa}$, (no sum $\kappa$)
for $V^{j\kappa} \hat{=} \ket{\mathrm{K}_\kappa}$, $\kappa = (1,2) \ \hat{=}$ (S,L), and regain
(\ref{eq:2.12}) in a different form:\\
\begin{equation}\label{eq:2.26}
\e^{- \i\Lz t} = \left(
\begin{array}{cc}
f_+ + 2 \den f_-  & -2\PLab f_-/\dlz \\
- 2 \PLba f_-/\dlz & f_+ - 2 \den f_-
\end{array}\right)
= \left(
\begin{array}{cc}
f_+ + 2 \den f_-  & -2 (\sigma - \epn) f_- \\
- 2 (\sigma + \epn) f_- & f_+ - 2 \den f_-
\end{array}\right)
\end{equation}\\
with
\begin{equation}\label{eq:2.27}
f_{\pm}(t) = \dfrac{\e^{-\i \lzs t }\pm \e^{-\i\lzl t }}{2}\;,
\end{equation}
\begin{subequations}\label{eq:2.28}
\begin{eqnarray}
\den   & \equiv & (\PLbb - \PLaa ) /(2 \dlz ) \,, \label{eq:2.28a}\\
\epn   & \equiv & (\PLba  -\PLab ) /(2 \dlz ) \,, \label{eq:2.28b}\\
\sigma & \equiv & (\PLba + \PLab ) /(2 \dlz ) \,. \label{eq:2.28c}
\end{eqnarray}
\end{subequations}
We have set
\begin{equation}\label{eq:2.29}
\lzls = \mls - \frac{\i}{2}\gls
\end{equation}
and
\begin{equation}\label{eq:2.30}
\dlz = \lzl -\lzs = \ml-\ms + \frac{\i}{2}(\gs-\gl)\equiv \dm + \frac{\i}{2}\dg =
\modulus{\dlz}\ \e^{\i(\frac{\pi}{2}-\fsw)} 
\end{equation}
with
\begin{eqnarray}\label{eq:2.31}
\dm \equiv \ml -\ms \;,\ \ \ 
\dg \equiv \gs -\gl\;,
\end{eqnarray}
and with \fsw\ defined by
\begin{equation}\label{phisw}
\tan(\fsw)=(2\dm/\dg) \ .
\end{equation}
%Hier bin ich 17. 5. 2004.***************************
%
The parameters in (\ref{eq:2.28}) satisfy the identity
\begin{equation}\label{ouridentity}
\sigma^2 - \epn^2 + \den^2  \equiv 1/4\ 
\end{equation}
which entails
\begin{equation}\label{modourid}
\zeta \equiv\ {\modulus{\sigma}}^2 + \modulus{\epn}^2 + \modulus{\den}^2 -1/4 \ \ge 0 \ .
\end{equation}
From Table \ref{tab:01} we can deduce that $\zeta $ signifies the violations of \Tz\ and \CPTz\ .\\\\
The positivity of the matrix \Gz\ requires the determinant $|\Gz|$ to be positive
\begin{equation}\label{ourpositivity}
0 < |\Gz| =
{\modulus{\dlz}}^2 (\frac{\gs\gl}{{\modulus{\dlz}}^2}-2\zeta)\ ,
\end{equation}
which needs
\begin{equation}\label{limmods}
\zeta\ <\
%({\modulus{\sigma}}^2  + \modulus{\epn}^2 + \modulus{\den}^2)-\frac{1}{4} \ < \
\frac{\gs\gl}{{2\modulus{\dlz}}^2} \approx \frac{\gl}{\gs} \ .
\end{equation}
The last approximation is valid for neutral kaons where, experimentally,
$2{\modulus{\dlz}}^2 \approx (\gs)^2$.\\
We see from
(\ref{limmods}), that the ratio \ $\gl/\gs\ (\approx 1.7\times 10^{-3})$ provides a general limit for
the violations of \Tz\ and \CPTz\ invariance.
\\\\
The eigenstates can now be expressed by the elements of $\Lz$
\begin{eqnarray}
\ket{\ks}=N_S \left(\PLab \ket{\kn} \ +\ (\lzs-\PLaa)\ket{\knb}\right)\;,\label{eq:2.34}\\ 
\ket{\kl}=N_L \left((\PLbb-\lzl)\ket{\kn} \ -\ \PLba \ket{\knb}\right)\;,\label{eq:2.35}
\end{eqnarray}
with suitable normalization factors $N_S$, $N_L$.\\
They develop in time according to 
\begin{equation*}
\ket{\kls} \rightarrow \e^{-\i\lzls t}\ket{\kls}\ .
\end{equation*}
\gls\ thus signify the decay widths of the eigenstates with mean lifes $\tls=1/\gls $, and \mls\
are the rest masses. \\
These quantities are directly measurable. The results show that \ $\tl\ \gg \ts$ and $\ml>\ms$\ .
We therefore have
\begin{eqnarray*}
0 \le \fsw\ \le \pi/2\ .\\
& &
\end{eqnarray*}
In the limit \ $\epn \rightarrow 0$, $\den \rightarrow 0$, the eigenstates are
\begin{eqnarray}
\ket{\ks}\rightarrow \ket{\mathrm{K}_1}=\frac{1}{\sqrt{2}} \left(\ket{\kn} \ +\ \ket{\knb}\right)\;,\notag\\ 
\ket{\kl}\rightarrow \ket{\mathrm{K}_2}=\frac{1}{\sqrt{2}} \left(\ket{\kn} \ -\ \ket{\knb}\right)\;.\notag
\end{eqnarray}

The differences \Mzaa\ $-$\ \Mzbb\ and \Gzaa\ $-$\ \Gzbb\ , which may be interpreted as \ (\CPTz\
violating) mass and decay width differences between the \kn\ and the \knb , 
are related to \den\ and to \dlz\ as follows: \\\\
Define the reals \ \dpar\ and \ \dper\ by
\begin{equation}\label{dpardsen}
\dpar + \i \dper\ = \den\ \e^{-\i \fsw}
\end{equation}
then
\begin{eqnarray}
\Mzaa\ -\ \Mzbb\ &=& 2\modulus{\dlz}\dper\label{eq:2.32}\\
\Gzaa\ -\ \ \Gzbb\ &=& 4\modulus{\dlz}\dpar\ .\label{eq:2.33}
\end{eqnarray}
We wish to remark that, given the constants $\modulus{\dlz}$\ and \fsw , the information contained in the mass and
decay width differences (\ref{eq:2.32},\ref{eq:2.33}) is identical to the one in \den .

%*******************19. 5. 2004***************
\subsection{Symmetry}\label{symmetry}

The measurement of particularly chosen transition rate asymmetries concerning the neutral kaon's time
evolution exploit properties of \Hwk\ in an astonishingly direct way.

To explain the principle of the choice of the observables we make the temporary assumption that the
detected decay products unambigously mark a relevant property of the kaon at the moment of its
decay: The decay into two pions indicates a \CPz\ eigenstate with a positive eigenvalue, a semileptonic
decay, (\ra\ e$\pi\nu$ \ or \ \ra\ $\mu\pi\nu$) , indicates the kaon's strangeness to be equal to the
charge of the lepton (in units of positron charge).

We will later show that previously unknown symmetry properties of the decay mechanism
($\Delta S = \Delta Q$ rule, $\CPTz$ violation 'in decay') or practical experimental
conditions (efficiencies, interactions with the detector material, regeneration of $\ks$
by matter) do not change the conclusions of this section.
\subsubsection{\Tz\ violation}\label{tviol}

Compare the probability for an antikaon to develop into a kaon, $| \bra{\kn }\e^{-\i\Lz t}\ket{\knb } |^2$ , with the one for
a kaon to develop into an antikaon, $| \bra{\knb }\e^{-\i\Lz t}\ket{\kn } |^2$, within the 
same time interval $t$. Intuition wants the probabilities for these mutually reverse processes to
be the same, if time reversal invariance holds.

We now show that the experimentally observed difference ~\cite{pen2} formally contradicts 
\Tz\ invariance in $\Hwk$.

Following ~\cite{leebook}, time reversal invariance, defined by $\Tz ^{-1} \ \Hwk\ \ \Tz = \Hwk $,
requires 
\begin{equation}\label{eq:2.36}
\Gzbas/\Gzba = \Mzbas/\Mzba
\end{equation}
which is equivalent to $\modulus{\PLab}^2 =\modulus{\PLba}^2$. This is measurable !\\
The normalized
difference of these quantities\\
\begin{equation}\label{eq:2.37}
\AT \equiv \frac{|\PLab |^2 - |\PLba |^2 }{|\PLab |^2 + |\PLba |^2}
\end{equation}\\
is a theoretical measure for time reversal violation ~\cite{kabirb}, and we find, using
(\ref{eq:2.14}), the identity
\begin{equation}\label{eq:2.38}
\AT  \equiv     \frac{| \bra{\kn }\e^{-\i\Lz t}\ket{\knb } |^2 -  
                      | \bra{\knb }\e^{-\i\Lz t}\ket{\kn } |^2} 
                     {| \bra{\kn }\e^{-\i\Lz t}\ket{\knb } |^2 + 
                      | \bra{\knb }\e^{-\i\Lz t}\ket{\kn } |^2} \ ,
\end{equation}
which expresses the different transition probabilities for the mutually reverse processes \
$\knb \Longleftrightarrow  \kn$ , as a formal consequence of the property of \Hwk \ not to commute
with \Tz .\\\\
The value of \AT\ is predicted as follows
\begin{equation}\label{eq:2.39}
\AT = \frac{-2\re (\epn\sigma^*)}{\modulus{\epn}^2+\modulus{\sigma}^2} \approx 4\ree
\ \ \ \ \ \ (\mathrm{for} \modulus{\epn}\ll \modulus{\sigma}\ \mathrm{and}\ \
\sigma \approx -\frac{1}{2}) \;.
\end{equation}
\\
We add some general remarks:\\
The directness of the relation between \AT \ and \Hwk \ rests partly on the fact that the
neutral kaons are described in a two dimensional space, (\kn, \knb), in which the corollary
(\ref{eq:2.14}) is valid. This is also the origin for the time independence of \AT ~\cite{gerberEPJ}.\\
\AT \ is a \Tz-odd quantity insofar as it changes its sign under the interchange of the initial and final states \
$\knb \Longleftrightarrow  \kn$.\\
Eqs. (\ref{eq:2.36}) and (\ref{eq:2.37}) describe time reversal invariance in an explicitly
phase transformation invariant form. In Eq. (\ref{eq:2.39}), both, the numerator and the
denominator, have this invariance. The approximations concerning
$\modulus{\epn}\ $ and $ \modulus{\sigma}\ $ correspond to the phase convention to be introduced later. (We will choose a phase angle $\vartheta $, neglecting
$\modulus{\epn}^2 \ll 1$ and $\modulus{\den}^2 \ll 1$, such that
 $\sigma  =  -1/2$). 
\\
Eq. (\ref{eq:2.36}) shows that the
present two dimensional system can manifest time reversal violation only, if \Gz \ is not the null
matrix, i. e. if there is decay. However, since the absolute value $\modulus{\Gzba}$ does not
enter Eq. (\ref{eq:2.36}), the definition of time reversal invariance would stay intact if the
decay rates $\gs$ and $\gl$ would (hypothetically) become the same, contradicting ~\cite{wolfe1,wolfe}.\\
\AT \ has been measured ~\cite{pen2} not to vanish, $\AT \ne 0$. Since only the relative phase
of \Gzba \ and \Mzba , $arg(\Gzba)-arg(\Mzba)$, and not the absolute values, determines time
reversal violation, Eq. (\ref{eq:2.36}) does not give any prescription as to what extent the
violation should be attributed to M or to \Gz.

\subsubsection{\CPTz\ invariance}\label{cpt}

The \CPTz invariance of \Hwk \ requires the equality of the probabilities, for a kaon and for an antikaon,
to develop into themselves.
\begin{equation}\label{eq:2.40}
\ACPT  \equiv \frac{| \bra{\knb} \e^{-\i \Lz t} \ket{\knb }|^2 -
                    | \bra{\kn } \e^{-\i \Lz t} \ket{\kn  }|^2} 
                   {| \bra{\knb} \e^{-\i \Lz t} \ket{\knb }|^2 +
                    | \bra{\kn } \e^{-\i \Lz t} \ket{\kn  }|^2}\
\end{equation}
is thus a measure for a possible \CPTz violation.
We note (from Ref.~\cite{leebook}), indicated in Table \ref{tab:01}, that \CPTz invariance entails $\PLaa =\PLbb $,
or \ $\den = 0$. Using (\ref{eq:2.12}), we obtain, with $\modulus{\den}\ll 1$,
\begin{equation}\label{eq:2.41}
\ACPT = \frac{4\red \sinh (\frac{1}{2}\dg t )+ 4\imd \sin (\dm t) } 
{\cosh(\frac{1}{2} \dg t ) + \cos (\dm t ) } \;,
\end{equation}
and confirm that $\ACPT \neq 0$ at any time, i. e. \den $\neq$ 0, would contradict the property of
\Hwk \ to commute with \CPTz \;.
\subsection{Decays}\label{decays}

We assume that the creation, the evolution, and the decay of a neutral kaon can be considered as
a succession of three distinct and independent processes. Each step has its own amplitude with its
particular properties. It determines the initial conditions for the succeeding one. (For a refined
treatment which considers the kaon as a virtual particle, see ~\cite{sachsd, lipu},\ with a comment
in \cite{enz}).

The amplitude for a kaon characterized by $\ket{\psi_0}$ at $t=0$, decaying at time $t$,
into a state $\ket{\psi_f}$ is given by
\begin{eqnarray}
{\cal A}^f = \bra{\psi_f}\Hwk \ket{\psi} = \bra{\psi_f}\Hwk \ \e^{-\i \Lz t}\ket{\psi_0} & = &
\bra{\psi_f}\Hwk \ket{s'}\bra{s'}\e^{-\i \Lz t}\ket{s} \sbraket{s}{\psi_0}\nonumber\\
& = & {\cal A}^f_{s'}\ (\e^{-\i \Lz t})_{s's}\ \psi_s(0).\label{eq:2.42}
\end{eqnarray}
The sum over $s'$, $s$ \ includes all existent, unobservable, (interfering) paths.\\
It is
\begin{equation}\label{eq:2.43}
{\cal A}^f_{s'} = \bra{\psi_f}\Hwk \ket{s'}
\end{equation}
the amplitude for the instantaneous decay of the state with
strangeness $s'$ into the state $\ket{\psi_f}$, and $\psi_s(0) = \sbraket{s}{\psi_0}$,
($s=\kn$, $\knb$) are the components of $\psi_0$. The $(\e^{-\i \Lz t})_{s's}$ are taken from
(\ref{eq:2.12}) or (\ref{eq:2.26}).\\
The probability density for the whole process becomes
\begin{equation}\label{eq:2.44}
\modulus{{\cal A}^f}^2 =
D^f_{s't'} \ (\e^{-\i \Lz t})^*_{s's}(\e^{-\i \Lz t})_{t't} \ \psi_s^*(0)\psi_t(0),
\end{equation}
or, for an initial $\kn (s=1)$ or \ $\knb (s=-1)$, it becomes
\begin{equation}\label{eq:2.45}
R^f_s = D^f_{s't'} \ \e_{s't',s}\;.
\end{equation}
Here, the contributions from the kaon's time development ($\e_{s't',s}$) and those from the
decay process ($D^f_{s't'}$) are neatly separated.\\
We have set
\begin{eqnarray}
D^f_{s't'} & = & {\cal A}^{*f}_{s'}{\cal A}^f_{t'}\label{eq:2.46}\\
\e_{s't',s} & = & (\e^{-\i \Lz t})^*_{s's}(\e^{-\i \Lz t})_{t's}\;.\label{eq:2.47}
\end{eqnarray}
\subsubsection{Semileptonic Decays}\label{semilep}

For the instant decay of a kaon to a final state $(\ell\pi\nu)$ we
define the four amplitudes
\begin{equation}\label{eq:2.48}
{\cal A}^q_{s'} = \bra{\ell\pi\nu}\Hwk \ket{s'}, \ \ \ q, s' = \pm 1 \ ,
\end{equation}
with
$q$: lepton charge (in units of positron charge), \ $s'$: strangeness of the decaying kaon.\\
We assume lepton universality. The amplitudes in (\ref{eq:2.48})  thus must not depend
on whether $\ell$ is an electron or a muon.

Known physical laws impose constraints on these amplitudes:\\
The $\DS = \DQ$ rule allows only decays where the strangeness of the kaon equals the
lepton charge, ${\cal A}^q_q$\;, and $\CPTz$ invariance requires (with lepton spins ignored)
${\cal A}^{-q}_{-s'} = {\cal A}^{*q}_{s'}$ \cite{lee2}. The violation of these laws will be
parameterized by the quantities \xx \ , \xb \ , and \rey , \ posing \\\\
${\cal A}^{1}_{-1}$ = \ \xx \ ${\cal A}^{1}_{1}$, \ \  ${\cal A}^{-1}_{1}$ = \
{\xb}* ${\cal A}^{-1}_{-1}$, \ \
$(\modulus{{\cal A}^{1}_{1}}^2 - \modulus{{\cal A}^{-1}_{-1}}^2 ) \ = \ -2\rey \
(\modulus{{\cal A}^{1}_{1}}^2 + \modulus{{\cal A}^{-1}_{-1}}^2 )$, \\\\
or by \ \ $\xp = (x+\xb)/2$ \ and \ $\xm=(x-\xb)/2$ \ .\\\\
$\xp $ describes the violation of the $\DS=\DQ$ rule in \CPTz 
-invariant amplitudes, $\xm $ does so in \CPTz -violating amplitudes.\\

The four probability densities for neutral kaons of strangeness $s = \pm 1$, born at $t=0$, to decay at time $t$
into $\ell\pi\nu$ with the lepton charge $q = \pm 1$ are, according to (\ref{eq:2.45}),
\begin{equation}\label{eq:2.49}
R^q_s = D^q_{s't'} \ \e_{s't',s}\;,\ \ \ \mathrm{with}
\ \ \ D^q_{s't'}= {\cal A}^{*q}_{s'}{\cal A}^q_{t'} \ .
\end{equation}
The decay rates, proportional to $R^q_s$, are given in \cite{pr}.\\\\
We discuss here the asymmetries \AT \ and \ACPT \ with possible, additional symmetry breakings
in the decay taken into account. The completed expressions are denoted by \ $\AT (t)$
and $\ACPT (t)$.
Using Eqs. (\ref{eq:2.49}) and (\ref{eq:2.12}), we obtain\\
\begin{eqnarray}
\AT   (t ) & = & \frac{R^{1}_{-1} (t ) - R^{-1}_{1} (t )}{R^{1}_{-1} (t )  + R^{-1}_{1} (t )}
            \notag \\
            & = & \AT - 2\re (y+x_{-})
             +  2 \frac{\rexm (\e^{-\frac{1}{2}\dg t } -
                         \cos(\dm t )) + \imxp \sin(\dm t )}
                  {\cosh(\frac{1}{2} \dg t ) - \cos(\dm t )}\label{eq:2.50}\\
 &\ra & \AT -2\re (y+x_{-})\; \text{for} \; t \gg \ts , \label{eq:2.51}\\
 &  & \notag \\
\ACPT (t ) & = &\frac{R^{-1}_{-1} (t ) - R^{1}_{1} (t )}{R^{-1}_{-1} t ) + R^{1}_{1} (t )} 
                    \notag\\
            & = &\frac{4\re (\delta+x_- /2) \sinh (\frac{1}{2}\dg \t )+4\im (\delta+x_+/2) \sin (\dm \t) }
                         {\cosh(\frac{1}{2} \dg \t ) + \cos (\dm \t )}
                  + 2\rey\ \label{eq:2.52}\\
 &\ra & 4\red + 2\re (y+x_{-}) \;\text{for}\; t \gg \ts .\label{eq:2.53}
\end{eqnarray}
We compare (\ref{eq:2.52}) with (\ref{eq:2.41}), and we recognize that, besides the
additional term \rey\ , the new expression has the same functional behaviour, just
with the old variable $\delta$\ replaced by 
\begin{equation}\label{eq:2.54}
\den\ \ra \den\ + \frac{1}{2}(\rexm + \i\ \imxp).
\end{equation}
This shows that the analysis of a measurement, based on Eq. (\ref{eq:2.52}) alone, 
can not distinguish a possible \CPTz\ violation in the time development from possible violations due to \rexm\ or \imxp\ in the decay.\\\\
We note that the combination
\begin{equation}\label{eq:2.55}
\AT(t \gg \ts) + \ACPT(t \gg \ts)\ =\ \AT + 4\red \approx 4\ree + 4\red
\end{equation}
yields a particular result on the kaon's time evolution that is free from symmetry
violations in the decay~!
\\\\
Eqs. (\ref{eq:2.50}) to (\ref{eq:2.54}) are valid for $\modulus{\xx} \ll 1 \ ,
\modulus{\xb}\ll 1, \modulus{\rey} \ll 1$, and $\modulus{\den} \ll 1$. For the last
term in (\ref{eq:2.55}) we also assume $\modulus{\epn} \ll 1$ and $\sigma \approx -1/2.$\\

Additional information on ${\re (y+x_{-})}$\ is gained by measuring the charge asymmetry
\dell\ in the semi-leptonic decays of `old kaons` \kn\ (or \knb\ )\ $\ra\ \ell^{\pm}\pi\nu$.\\
For \ $t \gg \ts$\ we obtain, up to first order in
$\modulus{\epn},\modulus{\den},\modulus{y}, \mathrm{and} \modulus{x_-}\ ,$
(and for $\sigma = -\frac{1}{2}$)
\begin{equation}\label{eq:2.56}
\dell\ = \frac{R^1_s - R^{-1}_s}{R^1_s + R^{-1}_s}=
2\re (\epn-\del)-2\re (y+x_{-})
\end{equation}
independent of \ $s = 1$ or $-1$\ .\\\\
An equivalent asymmetry $A_\mathrm{S}$ has been derived from the same rates integrated at short decay
time \cite{buch,misc}
\[A_\mathrm{S} = 2\re(\epn + \del) -2\re(y+x_{-})\ .\]

\subsubsection{Decays to two pions - Decay rate asymmetries}\label{twopi}

The amplitude ${\cal A}^{\pi\pi}$\ for the decay of the kaon $\ket{\psi_0}$\ into two pions
$\ket{\pi\pi}$ is, following (\ref{eq:2.42}),
\begin{equation}\label{2.57}
{\cal A}^{\pi\pi}\ = 
\bra{\pi\pi}\Hwk \ket{s'}(\e^{-\i \Lz t})_{s's}\ \psi_s(0)\ .
\end{equation}
We express $\ket{s'}$\ by the eigenstates $\ket{\mathrm{K}_\kappa}$,\ \
$\kappa =$ L or S,  using (\ref{eq:2.17}) to (\ref{eq:2.20}),
(\ref{eq:2.22}) and (\ref{eq:2.23}),\ and find $W^{\kappa s}\ =
\sbraket{\widetilde{\mathrm{K}_\kappa}}{s}$\ and\\
\begin{equation}\label{2.58}
\ket{s'}(\e^{-\i \Lz t})_{s's}\ =\ \ket{\mathrm{K}_\kappa}
\e^{-\i \lambda_\kappa t}\ \sbraket{\widetilde{\mathrm{K}_\kappa}}{s},
\end{equation}
and regain~\cite{enz}
\begin{equation}\label{2.59}
{\cal A}^{\pi\pi}\ = {\cal A}^{\pi\pi}_{\kappa}\ \e^{-\i \lambda_\kappa t}\
\sbraket{\widetilde{\mathrm{K}_\kappa}}{s}\ \psi_s(0)
\end{equation}
where
\begin{equation}\label{2.60}
{\cal A}^{\pi\pi}_{\kappa}=\bra{\pi\pi}\Hwk \ket{\mathrm{K}_\kappa},\ \ \ 
\kappa = \mathrm{L\ or\ S}\ .
\end{equation}\\
The decay rates are \ $\propto {\modulus{{\cal A}^{\pi\pi}}}^2\ .$\ For a
$\kn$ at\ \ $t=0$, we obtain\\
\begin{equation}\label{eq:2.61}
R^{\pi\pi}_{1}\ = \ {\modulus{{\cal A}^{\pi\pi}_{\kappa}\ 
\e^{-\i \lambda_\kappa t}\ W^{\kappa 1}  }}^2\ =
{\modulus{{\cal A}^{\pi\pi}_{\mathrm{S}}\ \e^{-\i \lzs t}\ W^{11}+
{\cal A}^{\pi\pi}_{\mathrm{L}}\ \e^{-\i \lzl t}\ W^{21}  }}^2\ .
\end{equation}
To calculate this expression it is convenient to use the following approximate
eigenvectors, derived from Eqs. (\ref{eq:2.34},\ \ref{eq:2.35}), with
$\sigma= -\frac{1}{2}$, and valid to
first order in \epn\ and \den\ ,
\begin{eqnarray}
\ket{\ks} &=& N^S \left((1+\epn\ +\den) \ket{\kn} \ +\ (1-(\epn\ +\den))\ket{\knb}\right)\;,\label{eq:2.62}\\ 
\ket{\kl} &=& N^L \left((1+\epn\ -\den)\ket{\kn} \ -\ (1-(\epn\ -\den)) \ket{\knb}\right)\;,\label{eq:2.63}
\end{eqnarray}\\
where $N^S = N^L \approx\ 1/\sqrt{2}$\ .\\
From Eqs. (\ref{eq:2.21}), (\ref{eq:2.62}), (\ref{eq:2.63})  we derive
\begin{equation}\label{eq:2.64}
W=\left(W^{ij}\right)\approx\ \frac{1}{\sqrt{2}}\left ( \begin{array}{cc}
(1-\epn\ +\den) & \ \ \ (1+\epn\ -\den) \\ (1-\epn\ -\den) & -(1+\epn\ +\den) \end{array} \right ),
\end{equation}\\
and obtain, from (\ref{eq:2.61}), the rates of the decays $\kn \ra  \pi\pi$ and
$\knb \ra  \pi\pi$,\\
%%%%%%%%%%%%%%%%%%%%%%%%%%%%%%%%%%%%%%%%%%%%% 1. Juni 2004. HIER weiter:
\begin{eqnarray}\label{eq:2.65}
R^{\pi\pi}_{\pm 1}\ (t )            &=& \; \frac{[1\mp 2\re (\epn -\den )]}{2}\ \mathrm{\Gamma_S^{\pi\pi}}\ \notag\\
            & &  \times \bigl [\e^{-\gs t } + {\modulus{\eta_f} }^{2}\ \e^{-\gl t } 
            \pm\ 2{\modulus{\eta_f} }\ \e^{-\frac{1}{2}(\gs +\gl )t }\cos (\dm t - \phi_f )
            \bigr ]\,, 
\end{eqnarray}
where $\mathrm{\Gamma_S^{\pi\pi}}$\ is the partial decay width of $\ks \ra  \pi\pi $, and
where $\eta_f$\ equals
\begin{equation}\label{eq:2.66}
\eta_f\ = \frac{{\cal A}^{\pi\pi}_{\mathrm{L}}}
             {{\cal A}^{\pi\pi}_{\mathrm{S}}} = \modulus{\eta_f}\e^{\i\phi_f}\ .
\end{equation}\\
In (\ref{eq:2.65}), terms of the order $\modulus{\epn}^2$, $\modulus{\den}^2$, and
$\modulus{\eta_f\ \den}$ are neglected.\\
The term $2\re (\epn - \del )$ may be of special use in some measurements, for example in
the CPLEAR experiment, see Section \ref{sec:exp_1}.\\

A difference between the rates of the decays of \kn\ and of \knb\ to the \CPz\ 
eigenstates \pipa\ is an indication of \CPz\ violation.\\\\
 For its study, the following rate asymmetries have been formed:
\begin{eqnarray}
\ACPf   (t ) & = & \frac{R^{f}_{-1}\ (t ) - R^{f}_1\ (t )}{R^{f}_{-1}\ (t ) + R^{f}_1\ (t )}
                                                                  \notag \\  
            & = &2\re (\epn - \del ) - 2\frac{
             |\ita _f|\e^{\frac{1}{2}(\gs -\gl)t }\cos(\dm t - \phi_f)}
                  {1+ |\ita _f| ^2 \e^{(\gs -\gl) t }}\,,\label{eq:2.67}
\end{eqnarray}
where $f=\pipi$ (and $\ita _f=\itapm $) or $f=\pinn $ (and $\ita _f=\itaoo $). \\

\subsubsection{Decays to two pions - Isospin analysis}\label{twopiISO}

The final states $f$ may alternatively be represented by states with a definite total 
isospin $I$. 
The following three physical laws are expressed in terms of $I$, and can then be applied to the
neutral kaon decay \cite{kall}:
(i) the Bose symmetry of the two-pion states,
(ii) the empirical \ $\Delta I = \frac{1}{2}$ rule, and
(iii) the final state interaction theorem.
\\\\
The Bose symmetry of the two-pion states forbids the pion pair to have $I=1$.
The $\Delta I = \frac{1}{2}$ rule in turn identifies the dominant transition \ \kn\ (or \knb) $\ra \ket{I=0}
$.
The final state interaction theorem, together with the assumption of
\CPTz\ invariance, relates the amplitudes \ $\bra{I}\Hwk\ket{\kn}$ and
$\bra{I}\Hwk\ket{\knb}$. It then naturally suggests a parametrization of \CPTz\ violation in the decay
process.
\\
The relations
\begin{eqnarray}
\bra{\pipi} &=& \sqrt{2/3}\ \bra{I=0}+\sqrt{1/3}\ \bra{I=2}\label{brapipi}\\
\bra{\pinn} &=&\sqrt{1/3}\ \bra{I=0}-\sqrt{2/3}\ \bra{I=2} \label{brapinn}
\end{eqnarray}
transfer the implications of the laws mentioned to the observable final pion states.\\

We can now calculate the expressions of $\eta_f$ in (\ref{eq:2.66}) for \pipi\ and for \pinn , in terms
of the decay amplitudes to the states with \ $I=0,2$.\\
If we denote $\bra{I}\Hwk\ket{\kn}=A_I\ \e^{\i\del_I}$, then the final state interaction theorem
asserts, that, if \CPTz\ invariance holds, the corresponding amplitude for the antikaon decay is
$\bra{I}\Hwk\ket{\knb}=A^*_I\ \e^{\i\del_I}$. \ $\del_I$ is the phase angle for \pipa\ elastic
scattering of the two pions at their center of momentum energy.\\
Following \cite{barmin} we violate \CPTz\ invariance by intruding the parameters
$B_I\ ,\ I=0,2\ ,$
\begin{eqnarray}
\bra{I}\Hwk\ket{\kn}  &=& (A_I+B_I)\ \e^{\i\del_I}\label{IHK} \\
\bra{I}\Hwk\ket{\knb} &=& (A^*_I-B^*_I)\ \e^{\i\del_I}.\label{IHKb}
\end{eqnarray}
With the relations (\ref{brapipi}) to (\ref{IHKb}), and using (\ref{eq:2.62}), (\ref{eq:2.63}),
we find the amplitudes $\bra{\pipa}\Hwk\ket{\kls}$, and the observables
\itapm\ and \itaoo\ \cite{buch}.
We give them as follows
\begin{eqnarray}
\itapm\ &\approx& \left[\epn\ + \i\ \frac{\imaa }{\reaa }\right] -\ \den' + \epn' \label{epm}   \\
\itaoo\ &\approx& \itapm\ -\ 3\ \epn' \label{enn}
\end{eqnarray}
with
\begin{equation}
\den'\ = \ \left[\den\ - \frac{\reba}{\reaa}\ \right] \label{dprim}, \ \ \ \ \  \ \ \ \ \ \ \
\end{equation}
and with
\begin{eqnarray}\label{eprm}
\epsp\ &\equiv& \ \ \ \e^{i\phi_{\epsp}} \ \times \modulus{\epsp}\ \notag \\  
&=& 
\e^{i{{\phi}^{\mathrm{CPT}}_{\epsp}
}} \times \
\frac{1}{\sqrt{2}}
%\e^{\i (\frac{\pi}{2}+\del _2 -\del _0)}
\frac{\reac }{\reaa }
         \left(\left[\frac{\imac }{\reac } - \frac{\imaa }{\reaa } \right]
           -\i \left[\frac{\rebc }{\reac } - \frac{\reba }{\reaa } \right] \right),
\end{eqnarray}
where
\begin{eqnarray*}
{\phi}^{\mathrm{CPT}}_{\epsp} \ = (\frac{\pi}{2}+\del _2 -\del _0) \ .\\
& &
\end{eqnarray*}
In (\ref{epm}) and (\ref{enn}), terms of second order in the \CPz\ and \CPTz\ parameters, and
of first order in \epsp\ 
multiplied by \reac/\ \reaa, are neglected \cite{kall}.\\ %HIER 21. Okt. 04.

\subsubsection{Decays to two pions - With focus on \CPTz\ invariance}\label{twopiCPT}

Equation (\ref{eq:2.08b}) relates the decay amplitudes $A_0$ and $B_0$ to the
elements of \Gz.\\
With approximating the decay rates by the dominating partial rates
into the \pipa\ states with $I=0$, we have
\begin{eqnarray}
\frac{\imaa }{\reaa }&\approx& -\frac{\im\ \Gzab}{\dg}\label{imadreao} \\
\frac{\reba}{\reaa}&\approx& \frac{\Gzaa\ -\ \Gzbb}{2\dg}\ , \label{rebdreao}
\end{eqnarray}
and we recognize that 
\begin{equation}\label{dprimphase}
\den'\ =\ \modulus{\den'}\ \e^{\i(\fsw\ \pm\ \frac{\pi}{2})}
=\ \imd\ (-\tan{(\fsw) + \i})\ ,
\end{equation}
and
\begin{equation}
\frac{\reba}{\reaa}= \den\ -\den'= \red\ +\ \imd\ \times \tan{(\fsw)}\label{rbdra}\ ,
\end{equation}
and that the terms 
$ \left[\epn\ + \i\ \frac{\imaa }{\reaa }\right] \ $ and \ $\den'$ in (\ref{epm})
are out of phase by $\pi/2$. \ See \cite{buch}, as well as \cite{pr,dmdg}, with
\cite{misc} and \cite{NA}, for a justification of the neglect of the other decay modes .\
(In Eq. (\ref{dprimphase}) we had made use
of $\im \den' = \im \den$) \ .\\\\
Equation (\ref{rbdra}) relates the \CPTz\ violating amplitude, of the dominating decay process into
\pipa , with the \CPTz\ violating parameter in the time evolution. Originally, these processes have been treated as
independent.\\

$\epsp$ measures \CPz\ violation in the decay process. From (\ref{eprm})\ we see that it is independent of the parameters of the time evolution.
It is a sum of two terms. One of them is made exclusively of the decay amplitudes
$A_0$ and $A_2$\ , and is thus \CPTz\ invariant. The other one contains the amplitudes $B_0$ and
$B_2$\ , and is thus \CPTz\ violating. They are out of phase by $\frac{\pi}{2}$\ .\\

The value of the phase angle of the \CPTz\ respecting part,
${\phi}^{\mathrm{CPT}}_{\epsp} =(\frac{\pi}{2}+\del _2 -\del _0)$,  happens to be \cite{cola} roughly equal to \fpm\
\begin{equation}\label{fpmfsw}
{\phi}^{\mathrm{CPT}}_{\epsp} - \fpm \approx\ \mathrm{few\ degrees}\ . % \ (-4.2 \pm 1.5)^\circ\ .
\end{equation}
From the sine theorem, applied to the triangle of Eq. (\ref{enn}),
\begin{equation}\label{sinth}
(\fpm-\foo)=\frac{3 \modulus{\epsp}}{\modulus{\itaoo}}(\phi_{\epsp}-\fpm)\ ,
\end{equation}
we conclude that \CPTz\ invariance in the decay process to two pions requires
\begin{equation}\label{fpmmoo}
\modulus{\fpm - \ \foo} \ll \mathrm{few\ degrees}\ .
\end{equation}
(We have used \ $\modulus{\phi_{\epn'}-\fpm} \ll 1$ \ and $\modulus{\epsp}/\modulus{\itaoo} \ll 1$).\\

On the other hand, the measured difference \ $\fpm -\foo\ $\ limits the \CPTz\ violating parameters
in $\epsp$ as follows.\\\\
From (\ref{eprm}) and (\ref{sinth}) we obtain
\begin{eqnarray}
\frac{\reac }{\reaa }\left[\frac{\rebc }{\reac } - \frac{\reba }{\reaa }\right]
&=&
- \mathrm{Im}(\sqrt{2}\modulus{\epsp}\e^{\i(\phi_{\epsp}- {\phi}^{\mathrm{CPT}}_{\epsp})})\notag
\\
&\approx&
\frac{\sqrt{2}}{3}\ \mitaoo(\foo -\fpm)\ ,\label{reb}
\end{eqnarray}
and finally, with the use
of the estimate \ $\frac{\reac }{\reaa }\approx \modulus{\frac{A_2}{A_0}}$
\ (see \cite{pr, dedi}), we arrive at
\begin{equation}
\frac{\rebc}{\reac}=\frac{\reba}{\reaa}\ +\frac{\sqrt{2}}{3}\
\modulus{\frac{A_0}{A_2}} \mitaoo(\foo -\fpm)\ .\label{B2A2}\\
\end{equation}
This equation, and (\ref{rbdra}), relate the \CPTz\ violating expressions in \epsp\ with
the measured quantities.\\

For any two complex numbers \itapm\ and \itaoo\ with similar phase angles,
$\modulus{\fpm - \foo} \ll 1$, we have to first order
\begin{equation}\label{epmmoo}
\itapm \mp \itaoo\ = \ \{\ \mitapm \mp\mitaoo\ \pm
\ \i \mitaoo(\fpm -\foo)\ \}\ \e^{\i \fpm} \ .
\end{equation}

If we allow for the approximation \ $\e^{\i(\fpm-\phi_\epn)}=1$, we obtain
from (\ref{enn}) and (\ref{epmmoo})
\begin{equation}\label{reepe}
\re{(\epsp/\epn)}=\frac{1}{3}\left(1-\frac{\mitaoo}{\mitapm}\right)
\frac{\mitapm}{\modulus{\epn}}\ .
\end{equation}
This quantity has been determined by a measurement of 
$\modulus{\itaoo/\ \itapm}^2$. See Section \ref{sec:exp_1}.\\\\
We apply (\ref{epmmoo}) to
\begin{equation}\label{et}
\ita\ \equiv \frac{2}{3}\itapm +\frac{1}{3}\itaoo
\end{equation}
and obtain \ (for $\modulus{\fpm - \foo} \ll 1$)
\begin{equation}\label{etx}
\ita = \ \{\ (\frac{2}{3}\mitapm\ +\frac{1}{3}\mitaoo)\ -
\ \frac{\i}{3} \mitaoo(\fpm -\foo)\ \}\ \e^{\i \fpm}=
\modulus{\ita}\e^{\i\phi_{\ita}}
\end{equation}
with
\begin{equation}\label{pheta}
\phi_{\ita}=\frac{2}{3}\fpm +\frac{1}{3}\foo\ .
\end{equation}
For the measured values see Section \ref{meas}.\\\\
%%%%%%%%%%%%%%%%%%%%
We eliminate now \epsp\ from (\ref{epm}) and (\ref{enn}):
\begin{equation}\label{path}
\epn\ + \i\ \frac{\imaa }{\reaa } - \den +\frac{\reba}{\reaa}\ =\ \ita \ ,
\end{equation}
and simplify this equation by setting the arbitrary phase angle $\vartheta$\ in
(\ref{eq:2.10}) to have
\begin{equation}
\Gzab\ =  \mathrm{real}\ , \label{phco}
\end{equation}
making \imaa\ negligible
\begin{equation}
\imaa\ \approx\ \ \ \ 0\ .\ \ \label{aconv}
\end{equation}
This allows one, given \ree\ $> 0$, to fix the phase angle of \epn\ to \fsw 
\begin{equation}
\epn\ =  \modulus{\epn}\ \e^{\i \fsw} \label{efsw}
\end{equation}
and to set $\sigma = -1/2$ 
(having neglected $\modulus{\epn}^2\ll 1$, $\modulus{\den}^2 \ll 1$).\\\\
\epn\ has now obtained the property to vanish, if \Tz\ invariance holds.\\\\
The term $\frac{\reba}{\reaa}\ $ in (\ref{path}) remains, as seen from (\ref{rebdreao}),
uninfluenced by the phase adjustment. We then obtain
\begin{equation}\label{pathconv}
\epn\ - \den +\frac{\reba}{\reaa}\ =\ \ita \ .
\end{equation}
Under \CPTz\ invariance, this relation would be
\begin{equation}\label{pathcpt}
\epn\ = \ita\ = \frac{2}{3}\itapm +\frac{1}{3}\itaoo\ ,
\end{equation}
with
\begin{equation}\label{fecpt}
\phi_{\epn}\ \equiv\ \fsw\ =\phi_{\ita}\ \approx\ \fpm\ .
\end{equation}
Applying (\ref{epmmoo}) to \ $\epn - \ita\ $\ in (\ref{pathconv}) yields, with (\ref{dpardsen}),
\begin{eqnarray}
\dpar + \i \dper\ &=& \den\ \e^{-\i \fsw}\notag \\
&=&  \modulus{\epn}-\mita\ +
\ \i \modulus{\ita}(\fsw-\frac{2}{3}\fpm -\frac{1}{3}\foo)+
\ \frac{\reba}{\reaa} \ \e^{-\i \fsw}\ ,
\end{eqnarray}
and with (\ref{eq:2.32}, \ref{rbdra})
\begin{eqnarray}
\Mzaa - \Mzbb
&=& 2\modulus{\dlz}\dper \notag \\
&=& 2\modulus{\dlz}
\{\modulus{\ita}
(\fsw- \phi_{\ita})
\ -\frac{\reba}{\reaa}\ \sin{(\fsw)}\}\notag \\ 
&=& 2\modulus{\dlz}
\{\ \modulus{\ita}
(\fsw- \phi_{\ita})
- (\ \red + \imd \tan(\fsw)\ )\sin{(\fsw)} \}. \label{mmmb} \ \ \ \ \ \ \ \
\end{eqnarray}
All terms on the rhs are deduced from measurements. \\\\
\CPTz\ invariance requires \ $\Mzaa - \Mzbb = \den = 0$,\ and thus
$\ (\fsw- \phi_{\ita}) = 0$~.\\
The comparison of the values of \ $\fsw $ and of $\ \phi_{\ita}\equiv \frac{2}{3}\fpm +\frac{1}{3}\foo$ ,
done in Section \ref{meas},
will confirm \CPTz~invariance.\\

Finally, combining (\ref{pathconv}) with the semileptonic charge asymmetry (\ref{eq:2.56}),
we obtain
\begin{equation}\label{etdl}
\re{\ (\frac{2}{3}\itapm\ +\frac{1}{3}\itaoo)}\ -\frac{\dell}{2}
=\frac{\reba}{\reaa}+\re{(y+x_-)} \ .
\end{equation}
The terms on the rhs are \CPTz\ violating. %HIER 2. Mal, 1. 11. 2004.
%%%%%%%%%%%%%%%%%%%%%%%%%%%%%%%%%%%%%%
\subsubsection{Unitarity}\label{uni}
%%%%%%%%%%%%%%%%%%%%%%%%%%%%%%%%%%%%%%

The relation between the process of decay of the neutral kaon and the non-hermitian
part \Gz\ of \Lz\ , expressed in the Eqs. (\ref{eq:2.07}) and (\ref{eq:2.08b}) offers
the study of certain symmetry violations of \Hwk\ .\\

The terms in the sum for \Gzij\ with \ $\alpha\ \neq\ \alpha'$,\ \ \ 
$\sum_{\beta}\sbraket{\alpha|\Hwk }{\beta}\sbraket{\beta|\Hwk }{\alpha^{\prime}}$,\ \
express simultaneous transitions from different states $\ket{\alpha'}\ \neq\ \ket{\alpha}$\
to one single final state $\ket{\beta}$. If the quantum numbers \ $\alpha'\ \neq\ \alpha$\
represent conserved quantities, then the transitions to the single final state
$\ket{\beta}\ $ would violate the conservation law in question.\\

Based on the fact that the occurence of decay products requires a corresponding decrease
of the probability of existence of the kaon, the following relation \cite{bell} holds 
\begin{equation} \label{bell1}
\ree - \i\ \imd = \frac{1}{2\i\dm + \gs  + \gl} \times \sum  \langle f | \Hwk | \kl 
\rangle \langle f | \Hwk | \ks \rangle ^*,
\end{equation}
where the sum runs over all the final decay states $f$. \\ 

This equation has several remarkable aspects:\\
(i) It is of great generality. Having admitted the time evolution to be of the general form
(\ref{eq:2.04}), its validity is not restricted to perturbation theory or to
\CPTz\ invariance.\\
(ii) The left-hand side (lhs) refers uniquely to the symmetry violations in the time evolution of the
kaon, before decay, while the right-hand side (rhs) consists of the measurements, which include the
complete processes.\\
(iii) The rhs is dominated by the decays to \pipi\ and \pinn\ . The other processes enter with
the reduction factor \gs /\ \gl \ $\approx 580$, and, given their abundances, can often be
neglected \cite{pr,misc,NA}. What remains of the sum, is approximately \ \gs\ \itapp\ , with \ \itapp\
defined by
\begin{equation}\label{etpp}
\itapp\ =\mitapp\ \e^{\i \phi_{\pipa}}
=\itapm\ \mathrm{BR^S _{\pip\pim}}+\itaoo\ \mathrm{BR^S _{\pin\pin}}
\approx\ \frac{2}{3}\ \itapm\ + \frac{1}{3}\ \itaoo
\equiv\ \eta\ ,
\end{equation}
where the BR denote the appropriate branching ratios.\\
The measurements show
\ $\itapp \approx \itapm \approx \itaoo \approx \eta $.\\
(iv) The factor $1/(2\i \dm + \gs + \gl)$ may be approximated by \ 
$\cos(\fsw)\ \e^{-\i \fsw}/\gs$\ ,
and thus
\begin{equation}\label{bellapp}
\ree - \i \imd \approx\ \mitapp \cos(\fsw)\ \e^{-\i (\fsw -\phi_{\pipa})}.
\end{equation}
Besides the results from the semileptonic decays, it is thus the phase in the decay to
\pipa\ which reveals the extent, to which the \CPz\ violation
in the time development, is a \Tz\ violation, and/\ or a \CPTz\ violation.\\ Since the
measurements yield $\phi_{\pipa}\approx \fsw$, the \CPz\ violation is a \Tz\ violation
with \CPTz\ invariance.\\
We will later consider the hypothetical outcome \ $\phi_{\pipa}\approx \fsw + \pi/2$, which would signal
a \ \CPz\ violation with \ \Tz\ invariance and \ \CPTz\ violation.\\
From Eq. (\ref{bellapp}) we note (since \ $\modulus{\fsw\ - \fpp} \ll 1$ and \ $\fpp \approx \phi_\eta$)
\begin{eqnarray}
\ree\ &\approx\ & \mitapp\ \cos(\fsw)\label{reappr} \\
\imd\ &\approx\ & \mitapp\ \cos(\fsw)\ (\fsw\ - \phi_\eta)\ \approx\ \ree\ (\fsw\ - \phi_\eta)\label{imappr}\ .
\end{eqnarray}
(v) It is straight forward to formally recognize that the measured value of
$\phi_{\pipa}\approx \fsw$ is in contradiction with $\Tz ^{-1} \ \Hwk\ \ \Tz = \Hwk\ $. However,
the experiment which measures $\phi_{\pipa}$ does not seem to involve any comparison of a process,
running forward, with an identical one, but running backward.\\
(vi) An analog mystery concerns \CPTz\ invariance, as the measurement of $\phi_{\pipa}$ also does
not obviously compare \CPTz\ conjugated processes.\\
(vii) Since the result of (\ref{bell1}) is independent on possible symmetry violations in the decay,
while the charge asymmetry \dell\ (\ref{eq:2.56}) contains such violations in the form of
$\re{(y+x_-)}$, we may combine Eqs. (\ref{eq:2.56}) and (\ref{bell1}) in view to evaluate this term.
\\\\
Details of the application of (\ref{bell1}) are found in \cite{pr,blois,phen2,NA}.
%%%%%%%%%%%%%%%%%%%%%%%%%%%%%%%%%%%%%%%
\subsection{\Tz\ violation and \CPTz\ invariance measured without assumptions on the decay processes}\label{ohneetwas}
%%%%%%%%%%%%%%%%%%%%%%%%%%%%%%%%%%%%%%%
Some following chapters explain the measurements of \AT ($t$)\ and of \ACPT ($t$),
performed in the CPLEAR experiment at CERN.
These quantities are designed as comparisons of processes with initial and final states
interchanged or, with particles replaced by antiparticles, and, as already shown above,
they are intimately related to the symmetry properties of \Hwk\ .
However, they include contributions from possible violations of symmetries in the decays, such as
of \CPTz\ invariance or of the $\Delta S = \Delta Q$ rule.\\
We will evaluate the sizes of \rey, $\re{(y+\xm)}$, and \imxp\ , which constrain such violations
to a negligible level.

As a preview, we recognize that the functions \AT ($t$)\ and \ACPT ($t$)\ consist of a
part which is constant in time, and of a part which varies with time. The varying parts are
rapidly decaying, and they become practically unimportant after $t\ \widetilde{>}\ 5\tau_S $.
The two parts depend differently on the unknowns.\\ The constant parts already, of \AT ($t$) and of
\ACPT ($t$), together with \dell\ , constitute three equations which show the feasibility
to evaluate \AT\ , \red\ , and $\re (y+x_{-})$, and thus to determine an \AT\ , which is independent
from assumptions on \CPTz\ symmetry or from the $\Delta S = \Delta Q$ rule in the semileptonic
decays.\\ This \AT\ depends uniquely on the time-reversal violation in the $evolution$ of the
kaon, and it is thus the direct measure for $\Tz ^{-1} \ \Hwk\ \ \Tz \neq\ \Hwk $ searched for.\\
The \red\ in turn is a limitation of a hypothetical violation of
$(\CPTz )^{-1} \ \Hwk\ \ (\CPTz ) = \Hwk $ , also uniquely concerning the time evolution.

\subsection{Time reversal invariance in the decay to \ $\pi\pi\e^+\e^-$ \ ?}\label{ppee}

The decay of neutral kaons into $\pip\pim\gamma\ $ has been studied in view of gaining
information on symmetry violations which could not be obtained from the decay into \pip\pim\ ,
especially on \CPz\ violation of a different origin than the kaon's time evolution 
\cite{dopo,sewo,coka,linv}.

The existence of a sizable linear $\gamma$ polarization and the possibility of its detection by internal
pair conversion \cite{dali,krwa}, as well as the presence of a \Tz\ noninvariant term have been
pointed out in \cite{dopo}.

Experiments have detected the corresponding \Tz- odd intensity modulation with respect to the
angle between the planes of (\pip\pim) and ($\e^+\e^-$) in the decay \kl \ra \pip\pim$\e^+\e^-$
\cite{ktev,wahl}. As expected, the decay \ks \ra \pip\pim$\e^+\e^-$ shows isotropy \cite{wahl}.
The data confirm a model \cite{svl}, where, as usual, the \CPz\ violation is also \Tz\ violating,
and localized entirely in the time evolution of the kaon.

We discuss this result here, because its interpretation as a genuine example of a time-reversal
noninvariance \cite{svl}, or as a first direct observation of a time asymmetry in neutral kaon
decay \cite{fnew} has triggered critical comments
\cite{gau1,elma,wolfe,gerberEPJ},
with the concurring conclusion that, in the absence of final state interactions, the KTEV experiment
at FNAL would find the same asymmetry when we assume there is no \Tz\ violation \cite{wolfe}.

The enigma is explained in Ref. \cite{elma}, whose authors remind us that a \Tz\ - odd term
does not involve switching `in` and `out` states, and so is not a direct probe of \Tz\ violation.

As a complement, we wish to show that the model of \cite{svl} is an example, that a
\Tz\ odd effect may well persist within \Tz\ invariance, even in the absence of final state
interactions.

The $\gamma$ radiation of \kls\ \ra\ $\pi\pi\gamma$ has basically only two contributions, allowed
by gauge invariance (up to third order in momenta) \cite{linv}, which we refer to as E and as M.
They have opposite space parity, and their space parity is opposite to their \CPz\ parity. Since
\CPz($\pi\pi$) $= +1$, we have \CPz($\pi\pi\gamma$) $= -\Pz(\gamma$).\\
In detail
\begin{equation}\label{cpppg}
\CPz(\pi\pi\gamma) = - \Pz(\gamma)=
\left\{ \begin{array}{c} 
                         +1 \ \ \mathrm{E \ \ radiation}\\
                     \ \ -1 \ \   \mathrm{M \ \ radiation}\ .
\end{array} \right.
\end{equation}
We thus see that the decays from the \CPz\ eigenstates 
$\mathrm{K}_1$ \ra\ $\pi\pi\gamma_\mathrm{E}$ \ and \
$\mathrm{K}_2$ \ra\ $\pi\pi\gamma_\mathrm{M}$
are allowed within \CPz\ invariance. A signal for \CPz\ violation is (e. g.) the simultaneous
occurence of E \ and M radiation from a decaying old \kn .

The variety of radiations is due to scalar factors, which multiply E and M, which are not
determined by gauge invariance. They have to be measured or calculated from models.

The experiment \cite{taur} at CERN has identified the $\gamma$ radiation from
\ks\ \ra\ $\pi\pi\gamma$ as pure low energy bremsstrahlung. This determines the scalar factor for 
the E radiation to be the one from soft photon emission, and fixes the phase of the 
$\pi\pi\gamma$ amplitude to be the one of the $\pi\pi$ amplitude \cite{low}. This will become an
important ingredient for the model \cite{svl} below.

The experiment \cite{carr} at BNL has found two similarly strong components in the $\gamma$ radiation
of \kl\ \ra\ $\pi\pi\gamma$, (i) the bremsstrahlung, which is now \CPz\ suppressed, and (ii) the
M radiation, whose energy spectrum is compatible with a rise $\propto$ E$^3_{\gamma}$. We can thus
naturally expect that there is a value of the gamma ray energy E$_{\gamma}$\ , where the two
components have equal intensity, and where thus the radiation shows a marked polarization due to interference.

The model of \cite{svl} calculates this polarization, and finds that the correponding observable
asymmetry in the distribution of the angle between the planes (e$^+$e$^-$) and (\pip\pim) is of
the form
\begin{equation}\label{A}
A \propto \mitapm\ \sin(\fpm + \phi_\mathrm{FSI})\ ,
\end{equation}
where $\phi_\mathrm{FSI}$\ is determined by the final state interaction theorem,
and where
\fpm\ , as mentioned above, is fixed by the soft photon emission law.

In order for (\ref{A}) to be a direct manifestation of \Tz\ violation, we would like to see
$A$ disappear, if \Tz\ violation is switched off, while \CPz\ violation remains present.\\
Doing this, following \cite{wolfe} or (\ref{bellapp}), we see that $A$ persists, if we set
\begin{equation*}
\mitapm\ \neq 0\ , \ \fpm\ = \ \fsw\ + \frac{\pi}{2}\ ,\ \ \mathrm{and}
\ \ \phi_\mathrm{FSI} = 0 . \notag
\end{equation*}\\
The model presents thus a \Tz -odd observable which, in the absence of final state interactions,
takes still a finite value when \Tz\ invariance holds.

\subsection{Pure and mixed states}\label{pure}

Until now we have implicitly assumed that a single neutral
kaon represents a {\it pure} state, described by a state vector 
whose components develop in time coherently according to 
Eq. (\ref{eq:2.04}).\\ 
The {\it ensemble} of kaons in a beam is formed most often in
individual reactions, and the kaons develop in time 
independently of each other. This ensemble represents a
{\it mixed} state, and its description needs two state vectors
and the knowledge of their relative intensity.

It is a deeply rooted property of quantum mechanics that the pure state of an 
isolated particle does not develop into a mixed state. Such a (hypothetical) 
transition would entail a loss of phase coherence of the amplitudes, and thus 
become detectable by the weakening of the interference patterns. It would also 
violate \CPTz\ invariance \cite{wald,page}, but in a different way than 
described in previous sections.

The various interference phenomena shown by neutral kaons 
have already been used as a sensitive detector in the search 
for coherence losses. As analysis tool in this search the density-matrix
formalism used to describe mixed states seems appropriate.

\subsubsection{Density matrix description}\label{dens}
The time development of mixed states, and the results of measurements can be compactly
described by the positive definite (hermitian) density matrix $\rho (t)$
\cite{vneu,pauli,fano}.

All density matrices (in Quantum Mechanics, QM) develop in time in the same way, i. e. like those of pure states. A pure
state $\psi$ (with components $\psi_\kappa (t), \ \kappa = $ \kn, \knb) has the density matrix
$\rho (t) = \psi \psi^\dagger$ (with components $\psi_\kappa (t) \psi^\ast_{\kappa'} (t)$).\\
Density matrices thus develop according to 
$\rho (t) = \e^{-\i\Lz t} \psi_0 \ (\e^{-\i\Lz t} \psi_0)^\dagger$, or, denoting
$U(t) = \e^{-\i\Lz t}$ and $\rho (0) = \psi_0 \psi^\dagger_0$, like
\begin{equation}\label{ruru}
\rho (t) = U(t)\rho (0) U^\dagger (t)\ , \ \ t \ge 0\ , \ \ \ \ \ \mathrm{(QM)}.
\end{equation}
The form of (\ref{ruru}) grants the conservation of \\
(i)  \  the \ $rank$,\ \ \ \ and\\
(ii)  the \ $positivity$ of $\rho (t)$.\\\\
Since the pure states have (by construction) density matrices of \ $rank = 1$, the development (\ref{ruru})
keeps pure states pure.\\
Since a matrix  $\rho$\ of $rank = 1$ can always be written as a tensor product of two vectors,
$\rho = \psi {\psi}^\dagger$, the developments by Eqs. (\ref{eq:2.04}) and (\ref{ruru}) become
equivalent for pure states.

Eq. (\ref{ruru}) does not automatically conserve the trace, tr\{$\rho\}$, since $U(t)$ is not unitary. In order to avoid that the probability of existence of a neutral
kaon does exceed the value one, we separately require, as a property of $U(t)$, that
\begin{equation}\label{trac}
1 \ge \mathrm{tr}\{\rho (0)\} \ge \mathrm{tr}\{\rho (t \ge 0)\}\ .
\end{equation}

The outcome of measurements can be summarized as follows:\\
The probability $W$ for a neutral kaon with the density matrix $\rho$\ , to be detected by an
apparatus, tuned to be sensitive to neutral kaons with the density matrix $\rho_f$ , is 
\begin{equation}\label{Wtrr}
W = \mathrm{tr}\{\rho_f\ \rho\} \ .
\end{equation}

\subsubsection{Transitions from pure states to mixed states ?}\label{mixd}

It has been suggested \cite{hawk} that gravitation might influence the coherence of wave functions and
thereby create transitions from pure states to mixed states. These could look like a violation
of Quantum Mechanics (QMV). In order to quantify observable effects due to such transitions, the authors of
\cite{ehns} have supplemented the Liouville equation of motion by a QM-violating term, linear in the density
matrix. They have provided relations of the QMV parameters to a set of observables, to which the
CPLEAR experiment has determined upper limits.\\
Our description includes extensions, specifications, and generalizations of the formalism.

In order to characterize the effects of QMV, it has been successful to introduce the Pauli
matrices $\sigma^\mu , \mu = 0 , \cdots , 3$ (with $\sigma^0 =$ unit matrix) as a basis for the
density matrices $\rho (t) \equiv \rho = R^\mu \sigma^\mu$, and
$\rho (0) \equiv \rho_0 = R_0^\mu \sigma^\mu$. $R^\mu$ and  $R_0^\mu$ are reals.
We note that the determinant equals $|\rho| = R^\mu R_\mu$, and find from (\ref{ruru}) that
\begin{equation}\label{lore}
R^\mu R_\mu = \|U\|^2 R_0^\mu R_{0\mu}\ .
\end{equation}
$\|U\|$ is the absolute value of det$(U)$. Indices are lowered with the $4 \times 4$ matrix \
$g =(g_{\mu \nu})=(g^{\mu \nu})$, with
$g^{00} = - g^{11} = - g^{22} = - g^{33} = 1 \ ,\ 
g^{\alpha \beta} = 0$ for $\alpha \neq \beta $\ . \\\\
Eq. (\ref{lore}) is a multiple of a $Lorentz$ transformation~\cite{geprl} between the
four-vectors $ R \equiv (R^\mu)$ and $ R_0 \equiv (R_0^\mu)$. We write its matrix as the
exponential $\e^{Tt}$, and the transformation as $R= \e^{Tt}R_0$,
where $T = (T^{\mu\nu}) = T^{00} {\bf 1}_{4 \times 4} + L$ , and where $L$ is an element of 
the Lie algebra of the Lorentz transformations, and therefore satisfies 
\begin{equation}\label{glgml}
g L g = - L^\mathrm{T} \ .
\end{equation}
$()^\mathrm{T}$ denotes the transpose of ().

Eq. (\ref{glgml}) characterizes the quantum mechanical time evolution, which conserves the
purity of the states, now expressed by $R^\mu R_\mu =0$ \ ($R$ light-like). An
obvious way to let the formalism create transitions from pure states to mixed states is, to
supplement the matrix $T$ above with a matrix $X$, which is $not$ an element
of the Lie algebra of the Lorentz transformations, e. g. which satisfies
\begin{equation}\label{gxgpx}
g X g = + X^\mathrm{T} .
\end{equation}
We will explicitly use
\begin{equation}\label{X}
X = \left(
\begin{array}{cccc}
0 & S^1 &	S^2	& S^3 \\		 
- S^1	& - J^1	& D^3	& D^2 \\		
- S^2 & D^3	& - J^2	& D^1	\\	 
- S^3	& D^2	& D^1	& - J^3 
\end{array}
\right) \ .
\end{equation}
The time evolution is now generated by the matrix
\begin{equation}\label{T}
T = T^{00} {\bf 1}_{4 \times 4} + L + X ,
\end{equation}
which is just a general \ $4 \times 4$ matrix. From (\ref{glgml}) and (\ref{T}) we see that QM
has 7 parameters, and (\ref{X}) shows that QMV has 9 parameters.

The probability for a neutral kaon, characterized by the four-vector
$R_0$ at time $t=0$ , to be detected, by an apparatus set to be sensitive to $R_f$ , is
\begin{equation}\label{W}
W(t) = \mathrm{tr}(\rho_f \rho (t)) = 2 R^\mu_f (\e^{Tt})^{\mu\nu} R_0^\nu \ \equiv
2\ \e^{T^{00}t} R_f^\mathrm{T} \e^{(L+X) t} R_0\ .
\end{equation}
Using~\cite{fano,feyn}
\begin{equation}\label{elxt}
\e^{(L+X) t}  = \e^{Lt}\ \e^{{\bf D}(t, - L, X)} = \e^{{\bf D}(t, L, X)}\ \e^{L t} =
({\bf 1}+{\bf D}+ \cdots + \frac{1}{n!}{\bf D}^n + \cdots)\ \e^{Lt}
\end{equation}
with
\begin{equation}\label{D}
{\bf D} \equiv {\bf{D}} (t, L, X) = \int\limits^t_0 d\tau\ \e^{L\tau} X \e^{-L\tau} =
-{\bf{D}} (-t, -L, X)\ , 
\end{equation} 
we obtain to first order in $X$
\begin{eqnarray}
W(t) &=& W_{QM}(t)+W^{(1)}_{QMV}(t)\label{Wapp} \\
W_{QM}(t) &=& 2\ \e^{T^{00}t}\ R_f^\mathrm{T}\ \e^{Lt}\ R_0 \label{WQM} \\
W^{(1)}_{QMV}(t) &=&
              2\ \e^{T^{00}t}\ R_f^\mathrm{T}\ {\bf D}\ \e^{Lt}\ R_0 \ .\label{WQMV}
\end{eqnarray}
%patch for 2.6.2

%%%%%%%%%%%%%%%%%%%%%%%%%%%%%%%%%%%%%%%%%%%%%%%%%%%%%%%%%%%%%%
These equations have an evident interpretation:  $R_0$ describes the kaon beam, $R_f$ describes
the detector, $\e^{Lt}$ describes the regular time evolution, and ${\bf D}$ describes
the decoherence.
$W_{QM}(t)$ represents the result within QM.\\
In order to calculate the expressions (\ref{WQM}), (\ref{WQMV}), we need $T^{00}$ and $L$.
\\
%%%%%%%%%%%%%%%%%%%%
For \ $T = (T^{\mu\nu}) = T^{00} {\bf 1}_{4 \times 4} + L$ \ we obtain, in terms of \Lz\ ,
\begin{equation}\label{TL}
T^{\mu\nu} = \im (\mathrm{tr}\{\sigma^\mu \Lz\ \sigma^\nu  \})\ ,
\end{equation}
and
\begin{equation}\label{T00}
T^{00} = -(\gs + \gl)/2 \ ,
\end{equation}
and, to the lowest order,
\begin{equation}\label{Lapp}
L = \left( 
\begin{array}{cccc}
0 &	- \dg/2	& 0 &	0 \\
- \dg/2 & 0	& 0	& 0 \\
0	& 0 & 0 &	\dm	\\
0 &	0 &	- \dm	& 0  \\
\end{array} \right)   \ \ ,
\end{equation}
and thus
\begin{equation}\label{eLap}
\e^{Lt} = \left(									
\begin{array}{cc}
\sigma^0 \cosh (\frac{\dg}{2} t) - \sigma^1 \sinh (\frac{\dg}{2} t) & (0) \\
(0) & \sigma^0 \cos (\dm t) + i \sigma^2 \sin (\dm t)
\end{array}
\right) \ .
\end{equation}
(\ref{eLap}) will be used in Eqs. (\ref{elxt}), (\ref{D}) as the starting point for the calculation of $W(t)$, to any order
in the QMV parameters in $X$, or in the small parameters in $L$.
(We note in passing that
$\ -L^{02}+ \i L^{13} = 2\epn\ \dlz = \i L^{13}$ (having used$ \ \ \phi_{\epn}\ = \fsw$) , and that \ $L^{12}- \i L^{03}=2\den\ \dlz $).\\

We are now able to list all possible experiments to search for QMV \cite{geprl}. The four dimensions make $R_0$
and $R_f$ capable to define four independent beams and four independent measurements, to give a total of 16 experiments.

It is a fortunate fact that ${\bf D}$ in Eq. (\ref{WQMV}) introduces a sufficiently rich time
dependence, which enables the existence of a specific set of four experiments \cite{geprl,geEPJCent},
that allows one to determine all 9 QMV parameters of (\ref{X}).\\

As an example we study the influence of QMV on the decay of an initially
pure \kn\ into two pions.
The expression for $R^{\pi\pi}_{+1}(t )$ in (\ref{eq:2.65}) corresponds to $W_{QM}(t)$.
We calculate the modification $W^{(1)}_{QMV}(t)$ due to QMV.\\\\
First, we verify that $R_f^\mathrm{T} = \frac{1}{2} (1\ 1\ 0\ 0)$ represents the state
$\bra{\mathrm{K}_1}$. We note that $\frac{1}{2} (1\ 1\ 0\ 0)\hat{=}\
\frac{1}{2}(\sigma^0+\sigma^1)
=\frac{1}{\sqrt{2}} \left ( \begin{array}{c} 1 \\ 1 \end{array} \right )
\frac{1}{\sqrt{2}} (1\ 1)$, and recognize that the last term is just the tensor product of
$\psi_{\mathrm{K}_1} = 
\frac{1}{\sqrt{2}} \left ( \begin{array}{c} 1 \\ 1 \end{array} \right )$
with $\psi^\dagger_{\mathrm{K}_1}$.\\
In the same way we find: 
$\mathrm{K}_2 \hat{=}\ \frac{1}{2} (1\ -1\ 0\ 0),\ $
$\kn          \hat{=}\ \frac{1}{2} (1\  0\ 0\ 1),\ $ and
$\knb         \hat{=}\ \frac{1}{2} (1\  0\ 0\ -1). $ \\\\
With \ $R_0 = \frac{1}{2}(1\ 0\ 0\ 1)^\mathrm{T} \hat{=} \ket{\kn}$ and \
${\bf D} = J^1 \ {\bf D} (t, L, \frac{\partial}{\partial J^1}X)$, we obtain from (\ref{WQMV})
\begin{equation}\label{WQMVpi}
W^{(1)}_{QMV}(t) = \frac{J^1}{4\dg} \ \e^{-\gl t} \ \ \ \ \ \mathrm{for} \ \ t \gg \ts\ .
\end{equation}
Eqs. (\ref{Wapp}) to (\ref{WQMV}) together with (\ref{WQMVpi}) and (\ref{eq:2.65}) yield the modified expression for the
decay rate
\begin{equation}\label{RQMVpipi}
R^{\pi\pi}_{+1\ QMV}\ (t ) \propto
\bigl [a_{QMV}\ \e^{-\gs t } + {\modulus{\eta_{QMV}} }^{2}\ \e^{-\gl t } 
            + \ 2{\modulus{\eta} }\ \e^{-\frac{1}{2}(\gs +\gl )t }\cos (\dm t - \phi )
            \bigr ]
\end{equation}
where
\begin{equation}\label{etaQMV}
{\modulus{\eta} }^{2} \ra\ {\modulus{\eta_{QMV}} }^{2}={\modulus{\eta} }^{2}+\frac{J^1}{2\dg}\ \ .
\end{equation}
The modification $1 \ra a_{QMV}$ of the short-lived term in (\ref{RQMVpipi}) will not be
considered further.\\
The outstanding features are the modification of ${\modulus{\eta} }^{2}$ in the long-lived term,
in contrast to ${\modulus{\eta}}$ in the interference term, and the fact that the $first$ order term
of QMV, $J^1$, combines with the $second$ order term of the \CPz\ violation, as seen in (\ref{etaQMV}).
This will allow one to determine an especially strict limit for $J^1$ \cite{ehns,huet}. \\
The parameters $J^2, J^3$, and $D^1$ are not present in (\ref{RQMVpipi}). Contributions from further
parameters in $X$ are presently ignored, and discussed below.\\

Eq. (\ref{W}), when $X \neq (0)$, does not guarantee positive values for $W(t)$, unless the
values of the parameters of $X$ fall into definite physical regions, since the time evolution
generated by a general $T$ of (\ref{T}) does not satisfy (\ref{ruru}).

We describe now the general law of time evolution of the density matrix and the ensuing
physical region for the values in $X$.\\
The intriguing mathematical fact \cite{choi,gori} is, that Eq. (\ref{ruru}) does guarantee the positivity of $W(t)$, not only 
for the evolution of a single kaon, but also for a (suitably defined) system of many kaons.
On the other hand, the precautions for the positivity (when $X \neq (0)$), tailored to the single-particle evolution, do,
in general, $not$ entail the positivity for the many-particle evolution, unless the single-particle evolution
has the property of $complete \ positivity$. For the application to neutral kaons, see
\cite{Ben97,Ben02,geEPJCent}.\\
We summarize three results.\\
(i) Complete positivity is a necessary condition \cite{geEPJCent}
for the consistent description of entangled neutral kaon pairs in a symmetric 
state, as
produced in the \PPb\ annihilation \cite{Ad97,pr}. A general law of evolution therefore
has to have this property.\\
The time evolution is completely positive if (and only if) it is given by (see \cite{choi})
\begin{equation}\label{rogen}
\rho (t) = U_i (t) \rho (0) U_i^+ (t) \ ,
\end{equation}
where the right-hand side is a sum over four terms, with suitably normalized $2 \times 2$
matrices $U_i$ \ \cite{alic}.\\\\
(ii) The physical region for the values of the QMV parameters in $X$ follows from (\ref{rogen})
\cite{Ben98,geEPJCent}. The most important conditions are:
\begin{equation}\label{phy1}
(D^i)^2 + (S^i)^2 \leq ( (J^i)^2 - (J^j - J^k)^2 )/4 \ ,
\end{equation}
\begin{equation}\label{phy2}
J^i \ge 0 \ \ \ \ \forall i = 1, \cdots ,\ 3 \ ,	
\end{equation}
\begin{equation}\label{phy3}
J^i \le J^j + J^k \ ,			
\end{equation}
\begin{center} $(ijk =$ permutation of 123)\ . \end{center}
We note:\\\\
If any one of the three diagonal elements, $J^i$, vanishes, then the other two ones are equal, and all 
off-diagonal elements, 
$D^i$ and $S^i$, vanish, and\\
if any two of the three diagonal elements, $J^i$, vanish, then all elements of $X$ vanish.\\\\
(iii) The condition $(1 \ge \mathrm{tr}\{\rho (0)\} \ge \mathrm{tr}\{\rho (t \ge 0)\})$ demands in 
addition
\begin{equation}\label{Tle0}
T^{00} \le 0 \ ,
\end{equation}\label{Tgeq}
\begin{equation}
(T^{00})^2 \geq ( {L}^{01} + S^1)^2 + ( {L}^{02} + S^2)^2 + ( {L}^{03} + S^3)^2 \ .
\end{equation}
This shows that, with the properties of neutral kaons, especially since \gs\ $\gg\ $\gl\ ,  there 
is little room for the values of the small parameters.\\\\

\subsection{Entangled kaon pairs}\label{enta}
Particles with a common origin may show a causal behaviour, still when they have become
far apart, that is unfamiliar in a classical description.

Pairs of neutral kaons \cite{gold,lipk} from the decay $\phi \ra \kn \knb $, with the kaons 
flying away in opposite directions in the $\phi$'s rest system, have a number of remarkable 
properties \cite{huet,enz}:\\\\
They are created in the entangled, antisymmetric $J^{PC} = 1^{--}$ state
\begin{eqnarray}
\ket{\psi_-} &=& \frac{1}{\sqrt{2}}(\ket{\kn}\ket{\knb}-\ket{\knb}\ket{\kn})\label{str}\\
             &=& \frac{1}{\sqrt{2}}\ (\ket{\kk_2}\ket{\kk_1}-\ket{\kk_1}\ket{\kk_2})\label{cp}\\
      &\approx & \frac{1}{\sqrt{2}}\ (\ket{\kl}\ket{\ks}-\ket{\ks}\ket{\kl})\label{mass}.
\end{eqnarray}

In each of these representations the two particles have opposite properties: opposite
strangeness (\ref{str}), opposite CP parity (\ref{cp}), or opposite shifts of the eigenvalues 
of \Lz\ (\ref{mass}).\\
This allows the experimenter to define an ensemble of neutral kaons which has one of these
properties with high purity \cite{dafn,buch}.\\
The identification of the particular quantum number shown by the
particle which decays first, assures the opposite value for the surviving one.\\
This intriguing feature is not merely a consequence of conservation laws, since, at the moment
of the pair's birth, there is nothing which determines, which particle is going to show what
value of what observable, and when.\\

From (\ref{mass}), we see that $\ket{\psi_-}$ develops in time just by a multiplicative factor,
$\ket{\psi_-} \ra \ket{\psi_- (t)} = \e^{-\i (\ml + \ms)\gamma t}\
\e^{-\frac{1}{2}(\gs +\gl)\gamma t}\ket{\psi_-}$ which is independent of the symmetry violations
($\gamma t$ = eigentime of the kaons).
$\ket{\psi_- (t)}$ and $ \ket{\psi_-}$ have thus the same decay properties, e. g. the kaon pair cannot
decay simultaneously into two \pipa\ pairs or into the same \pen\ triplets, at all times. It 
is due to this simplicity of the time evolution, and due to the antisymmetry of $\ket{\psi_-}$,
that the kaons from $\phi\ $ decay are so well suited to explore symmetry violations in the
decay processes, or to search for QMV, which ignores the states' antisymmetry.\\

For the formal description of a pair of neutral kaons in a general (mixed) state, we use the positive
definite $4\times 4$ density matrix $\rho (t_1 ,t_2)$. The times $t_1, t_2$ indicate the moments when
later measurements on the individual particles will be performed. As a basis we use
$(\kn \kn, \ \kn \knb, \ \knb \kn, \ \knb \knb)$, and we assume (with \cite{enz}), that $\rho (t_1 ,t_2)$
evolves like $\rho (t_1) \otimes \rho (t_2)$. The two-particle evolution is thus uniquely determined by the
one-particle evolutions, and thus the introduction of QMV becomes obvious.\\
Similar to the one-particle $2 \times 2$ density matrices, we develop the $4 \times 4$ density
matrix $\rho (t_1 ,t_2)$ in terms of the products $(\sigma^\mu \otimes \sigma^\nu)$ with
coefficients $R^{\mu\nu} \equiv R^{\mu\nu} (t_1 , t_2) , \  \ R_0^{\mu\nu} \equiv R^{\mu\nu} (0,0)$,
as 
\begin{equation}
\rho (t_1 ,t_2) = R^{\mu\nu} (\sigma^\mu \otimes \sigma^\nu)
\label{eq:rho2}
\end{equation}
and obtain
\begin{equation}
R^{\mu\nu} =  (\e^{T t_1})^{\mu\alpha} R_0^{\alpha\beta} (\e^{T t_2})^{\nu\beta}
\label{Rmunu}.
\end{equation}
The generator \ $T$ may, or may not, contain QMV terms.\\
The probability density that an apparatus, tuned to $\rho_f = R_f^{\mu\nu} (\sigma^\mu \otimes \sigma^\nu)$
detects the particles at the times $t_1, t_2$ is
\begin{eqnarray}
W(t_1 ,t_2) & = & \mathrm{tr}( \rho_f  \rho (t_1 ,t_2) ) \ = \ 4 R_f^{\mu\nu} R^{\mu\nu} = \ 4 R_f^{\mu\nu} (\e^{T t_1})^{\mu\alpha}
R_0^{\alpha\beta} (\e^{T t_2})^{\nu\beta}  \nonumber \\
& = & 4 R_f^{\mu\nu} (\e^{(L+X) t_1})^{\mu\alpha} R_0^{\alpha\beta} (g \ \e^{( -L+X) t_2}\ g)^{\beta \nu} \e^{T^{00}(t_1 + t_2)}
\nonumber \\ 
& \equiv & 4\ \mathrm{Tr} \{\ {\bf{R}}_f^\mathrm{T}\ \e^{(L+X) t_1}\ {\bf{R}}_0 \ \ g \ \e^{( - L+X) t_2}\ g\ \}\ \e^{T^{00}(t_1 + t_2)}
\label{Wt1t2}.
\end{eqnarray}
The last expression in (\ref{Wt1t2}) is identical to the one before. The elements of the
$4 \times 4$ matrices $\bf{R}_0$ and $\bf{R}_f$~, e.g., are respectively,
$R_0^{\alpha\beta}$ and $R_f^{\nu\mu}$. Tr acts on their superscripts.\\
Eq. (\ref{Wt1t2}) is the general expression for the measurements of all the parameters of 
the (general) \kn \knb pair.\\
In the deduction of (\ref{Wt1t2}), the equations (\ref{glgml}) and (\ref{gxgpx}) have been used.
The different relative signs of $L$ and $X$ in the two exponents in (\ref{Wt1t2}) mark the
difference of the effects due to QM or to QMV.\\
As a benefit of complete positivity, $W(t_1 ,t_2)$, for $t_1, t_2 \ge 0$, will be positive, if $X$ satisfies the 
correspondig single-particle criteria.\\

We now give the general expression for the results of measurements on the pair (\ref{str}).
Its density matrix is
\begin{eqnarray}\label{ro00}
\rho (0 ,0)_-\
=&\frac{1}{\sqrt{2}} \left ( \begin{array}{c} \phantom{-}0 \\ \phantom{-}1 \\ -1 \\ \phantom{-}0 \end{array} \right )
\frac{1}{\sqrt{2}} (0\ 1 -1\ 0) 
= \frac{1}{2}
\left ( \begin{array}{cccc} 0 & \phantom{-}0 &\phantom{-} 0 &\phantom{-} 0 \\
                            0 & \phantom{-}1 &           -1 &\phantom{-} 0 \\ 
                            0 &           -1 &\phantom{-} 1 &\phantom{-} 0 \\ 
                            0 & \phantom{-}0 &\phantom{-} 0 &\phantom{-} 0 
\end{array} \right )\notag\\
& &  \notag \\
=& \frac{1}{4}(\sigma^0 \otimes \sigma^0 - \sigma^m \otimes \sigma^m)\ \ \
=\ \ (g^{\mu \nu}/4)\ (\sigma^\mu \otimes \sigma^\nu)\ ,
\end{eqnarray}
and thus
\begin{equation}\label{R0SI}
{\bf{R}}_0={\bf{R}}_0^{-} = g/4 \ .
\end{equation}
Inserting this interesting result, and (\ref{elxt}), into (\ref{Wt1t2}), we obtain
\begin{equation}\label{Wkkb}
W_{(\kn \knb)}(t_1 ,t_2) = \mathrm{Tr} \{\ {\bf{R}}_f^\mathrm{T}\
\e^{{\bf{D}}(t_1, L, X)} \e^{L(t_1 - t_2)} \e^{{\bf{D}}(t_2, L, X)}
 g\ \}\ \e^{T^{00}(t_1 + t_2)}\ .
\end{equation}
Again, all the terms in this expression, apart perhaps of $g$, have an obvious physical
interpretation.\\
Developping the exponentials $\e^{\bf{D}}$, Eq. (\ref{Wkkb}) allows one to calculate the
frequency of occurence of the events detected by the apparatus tuned to ${\bf{R}}_f$ as a
function of all the 16 paramters to any order in the small ones.\\
Explicit expressions have been
published for 3 QMV parameters \cite{huet,ellis96}, for 6 ones \cite{Ben98,Ben971}, and
for 9 ones \cite{geEPJCent}.\\

Finally, we consider neutral-kaon pairs in the symmetric state
\begin{equation}
\ket{\psi_+} = \frac{1}{\sqrt{2}}(\ket{\kn}\ket{\knb}+\ket{\knb}\ket{\kn})\ .\label{psip}
\end{equation}
They have the density matrix \cite{enz}
%%%%%%%
\begin{equation}\label{rotri}
\rho (0 ,0)_{+}\
= \frac{1}{2}
\left ( \begin{array}{cccc} 0 & 0 & 0& 0 \\ 0& 1& 1& 0 \\ 0& 1& 1&  0 \\ 0& 0& 0& 0 
\end{array} \right )
=\ \frac{1}{4}(\sigma^0 \otimes \sigma^0 + \sigma^m \otimes \sigma^m
-2\ \sigma^3 \otimes \sigma^3)\ ,
\end{equation}
and
\begin{equation}\label{R0TR}
{\bf{R}}_0 =
{\bf{R}}_0^{+}
= \frac{1}{4}
\left ( \begin{array}{cccc}\phantom{} 1 &\phantom{-} 0 &\phantom{-} 0& \phantom{-} 0\\
                           \phantom{} 0 &\phantom{-} 1 &\phantom{-} 0& \phantom{-} 0\\ 
                           \phantom{} 0 &\phantom{-} 0 &\phantom{-} 1& \phantom{-} 0\\
                           \phantom{} 0 &\phantom{-} 0 &\phantom{-} 0&            -1\\
\end{array} \right ) \ ,
\end{equation}
to be inserted into (\ref{Wt1t2}).\\
The explicit expression for $W(t_1 ,t_2)$ has been given, for the special case of QM, in
\cite{enz}.\\
$\ket{\psi_+}$, in contrast to $\ket{\psi_-}$, is allowed under QM, to evolve into
$\ket{\kn}\ket{\kn}$ and into $\ket{\knb}\ket{\knb}$.\\
\CPTz\ forbids that \ $\ket{\psi_+}$ evolves into \ \
$\sim (\ket{\ks}\ket{\kl}+\ket{\kl}\ket{\ks})$.\\

%%%%%%%%
The treatment presented here is based on the description of the time evolution of the density
matrix, generated by a general $4 \times 4$ matrix. An important difference to the regular
quantum-mechanical time evolution is, that conservation laws do not follow anymore from
symmetry properties, and that their existence is no more compulsory \cite{ehns,huet,ellis96}.
The question has to be left to the models, which enable QMV, whether the creator of QMV may
also be the supplier of the otherwise missing conserved quantities.

\section{Measuring neutral kaons}\label{sec:exp_1}

Many measurements concerning the neutral-kaon
system have been carried out with beams containing a mixture
of \kn\ and \knb\ \cite{blois}. Neutral kaons are
identified, among other neutral particles, by their masses, 
as obtained  from measurements of their decay products.\\
The relative proportion of the numbers of \kn\ and \knb\  
particles at the instant of production is measured separately, 
and taken into account in the analysis.
If the beam crosses  matter, regeneration effects take place,
and the measured ratio has to be corrected \cite{good1,klkn}.\\
Alternatively, \ks\ are separated from \kl\
taking advantage of the fact that a beam containing
\kn\ and \knb\ decays as a nearly pure \ks\ or \kl\ beam
depending on whether it decays very near or far away
from the source. This property was exploited at  CERN
in a precision measurement of the double ratio
\begin{center}
$\dfrac{
\Gamma\ (\kl \ra 2\pin)/
\Gamma\ (\kl \ra \pip\pim)
}{
\Gamma\ (\ks \ra 2\pin)/
\Gamma\ (\ks \ra \pip\pim)
}
=\modulus{\itaoo/\itapm}^2
\approx\ 1-6\re(\epsp/\epn)\approx\ 1-6\epsp/\epn
$~\ ,
\end{center}
which led to the discovery \cite{bu}
of $\epsp\ \ne 0$.

In a different approach neutral kaons have been identified and measured 
at their birth through the accompanying particles, in a 
convenient exclusive reaction. Conservation of energy and momentum allows 
neutral particles with the neutral-kaon mass to be selected. Differentiation 
between \kn\ and \knb\ is achieved taking advantage of the conservation of 
strangeness in strong and electromagnetic interactions, through which kaons 
are produced. This dictates that the strangeness of the final state is equal 
to that of the initial state.

Thus, opposite-sign kaon beams have been used to produce \kn\ and \knb\ by 
elastic 
charge-exchange in carbon, in order to compare  \kn\ and \knb\ decay 
rates to \pipi\ \cite{bann73}. \\
Similarly, Ref.~\cite{nieb74} reports 
measurements on \pen\ decays from old  \kn\ obtained by inelastic 
charge-exchange of positive kaons in hydrogen.\\ 
CPLEAR \cite{pr} produced concurrently \kn\ and \knb\ starting 
from \PPb\ annihilations, by selecting two charge-conjugate annihilation
channels, \kn\km\pip\  and \knb\kp\pim . 

Another interesting \PPb\ annihilation channel is \kn\knb , see \cite{pr}.
The same state  is exploited, as decay channel of the $\phi$, at the $\phi$ 
factories, like 
the KLOE experiment does at DA$\Phi$NE \cite{misc}. Here, the speciality is, 
that neutral kaon pairs are created in entangled states.
KLOE defines a neutral kaon as a \ks\ or a \kl\ according to the decay mode, 
or to the interaction, of the other neutral kaon of the pair.    
   
In the course of the neutral-kaon time evolution, pionic and semileptonic 
decays may be used to define a fixed time $t$ subsequent to the production 
time ($t=0$). \\
Pionic (\pipa\ and \pipb ) final states (which are \CPz\ 
eigenstates or a known superposition of them) are suitable for \CPz\ 
studies.\\ 
Semileptonic (\pen\ and \pmn ) final states allow \kn\ to be 
differentiated from \knb\ at the decay time, and  are convenient for \Tz\ and 
\CPTz\ studies. 
 
Alternatively, in order to identify the strangeness at a time $t$, 
neutral kaons could be observed to interact in a thin slab of matter (in 
most cases bound nucleons), in a two-body reaction like $\kn \p  \ra \kp\nn $
and  $\knb\nn \ra \km\p $ or $\knb\nn \ra \pin\Lz (\ra \pim\p )$, where the
charged products reveal the strangeness of the neutral kaon.

As a case study we shall focus on the CPLEAR measurements, which yield 
results on \Tz\ violation and on \CPTz\ invariance. Our presentation 
follows closely the description given by the CPLEAR group, summarized in Ref. 
\cite{pr}.

\subsection{CPLEAR experiment}\label{sub:cpl} 

\subsubsection*{\hspace{1cm}Experimental method}\label{sub:meth}

\noindent
% The method chosen by CPLEAR \cite{pr,erice} was
Developping the ideas discussed in Ref. \cite{erice}, CPLEAR chose 
to study the neutral-kaon time evolution by labelling (tagging) the 
state with its strangeness, at two subsequent times, see \cite{pr}. 

Initially-pure \kn\ and \knb\ states were produced concurrently by
antiproton annihilation at rest
in a hydrogen target, via the {\em golden} 
channels:
\begin{equation}\label{eq:3_1}
\PPb  \ra  \begin{array}{c}\km \pip \kn \\
\kp \pim \knb \end{array}\,,
\end{equation}
each having a branching ratio of $\approx  2  \times  10^{-3}$. The 
conservation of strangeness in the strong interaction dictates that a 
\kn\ is accompanied by a \km , and a \knb\ by a \kp . Hence, the 
strangeness of the neutral kaon at production was tagged by measuring
the charge sign of the accompanying charged kaon, and was therefore 
known event by event. The momentum of the produced \kn (\knb ) was 
obtained from the measurement of the $\pi^{\pm} {\rm K}^{\mp}$ pair 
kinematics. If the neutral kaon subsequently decayed to \pen , its
strangeness could also be tagged at the decay time by the charge of 
the decay electron: in the limit that only transitions with 
$\DS = \DQ$ take place, neutral kaons decay to \elp\ if the strangeness 
is positive at the decay time and to \elm\ if it is negative. This clearly
was not possible for neutral-kaon decays to two or three pions.

For each initial strangeness, the number of neutral-kaon decays was 
measured as a function of the decay time $t$. These numbers
% , \nrf ($t$) 
% and \nrfb ($t$) for a non-leptonic final state $f$, or \nrpm ($t$) and
% \nrpmb ($t$) for an \pen\ final state, 
were combined to form asymmetries -- 
thus dealing mainly with  ratios between measured quantities.
However, the translation of measured numbers of events into decay rates
requires (a) acceptance factors which do not cancel in the asymmetry, (b)
residual background, and (c) regeneration effects to be taken into account.
These experimental complications were handled essentially with the same 
procedure in the different asymmetries. Here we exemplify the procedure
referring to \pen\ decays.

\renewcommand{\theenumi}{(\alph{enumi})}
\renewcommand{\labelenumi}{\theenumi}
\medskip
\begin{enumerate}
\item
Detecting and strangeness-tagging neutral kaons at production and decay
relied on measuring, at the production (primary) vertex, a \kpm\pimp\ 
track-pair and the corresponding  momenta $\pvec_{\kpm}$ and $\pvec_{\pimp}$,
and, at the decay (secondary) vertex, an \elmp\pipm\ track-pair and the 
corresponding momenta $\pvec_{\elmp}$ and $\pvec_{\pipm}$. The detection 
(tagging) efficiencies of the \kpm\pimp\  track-pairs depend on the pair 
charge configuration and momenta, and are denoted by $\eps(\pvec_{\kpm },
\pvec_{\pimp })$. A similar dependence exists
for the detection efficiencies of the \elmp\pipm\ track-pairs, 
$\eps(\pvec_{\elmp },\pvec_{\pipm })$.
Since the detection efficiencies of primary and secondary track-pairs were
mostly uncorrelated, the acceptance of a signal (\pen ) event was factorized 
as $\varrho_S\times \eps(\pvec_{\kpm }, \pvec_{\pimp })\times 
\eps(\pvec_{\elmp },\pvec_{\pipm })$. The factor $\varrho_S$ represents
the portion of the acceptance which does not depend on the charge
configuration of either primary or secondary particles.  
The acceptances of the events corresponding to different
charge configurations were then equalized (or normalized) by introducing 
two functions:
\begin{subequations}\label{eq:norm1}
\begin{eqnarray}
\xi(\pvec_{\kk }, \pvec_{\pi }) & \equiv &
                        \frac{\eps(\pvec_{\kp}, \pvec_{\pim})}
                        {\eps(\pvec_{\km}, \pvec_{\pip})},\label{eq:norm1a}\\
\eta(\pvec_{\e }, \pvec_{\pi}) & \equiv &
                        \frac{\eps(\pvec_{\elm}, \pvec_{\pip})}
                        {\eps(\pvec_{\elp}, \pvec_{\pim})}.\label{eq:norm1b}
\end{eqnarray}
\end{subequations}
These functions, referred to as  {\em primary-vertex normalization factor} and
{\em secondary-vertex normalization factor}, respectively, are weights
applied event by event, $\xi$ to \kn\ events and  $\eta$ to the events
with a neutral kaon  decaying to \elp \pim .\\
%%%%%%%%%%%%%%%%%%%%%%%%%%%%%%%%%%%%%%%%%%%%%%%%%%%%%%%%%%%%%%%%%%%%%%%%%%%%%
\item
The background events mainly consist of  neutral-kaon decays to final states
other than the signal. Their number depends on the decay time $t$. To a
high degree of accuracy the amount of background is the same for initial
\kn\ and \knb\ and hence cancels in the numerator but not in the denominator
of any asymmetry: thus it is a dilution factor of the asymmetry.
To account for these events, the analytic expressions of the asymmetries
were modified by adding to the signal rates \rr\ and \rrb\ the corresponding
background rates \brr\ and \brrb :
\begin{eqnarray}\label{eq:3_2} 
\brr  (t) = \sum_{i} R_{Bi} \times \varrho_{Bi}/\varrho_S \,,&\quad &
\brrb (t) = \sum_{i} \overline{R}_{Bi} \times \varrho_{Bi}/\varrho_S\,,
\end{eqnarray}
where $ R_{Bi}, \overline{R}_{Bi} $ are the rates of the background
source $i$ for initial \kn\ and \knb , respectively, $\varrho_S$ is
defined above and $\varrho_{Bi}$ is the corresponding term for the 
acceptance of events from the  background source $i$. The quantities
$\varrho_{Bi}$
and $\varrho_S $ were obtained by Monte Carlo simulation.
Experimental asymmetries were formed from event rates including signal
and background: $\rr ^* =\rr +\brr $ and $\rrb^* =\rrb +\brrb $.
These asymmetries were then fitted to the asymmetries of the {\em 
measured rates} (see below), which included residual background.\\
%%%%%%%%%%%%%%%%%%%%%%%%%%%%%%%%%%%%%%%%%%%%%%%%%%%%%%%%%%%%%%%%%%%%%%%%
\item
The regeneration probabilities of \kn\  and \knb\ propagating through the 
detector material are not the same, thus making  the measured ratio of
initial \knb\ to \kn\  decay events at time $t$  different from that
expected in vacuum \cite{regen1}. A correction was performed by giving 
each \kn\ (\knb ) event a weight \wra\ (\wrb ) equal to the ratio of the 
decay probabilities for an initial \kn\ (\knb ) propagating in vacuum and 
through the detector.\\
\end{enumerate}
%%%%%%%%%%%%%%%%%%%%%%%%%%%%%%%%%%%%%%%%%%%%%%%%%%%%%%%%%%%%%%%%%%%%%%%%%%%
Finally, when \pen\ decays were considered, each initial-\kn\ event was 
given a total weight $\wt _+ =\xi\times\eta\times\wra $ or $\wt _-=\xi
\times\wra $ if the final state was $\elp\pim\net$ or $\elm\pip\netb$, 
respectively. The summed weights in a decay-time
bin are \nwp ($t$) and \nwm ($t$). In the same way, each initial-\knb\ event
was given a total weight $\wtb _+=\eta\times\wrb $ or $\wtb _-=\wrb $ if 
the final state was $\elp\pim\net$ or $\elm\pip\netb$. The corresponding
summed weights are \nwpb ($t$) and \nwmb ($t$). In the case 
of decays to two or three pions, each initial-\kn\ event was given a total
weight $\wt = \xi\times\wra $, and each initial-\knb\ event a total weight
$\wtb = \wrb $. The corresponding summed weights are \nw ($t$) and \nwb ($t$).
In the following the summed weights are referred to as the {\em measured
decay rates}. With these quantities are formed the {\em measured asymmetries}.

The measured asymmetries of interest here are
\begin{subequations}\label{eq:asym_1}
\begin{eqnarray}
\ATexp  (t) & = & 
       \frac{\nwpb (t) - \nwm (t)}{\nwpb (t) + \nwm (t)}\,,
       \label{eq:asym_1a} \\
\Adexp  (t)& = &
       \frac{\nwpb (t) - \alpha\nwm (t)}{\nwpb (t) +\alpha\nwm (t)} +  
       \frac{\nwmb (t) - \alpha\nwp (t)}{\nwmb (t) +\alpha\nwp (t)}\,,
       \label{eq:asym_1b} \\
\Apmexp (t)  &=&
       \frac{\nwb (t) - \alpha\nw (t)}{\nwb (t) + \alpha\nw (t)}\,,
       \label{eq:asym_1c} 
\end{eqnarray}
\end{subequations}

The quantity $\alpha = 1 + 4\reel$ is related to the primary vertex 
normalization procedure, see below.
The phenomenological asymmetries to be fitted to each of the above 
expressions include background rates. Explicit expressions of the 
phenomenological asymmetries, in the limit of negligible background, 
% were already given.  
can be written using (\ref{eq:2.50}), (\ref{eq:2.52}) and (\ref{eq:2.67}).
For \ATexp , as we shall see, Eq. (\ref{eq:2.56}) is also used.

(Two points are worth mentioning with regard to this method. Effects
related to a possible violation of charge asymmetry in the reactions of 
Eq.~(\ref{eq:3_1}) are taken into account by the weighting procedure at 
the primary vertex. When comparing the measured asymmetries with the 
phenomenological ones we take advantage of the fact that those reactions 
are strangeness conserving. A small strangeness violation (not expected 
at a level to be relevant in the CPLEAR experiment) would result in a 
dilution of the asymmetry and affect only some of the parameters.)

\subsubsection*{\hspace{1cm}The detector}\label{sub:detec1}

\noindent 
The layout of the CPLEAR experiment is shown in Fig. \ref{fig:detec}; a
comprehensive description of the detector is given in Ref.~\cite{det}.

The detector had a typical near-4$\pi$ geometry and was embedded in a (3.6 m
long, 2 m diameter) warm solenoidal magnet with a 0.44 T uniform
field (stable in a few parts in $10^4$). The 200 MeV/$c$ \Pb\ provided 
at CERN by the Low Energy Antiproton Ring (LEAR) \cite{ba} were stopped in a 
pressurized hydrogen gas target, at first a sphere of 7 cm radius at 16 bar  
pressure, later a 1.1 cm radius cylindrical target at 27 bar pressure.

A series of cylindrical tracking detectors provided information about the
trajectories of charged particles. The spatial  resolution $\sigma  \approx
300~\mu \mathrm{m}$ was sufficient to locate the annihilation vertex, as well
as the decay vertex if \kn\  decays  to charged particles, with a precision 
of a few millimetres in the transverse plane. Together with the momentum
resolution $\sigma_p/p \approx 5 ~\mathrm{to}~10$\% this enabled a lifetime
resolution of $\sigma \approx (5-10)\times 10^{-12}$~s. 
 
The tracking detectors were followed  by the particle identification detector
(PID), which comprised
a threshold Cherenkov detector, mainly effective for K/$\pi$
separation above 350 MeV/$c$ momentum ($> 4\sigma$), and scintillators
which measured the energy loss (d$E$/d$x$) and the time of flight of 
charged particles. The PID was also used to separate \e\ from
$\pi$ below 350~MeV/$c$.
 
The outermost detector was a  lead/gas sampling calorimeter
designed to detect the photons of the $\kn  \to 2\pi^0$ or $3\pi^0$ decays.
It also provided e/$\pi$ separation at higher momenta ($p> 300$ MeV/$c$). 
To cope with the branching ratio for reaction (3)
and the high annihilation rate (1 MHz), a set of hardwired
processors (HWP) was specially designed to  provide full event
reconstruction and selection in a few microseconds.

\subsubsection*{\hspace{1cm}Selection of \pen\ events}\label{sub:pen_sel}

\noindent
The \PPb\ annihilations followed by the decay of the neutral
kaon into \pen\  are first selected by topological criteria and by
identifying one of the decay tracks as an electron or a positron, 
from a Neural Network algorithm containing the PID information. The electron 
spectrum and identification efficiency are shown in Fig. \ref{fig:eleff}a. 
%the probability to identify a $\mu$ as an %\e\ is $\approx 15\%$.

The method of kinematic constrained fits was used to further reduce
the background and also determine the neutral-kaon lifetime with an
improved  precision (0.05 \ts\ and 0.2--0.3 \ts\ for short and long
lifetime, respectively). The decay-time resolution was known to
better than  $\pm 10\%$. In total $1.3\times 10^6$ events were selected,
and  one-half of these entered the \ATexp asymmetry.    

The residual background is shown in Fig.~\ref{fig:eleff}b. The 
simulation was controled by relaxing some of the selection cuts to increase 
the background contribution by a large factor. Data and simulation agree 
well and a conservative estimate of 10\% uncertainty was made. The background 
asymmetry arising from different probabilities of misidentifying \pip\ 
and \pim , was determined to be $0.03\pm 0.01$ by using \PPb\  multipion 
annihilations. 

Each event selected, labeled  by the initial kaon strangeness and the 
decay electron charge, was then properly weighted before forming 
the numbers $N$  of events entering the asymmetries \ATexp\ and \Adexp ,
see Eqs. (\ref{eq:asym_1a}) and (\ref{eq:asym_1b}).

\subsubsection*{\hspace{1cm}Weighting \pen\ events and building measured
                 asymmetries}\label{sub:pen_wei}

\noindent
Regeneration was corrected on an event-by-event basis using the amplitudes 
measured by CPLEAR \cite{regen2}, depending on the momentum of the neutral 
kaon and on the positions of its production and decay vertices. Typically, 
this correction amounts to a positive shift of the asymmetry \ATexp  of 
$0.3 \times 10^{-3}$ with an error dominated by the amplitude measurement.
                                                            
The detection efficiencies common to the classes of events being compared 
in the asymmetries cancel; some differences in the geometrical acceptances are 
compensated to first order since data were taken with a frequently reversed  
magnetic field. 

No cancellation takes place for the detection probabilities of the charged 
(K$\pi$) and (\e$\pi$) pairs used for strangeness tagging, thus the two 
normalization factors $\xi$ and $\eta$ of Eqs. (\ref{eq:norm1a}) and 
(\ref{eq:norm1b}) were measured as a function of the kinematic configuration.

The factor $\xi$, which does not depend on the decay mode, was obtained from 
the data set of \pipi\ decays between 1 and 4~\ts\ , where the number of events
is high and where the background is very small, see Ref. \cite{pipm}.
At any time $t$ in this interval, after correcting for regeneration, and 
depending on the phase space configuration, the ratio between the numbers 
of decays of old \kn\ and old \knb , weighted by $\xi$, is compared to the
phenomenological ratio obtained from (\ref{eq:2.67}):  
\begin{eqnarray}\label{eq:xi}
\frac{\xi N(\kn \rightarrow \pi^+\pi^-)}
         {N(\knb\rightarrow \pi^+\pi^-)} = 
(1 -  4\reel)\times \biggr (1 + 4|\eta_{+-}|\cos(\dm\ t -\phi_{+-})
\e^{\frac{1}{2}\gs\ t } \biggl ) \; .
\end{eqnarray}

Thus, the product $\xi\times (1+4\reel)$ can be evaluated. 
The oscillating term on the right-hand side is known with a precision
of $\approx 1 \times 10^{-4}$ (with the parameter values from Ref. 
\cite{pdg}), and remains $< 4\times 10^{-2}$. The statistical error 
resulting from the size of the \pipi\ sample is $\ \pm 4.3\times 10^{-4}$. 
 
The effectiveness of the method is illustrated in Fig. \ref{fig:norm}.
For the order of magnitude of $\xi$, as given by its average $\mean{\xi}$,
CPLEAR quotes $\mean{\xi} = 1.12023 \pm 0.00043$, with $2\reel \approx \dell 
= (3.27 \pm 0.12) \times 10^{-3}$ \cite{pdg}. 

Some of the measured asymmetries formed by CPLEAR, (\ref{eq:asym_1b}) and  
(\ref{eq:asym_1c}), contain just the product $\xi\times [1+4\reel]$, 
which is the quantity measured. However, for \ATexp , (\ref{eq:asym_1b}),
$\xi$ alone was needed.  The analysis was then performed taking \reel\ 
from the measured \kl\ charge asymmetry, $\dell = 2\reel - (\rexm + \rey )$. 
As a counterpart, the possible contribution to \dell\ of direct \CPTz\ 
violating terms had to be taken into account. 

The factor $\eta$ was measured as a function of the pion momentum, using \pip\ 
and \pim\ from  \PPb\ multipion annihilations. The dependence on  the electron 
momentum was determined using $\e^+ \e^-$ pairs from $\gamma$ conversions, 
selected from decays $\kn (\knb )\rightarrow 2\pin$, with a $\pin \rightarrow 
2\gamma$. 

The value of $\eta$, averaged over the particle momenta, is $\mean{\eta} = 
1.014\pm 0.002$, with an error dominated by the number of events in the \elpm\ 
sample.
 
The factors $\xi$ and $\eta$ are the weights applied event by event, which 
together with the regeneration weights, allowed CPLEAR to calculate the 
summed weights, in view of forming the measured asymmetries. The power of 
this procedure when comparing \kn\ and \knb\ time evolution is illustrated 
in Figs. \ref{fig:twopi_a} and \ref{fig:twopi_b} for the \pipi\ decay case. 

The comparison of the measured asymmetries with their phenomenological 
expressions allows the extraction of the physics parameters, as reported 
in Section \ref{meas}.
     
\section{Measurements}\label{meas}
\subsection{\CPTz\ invariance in the time evolution}
%%%%%%%%%%%%%%%%%%%%%%  Results for T and CPT %%%%%%%%%%%%%%%%%%%%%%%%%%%%%%%%%
What one measures is the parameter \den\ defined in (\ref{eq:2.28a}).

\noindent As for \red ,
exploring the fact that $\re (y+x_{-})$, which expresses \CPTz\ violation in the semileptonic
decay process, cancels out in the sum \AT ($t$)\ + \ACPT ($t$)\ , the CPLEAR group has formed
a data set $A_\delta^{\mathrm{exp}}(t)$ which measures this sum \cite{pen3}. 
Using Eqs. (\ref{eq:2.50}), (\ref{eq:2.52}), and (\ref{eq:2.39}), for $\ACPT (t)$, $\AT (t)$
and \AT , respectively, the measured quantity is shown to become
\begin{eqnarray}
A_\delta^{\mathrm{exp}}(t) &=& \ACPT (t) + \AT (t) - 4\re (\epn -\den )\label{adexp}\\
                           &=& 8\red\ + f(t,\imxp,\rexm,\red,\imd)\ .\label{adexp2}
\end{eqnarray}
The term $4\re (\epn -\den )$ follows from the normalization procedure, and the use of 
the decay rates to two pions (\ref{eq:2.65}). It does not require, however, a measurement 
of \dell\ (\ref{eq:2.56}). \\ 
The function $f$ is given in \cite{pen3}. It is negligible for \ $t\ \widetilde{>}\ 5\tau_S $\ .\\

Fig. \ref{fig:ad_mod} shows the data, together with the fitted curve
$A_\delta^{\mathrm{exp}}(t)$,
calculated from the corresponding parameter values.\\
The main result is \cite{pen3}
\begin{eqnarray*}
\red   = (0.30 \pm 0.33_{\mathrm{stat}}  
                          \pm 0.06_{\mathrm{syst}})\times 10^{-3}~.
\end{eqnarray*}
The global analysis \cite{phen2} gives a slightly smaller error
\begin{equation}\label{Redelta}
\red\ = (0.24 \pm 0.28)\times 10^{-3}\ .
\end{equation}
It confirms \CPTz\ invariance in the kaon's time evolution, free of assumptions on the semileptonic
decay process (such as \CPTz\ invariance, or the \ $\Delta S = \Delta Q$ rule ).\\\\
As for \imd , the most precise value 
\begin{eqnarray*}
\imd &=& (0.000 \pm 0.020)\times 10^{-3}
\end{eqnarray*}
is  obtained by inserting
\ $\mitapp\ =(2.284 \pm 0.014)\times 10^{-3},\  \fsw\ = (43.51 \pm 0.05)^\circ$,
and \ $\phi_\eta\ =(43.5 \pm 0.7)^\circ$, all from \cite{pdg4},
into (\ref{imappr}).\\
A more detailed analysis \cite{NA} yields (within the statistical error) the same result.\\
The formula (\ref{imappr}) also shows that the uncertainty of \imd\ is, at present, just a multiple of the one of $\phi_\eta$.\\\\
Using the Eqs. (\ref{phisw}) and (\ref{dpardsen}) with the values of \imd\ given above, and of
\red\ in (\ref{Redelta}),
we obtain
\begin{eqnarray*}
\dpar = ( 0.17 \pm 0.20)\times 10^{-3} ~,\quad
\dper = (- 0.17 \pm 0.19)\times 10^{-3}~.
\end{eqnarray*}
The mass and decay-width differences then follow from Eqs. (\ref{eq:2.32}) and (\ref{eq:2.33}). \\
With
$\dm = (3.48  \pm 0.01 )\times10^{-15}\ \geve$ and \
$\dg = (7.335 \pm 0.005)\times10^{-15}\ \geve$,
we obtain
\begin{equation}\label{dGdM}
\begin{split}
\Gamma_{\kn\kn}\ -\  \Gamma_{\knb\knb} =& \phantom -(\phantom - 3.5 \pm 4.1)
                                      \times 10^{-18}\; \geve ~,\\ 
{\mathrm M}_{\kn\kn}-{\mathrm M}_{\knb\knb} =& \phantom -(- 1.7 \pm 2.0)
                                      \times 10^{-18}\; \geve ~.
\end{split}
\end{equation}
\imd\ is constrained to a much smaller value than is \red\ . \imd\ could thus well be neglected.
The results (\ref{dGdM}) \ are then, to a good approximation, just a multiple of \red\ .

\subsection{\CPTz\ invariance in the semileptonic decay process}

We can combine the result on \red\ with the measured values for \dell\ ,
Eq.\ (\ref{eq:2.56}), and \ree , Eq.~\ (\ref{reappr}), and obtain
\begin{equation}\label{ryx}
\re (y+x_{-}) = -\red\ + \ree\ - \dell/2\ .
\end{equation}
For \dell\ the value 
\begin{equation}
\dell = \ (3.27 \pm 0.12)\times 10^{-3}
\end{equation}
can been used \cite{dorf, benn}.\\
For \ree , when entering Eq.\ (\ref{reappr}) with the values of \mitapp\ and with \fsw\  
given above, we have
\begin{equation}
\ree = \ (1.656 \pm 0.010)\times 10^{-3}.
\end{equation}
The value of $\re (y+x_{-})$  thus obtained
is in agreement with the
one reached  
in a more sophisticated procedure by the CPLEAR group
\cite{phen2}:
\begin{equation}\label{ryxval}
\re (y+x_{-})   = (-0.2 \pm 0.3)\times 10^{-3} .
\end{equation}
\\
This result confirms the validity of \CPTz\ invariance in the semileptonic decay
process, as defined in \ref{semilep}. (The new, more precise values of
\dell \cite{ah2,na48ru} do not change this conclusion).\\
We note in passing \cite{phen2}
\begin{eqnarray*}
\rey   = (0.5 \pm 3.0)\times 10^{-3} .
\end{eqnarray*}

\subsection{\Tz\ violation in the kaon's time evolution}
The measured asymmetry between the rates of \ \knb\ \ra\ $\e^+\pi^-\nu$ \ and
of \ \kn\ \ra\ $\e^-\pi^+\nu$
shall now be identified as an asymmetry between the rates of the mutually inverse
processes \ \knb\ \ra\ \kn \ and \ \kn \ra \knb, and thus be a demonstration of time reversal
violation in the kaon's time evolution, revealing a violation of
\ $\Tz ^{-1} \ \Hwk\ \ \Tz\ =\ \Hwk $\ .\\\\
Based on Eq. (\ref{eq:2.50}), whose time independent part (\ref{eq:2.51})\ becomes slightly
modified by the normalization procedure \cite{pen2}, the CPLEAR data is expected to follow
\begin{equation}\label{atexp}
\AT^{exp} (t) = \AT - 4\re (y+x_{-})\ + g(t,\rexm,\imxp)\ .
\end{equation}
The function $g$ is given in \cite{pen2}. It is negligible for \ $t\ \widetilde{>}\ 5\tau_S $\ .\\
Fig. \ref{fig:at_mod} shows \ $\langle A_\mathrm{T}^{\mathrm{exp}}\rangle
=6.6\times10^{-3}$, $-\ 4\re (y+x_{-})$, and \ $g(t,\rexm,\imxp)$, calculated
with the values for \rexm\ and \imxp\ , given below, but increased by one standard deviation.\\\\
The main result is
\begin{equation*}
\langle A_\mathrm{T}^{\mathrm{exp}}\rangle   = (6.6 \pm 1.3_{\mathrm{stat}}  
                          \pm 1.0_{\mathrm{syst}})\times 10^{-3},
\end{equation*}
in agreement with its theoretical prediction (\ref{eq:2.39}). \\\\
This is the only occasion in physics where a different transition rate of a subatomic process,
with respect to its inverse, has been observed.

\subsection{Symmetry in the semileptonic decay process ($\Delta S = \Delta Q$ rule)}
The measurements discussed above allow one also to confirm the $\Delta S = \Delta Q$ rule
for the semileptonic decay processes.\\Much of the information is contained in the time
dependent parts $f(t,\imxm,\rexm,\red,\imd)$ and $g(t,\rexm,\imxp)$.\\
CPLEAR has found \cite{pr},\cite{phen2}
\begin{align*}
\rexm  &= (          - 0.5 \pm  3.0_{{\rm stat}} \pm 0.3_{{\rm syst}})
\times 10^{-3}~, \notag\\
\imxp  &= (          - 2.0 \pm  2.6_{{\rm stat}} \pm 0.5_{{\rm syst}})
\times 10^{-3}~. \notag 
\end{align*}

\subsection{\CPTz\ invariance in the decay process to two pions}
%%%%%%%%%%%%%%%%%%%%%%%%%%%%%Begin new
A contribution to the study of \CPTz\ invariance in the two-pion decay
is the measurement \cite{pdg4}\\
\begin{equation*}
\foo\ - \fpm = (0.2 \pm\ 0.4)^\circ\ ,\\
\end{equation*}
which is
in agreement with the request (\ref{fpmmoo}) that  
\begin{equation*}
\modulus{\fpm\ -\ \foo } \approx\ (\frac{1}{50})^\circ\ .
\end{equation*}
The following \CPTz-violating
quantities have been given values,
using Eqs. (\ref{rbdra}) and (\ref{B2A2}),
\begin{eqnarray}
\frac{\reba }{\reaa } &=& (0.24 \pm 0.28)\times 10^{-3} ~,\quad
\frac{\rebc }{\reac } = (0.32 \pm 0.32)\times 10^{-3} ~.\label{bodao}
\end{eqnarray} 
In addition to the above value for $(\fpm\ -\foo\ )$, we have entered 
\red\ and \imd\ from Section 4.1, $|A_2/A_0| = 0.0448 \pm 0.0002$ from 
\cite{dedi},
\mitaoo\ and \fsw\ from \cite{pdg4}. See also \cite{pr}.

All together \CPTz\ invariance is confirmed. The measurements \cite{pdg4,ah3} 
below fulfil (\ref{fecpt}):
\begin{eqnarray*}
\fpm &=& (43.4 \phantom{1}\pm 0.7\phantom{9})^{\circ},\\
         \phi_{\ita} = 2/3 \ \fpm + 1/3 \ \foo &=& (43.5\phantom{1} \pm
                                            0.7\phantom{9})^{\circ} ,\\
 \fsw  &=& (43.51 \pm 0.09)^{\circ}.\\
& &
\end{eqnarray*}

In terms of the \ \kn \knb\  mass difference we note that all the terms on 
the rhs of (\ref{mmmb}) are negligible with respect to \red\ and we regain, 
in good approximation, Eq. (\ref{dGdM})
\begin{equation*}
\Mzaa\ -\ \Mzbb\ \approx -2\modulus{\dlz}\red\ \sin(\fsw) = (-1.7 \pm 2.0)
                                                        \times 10^{-18}\ \geve.
\end{equation*}
Authors \cite{pdg4wt,NA} have considered the model, which assumes $\dpar =0 $ \
($\hat{=}\  \Gzaa -\Gzbb = 0$), entailing \
$\den = -\den'=\imd\ (\tan{(\fsw) - \i})$, and thus leading to a roughly ten 
times more precise constraint. See also \cite{pr}.\\

\subsection{Further results on \CPTz\ invariance}

Each of the two terms on the rhs of Eq. (\ref{etdl}), $\reba/\reaa$ and 
$\re{(y+x_-)}$,   vanishes under \CPTz\ invariance. This is confirmed by the
experimental results (\ref{ryxval}) and (\ref{bodao}).\\
The experiment \cite{ah2} has allowed one, in addition, to confirm the 
vanishing of their sum by the experimental determination of the lhs of 
Eq. (\ref{etdl})
\begin{equation}\label{etdllhs}
\re{\ (\frac{2}{3}\itapm\ +\frac{1}{3}\itaoo)}\ -\frac{\dell}{2}
= (-0.003 \pm 0.035)\times 10^{-3}=\frac{\reba}{\reaa}+\re{(y+x_-)} \ .
\end{equation}
Although these two terms represent two hypothetical \CPTz\ violations of very 
unlike origins, we can see from Eqs. (\ref{rbdra}) and (\ref{ryx}), that, with 
to-day's uncertainties on the
values of \imd, \ree, and \dell\ , their possible sizes are roughly equal to 
\red:
\begin{equation}\label{bravo}
\frac{\reba}{\reaa}\approx -\re{(y+x_-)}\approx \red \ .
\end{equation}

\subsection{Transitions from pure states of neutral kaons to mixed states ?}
The authors of \cite{ehns} assume, for theoretical reasons, $J^3=0$. Taking complete 
positivity into account, we obtain $J^1=J^2$, and all other elements of $X$ vanish. 
This allows one to use (\ref{WQMVpi}) to (\ref{etaQMV}).
For $\modulus{\eta_{QMV}}$, the measurement  of $R^{\pi \pi}_s (t)$ by \cite{CHDG}, 
with the result $\modulus{\eta } = (2.30 \pm 0.035) \times 10^{-3}$, is well suited, 
since it would include effects of QMV. For $\modulus{\eta}$ in (\ref{etaQMV}) we take 
the value of $\modulus{\epn}$ reported in \cite{Ad95} from a first fit to CPLEAR data
(mainly to the asymmetry $\ACPf (t )$ of the decay rates to $f=\pipi$), with three of
the QMV parameters left free. One could then evaluate that this result $\modulus{\epn} 
= (2.34 \pm 0.08) \times 10^{-3}$ corresponds to the quantity $\modulus{\eta }$, 
which is free of QMV effects.
Inserting the values above into (\ref{etaQMV}), we obtain \ $\frac{1}{2}J^1 = (-1.4 
\pm 2.9) \times 10^{-21}\ \geve$. The analysis by \cite{Ad95} (for $\frac{1}{2}J^1 
\equiv \gamma$, $\frac{1}{2}J^2 \equiv \alpha$, $\frac{1}{2}D^3 \equiv \beta$, all 
other QMV parameters $= 0$, and without the constraint of complete positivity)
has yielded an upper limit (with 90 \% CL) of
\begin{equation}
\frac{1}{2}J^1 <  3.7 \times 10^{-21}\ \geve.
\end{equation}
We note that this value is in the range of ${\cal O}(m_\mathrm{K}^2/m_\mathrm{Planck})
= 2\times10^{-20}$ \geve . \\\\ 

%%%%%%%%%%%%%%%%%%%%%% END Results for T and CPT %%%%%%%%%%%%%%%%%%%%%%%%%%%%%%%%%
%%%%%%%%%%%%%%%%%%%%%%%%%%%%%CONCLUSIONS%%%%%%%%%%%%%%%%%%%%%%%%%%%%%%%%%%%%%%%%%%
\section{Conclusions}\label{conc}

Measurements of interactions and decays of neutral kaons, which have been produced in well
defined initial states, have provided new and detailed information on \Tz\ violation and
on \CPTz\ invariance in the time evolution and in the decay.

\Tz\ violation in the kaon evolution has been demonstrated by measuring that \ \knb\ \ra\ \kn\
is faster than  \ \kn\ \ra\ \knb\ , and by proving that this result is in straight conflict
with the assumption, that \ \Tz\ and \Hwk\ would commute. \\
Complementary measurements have confirmed that hypothetical violations of the
$\Delta$S=\ $\Delta$Q rule or of \CPTz\ invariance in the semileptonic decays, could not have
simulated this result. See Fig. \ref{fig:at_mod}.

\CPTz\ invariance is found intact. The combination of measurements on the decays to \pipa\ and
to \pen\ yields constraints on parameters, which have to vanish under \CPTz\ invariance, as well
concerning the evolution as the decay processes.

The interplay of results from experiments at very high energies (CERN, FNAL) and at medium ones
(CERN) has been displayed. A typical constraint on a hypothetical \kn\knb\ mass or decay width
difference is a
few times $10^{-18}$ \geve, resulting from the uncertainty of (the time evolution parameter)
\red\ .\\

In the future, more experiments with entangled neutral kaon-antikaon
pairs, in an antisymmetric
(\ref{str}, \ref{cp}, \ref{mass}) \ or in a symmetric (\ref{psip}) state,
will be performed.\\
The $\phi$ decay is a source of pairs in an antisymmetric state, which
allows one to select a
set of particles with precisely known properties. We wish to remind that
pairs in the symmetric
state have a complementary variety of phenomena, and also allow for a
particular \ \CPTz\ test.

The experiments have achieved precisions which may open the capability to explore the validity
of some of the often tacitly assumed hypotheses.\\
Some examples are: (i) the perturbation
expansion of the Schr\"odinger equation \cite{dagr,urba} leading to the two dimensional spinor
representation with the exponential decay law, (ii) the unitarity relations \cite{tanner}, (iii) the
conservation of the purity of states of isolated particles, manifested by the long time
coherence of the kaon matter wave \cite{hawk,ehns}.\\
A dedicated measurement on this subject has been performed at CERN. Data from 
the CPLEAR Collaboration in combination with earlier data from the CERN-HEIDELBERG 
Collaboration achieve a sensitivity of \ $\approx 10^{-21}$ \geve.\\ Comparisons of 
the long time coherence among kaons, neutrons and neutrini \cite{bm} are already 
possible.\\

Neutral kaons might well bring even more surprises. Probably the best probes in the 
world.\\\\

\section*{Acknowledgements}\label{memo}%%%%%%%%%%%%%%%%%%%%%%%%%%%

We thank the many colleagues, students and staff we have met at CERN 
and at ETH, who, with their work, their questions and their advice, 
helped us to  gain the insights about which we report in this article.

%%%%%%%%%%%%%%%%%%%%%%%%%%%%%%%%%%%%%%%%%%%%%%%%%%%%%%%%%%%%%%%%%

\include{captions}

\include{new_biblio}
\end{document}

%% file: preamble.tex
%%%%%%%%%%%%%%%%%%%%%%%%%%%%%%%%%%%%%%%%%%%%%%%
%% \documentclass [11pt]{cernart}            %%
%% \usepackage{times}                        %%
%% \usepackage{epsfig}                       %%
%% \usepackage{amsmath}                      %% 
%% \usepackage{graphicx}                     %% 
%%%%%%%%%%%%%%%%%%%%%%%%%%%%%%%%%%%%%%%%%%%%%%%
%% \setlength{\oddsidemargin}{-10mm}         %%
%% \setlength{\topmargin}{-15mm}             %%
%% \setlength{\headsep}{0mm}                 %%
%% \setlength{\textwidth}{183mm}             %%
%% \setlength{\textheight}{270mm}            %%
%% \pagestyle{plain}                         %%
%% \begin{document}                          %%
%%%%%%%%%%%%%%%%%%%%%%%%%%%%%%%%%%%%%%%%%%%%%%%
\renewcommand{\t}{\ensuremath{\tau }}
\renewcommand{\i}{\ensuremath{\mathrm{i}}}
\renewcommand{\a}{\ensuremath{\alpha }}
%%%%%%%%%%%%%%%%%%%%%%%%%%%%%%%%%%%%%%%%%
\newcommand{\PL}{Phys.\ Lett.\ }
\newcommand{\EPJ}{Eur.\ Phys.\ J.\ }
\newcommand{\re}{\ensuremath{\mathrm{Re}}}
\newcommand{\im}{\ensuremath{\mathrm{Im}}}
\newcommand{\e}{\ensuremath{\mathrm{e}}}
\newcommand{\s}{\ensuremath{\mathrm{s}}}
\newcommand{\nn}{\ensuremath{\mathrm{n}}}
\newcommand{\p}{\ensuremath{\mathrm{p}}}
\newcommand{\gev}{\ensuremath{\mathrm{GeV}/c}}
\newcommand{\mev}{\ensuremath{\mathrm{MeV}/c}}
\newcommand{\geve}{\ensuremath{\mathrm{GeV}}}
\newcommand{\br}{\ensuremath{\mathrm{BR}}}
\newcommand{\brsf}{\ensuremath{\mathrm{BR}_f^{\mathrm{S}}}}
\newcommand{\brlf}{\ensuremath{\mathrm{BR}_f^{\mathrm{L}}}}
\newcommand{\cs}{\ensuremath{\mathrm{cos}}}
\newcommand{\sn}{\ensuremath{\mathrm{sin}}}
\newcommand{\sbraket}[2]{\langle #1 | #2 \rangle}
\newcommand{\ra}{\ensuremath{\rightarrow}}
\newcommand{\pvec}{\ensuremath{\vec{p}}}
\newcommand{\pt}{\ensuremath{p_{\mathrm T}}}
\newcommand{\dt}{\ensuremath{d_{\mathrm T}}}
\newcommand{\gspm}{\ensuremath{\mathrm{\Gamma_S^{\pipi}}}}
\newcommand{\gsppp}{\ensuremath{\mathrm{\Gamma_S^{\pipb}}}}
\newcommand{\glppp}{\ensuremath{\mathrm{\Gamma_L^{\pipb}}}}
\newcommand{\gspmn}{\ensuremath{\mathrm{\Gamma_S^{+-0}}}}
\newcommand{\glpmn}{\ensuremath{\mathrm{\Gamma_L^{+-0}}}}
%%%%%%%%%%%%%%%%%%%%%%%%%%%%%%%%%%%%%%%%%%%%%%%%%%%%%
\newcommand{\eps}{\ensuremath{\epsilon}}
\newcommand{\epn}{\ensuremath{\varepsilon}}
\newcommand{\epnb}{\ensuremath{\bar{\varepsilon}}}
\newcommand{\et}{\ensuremath{\mathrm{\epn_T}}}
\newcommand{\el}{\ensuremath{\mathrm{\epn\ - \den}}}
\newcommand{\es}{\ensuremath{\mathrm{\epn_S}}}
\newcommand{\rl}{\ensuremath{r_L}}
\newcommand{\rs}{\ensuremath{r_S}}
\newcommand{\tel}{\ensuremath{\mathrm{\tilde{\epn}_L}}}
\newcommand{\tes}{\ensuremath{\mathrm{\tilde{\epn}_S}}}
\newcommand{\epsp}{\ensuremath{\epn^{\prime}}}
\newcommand{\ree}{\ensuremath{\re (\epn )}}
\newcommand{\reel}{\ensuremath{\re (\el )}}
\newcommand{\rees}{\ensuremath{\re\ (\es )}}
\newcommand{\del}{\ensuremath{\delta}}
\newcommand{\delb}{\ensuremath{\bar{\delta}}}
\newcommand{\delcpt}{\ensuremath{\delta_{\mathrm CPT}}}
\newcommand{\den}{\ensuremath{\delta}}
\newcommand{\dell}{\ensuremath{\delta_{\ell}}}
\newcommand{\dper}{\ensuremath{\del _{\perp}}}
\newcommand{\dpar}{\ensuremath{\del _{\parallel}}}
\newcommand{\red}{\ensuremath{\re (\del )}}
\newcommand{\imd}{\ensuremath{\im (\del )}}
\newcommand{\itapm}{\ensuremath{\eta_{+-}}}
\newcommand{\mitapm}{\ensuremath{|\eta_{+-}|}}
\newcommand{\fpm}{\ensuremath{\phi_{+-}}}
\newcommand{\itaoo}{\ensuremath{\eta_{00}}}
\newcommand{\mitaoo}{\ensuremath{|\eta_{00}|}}
\newcommand{\foo}{\ensuremath{\phi_{00}}}
\newcommand{\ita}{\ensuremath{\eta}}
\newcommand{\mita}{\ensuremath{|\eta|}}
\newcommand{\fsw}{\ensuremath{\phi_\mathrm{SW}}}
\newcommand{\reitapmo}{\ensuremath{\re (\eta_{+-0})}}
\newcommand{\imitapmo}{\ensuremath{\im (\eta_{+-0})}}
\newcommand{\reitaooo}{\ensuremath{\re (\eta_{000})}}
\newcommand{\imitaooo}{\ensuremath{\im (\eta_{000})}}
\newcommand{\itapmo}{\ensuremath{\eta_{+-0}}}
\newcommand{\mitapmo}{\ensuremath{|\eta_{+-0}|}}
\newcommand{\fpmo}{\ensuremath{\phi_{+-0}}}
\newcommand{\itaooo}{\ensuremath{\eta_{000}}}
\newcommand{\mitaooo}{\ensuremath{|\eta_{000}|}}
\newcommand{\fooo}{\ensuremath{\phi_{000}}}
\newcommand{\itapp}{\ensuremath{\eta_{\pipa}}}
\newcommand{\mitapp}{\ensuremath{|\eta_{\pipa}|}}
\newcommand{\fpp}{\ensuremath{\phi_{\pipa}}}
\newcommand{\itappp}{\ensuremath{\eta_{\pipa\pi}}}
\newcommand{\mitappp}{\ensuremath{|\eta_{\pipa\pi}|}}
\newcommand{\fppp}{\ensuremath{\phi_{\pipa\pi}}}
%%%%%%%%%%%%%%%%%%%%%%%%%%%%%%%%%%%%%%%%%%%%%%%%%%%%%%%%%%%%%%
\newcommand{\amz}{\ensuremath{{\cal A}}}
\newcommand{\amb}{\ensuremath{\overline{{\cal A}}}}
\newcommand{\afz}{\ensuremath{{\cal A}_f}}
\newcommand{\afb}{\ensuremath{\overline{{\cal A}}_f}}
\newcommand{\afzc}{\ensuremath{{\cal A}_f^*}}
\newcommand{\afbc}{\ensuremath{\overline{{\cal A}}_f^*}}
\newcommand{\aiz}{\ensuremath{{\cal A}_I}}
\newcommand{\aib}{\ensuremath{\overline{{\cal A}}_I}}
\newcommand{\ailz}{\ensuremath{{\cal A}_{I,l}^{3\pi}}}
\newcommand{\ailb}{\ensuremath{{\bar{\cal A}}_{I,l}^{3\pi}}}
\newcommand{\apz}{\ensuremath{{\cal A}_I^{\pipa}}}
\newcommand{\apb}{\ensuremath{\overline{{\cal A}_I^{\pipa}}}}
\newcommand{\anp}{\ensuremath{{\cal A_+}}}
\newcommand{\abp}{\ensuremath{\overline{{\cal A}}_+}}
\newcommand{\anm}{\ensuremath{{\cal A_-}}}
\newcommand{\abm}{\ensuremath{\overline{{\cal A}}_-}}
\newcommand{\ams}{\ensuremath{{\cal A}_{f{\mathrm S}}}}
\newcommand{\amsc}{\ensuremath{{\cal A}_{f{\mathrm S}}^*}}
\newcommand{\aml}{\ensuremath{{\cal A}_{f{\mathrm L}}}}
%%%%%%%%%%%%%%%%%%%%%%%%%%%%%%%%%%%%%%%%%%%%%%%%%%%%%%%%%%%
\newcommand{\Aa}{\ensuremath{\mathrm{A}_0}}
\newcommand{\Ab}{\ensuremath{\mathrm{A}_2}}
\newcommand{\Ba}{\ensuremath{\mathrm{B}_0}}
\newcommand{\Bb}{\ensuremath{\mathrm{B}_2}}
\newcommand{\aab}{\ensuremath{A_0^*\overline{A_0}}}
\newcommand{\reaa}{\ensuremath{\re (A_0)}}
\newcommand{\imaa}{\ensuremath{\im (A_0)}}
\newcommand{\reba}{\ensuremath{\re (B_0)}}
\newcommand{\reab}{\ensuremath{\re (A_1)}}
\newcommand{\rebb}{\ensuremath{\re (B_1)}}
\newcommand{\reac}{\ensuremath{\re (A_2)}}
\newcommand{\imac}{\ensuremath{\im (A_2)}}
\newcommand{\rebc}{\ensuremath{\re (B_2)}}
\newcommand{\rebi}{\ensuremath{\re (B_I)}}
\newcommand{\reai}{\ensuremath{\re (A_I)}}
\newcommand{\imbi}{\ensuremath{\im (B_I)}}
\newcommand{\yy}{\ensuremath{y}}
\newcommand{\xp}{\ensuremath{x_{+}}}
\newcommand{\xm}{\ensuremath{x_{-}}}
\newcommand{\xx}{\ensuremath{x}}
\newcommand{\xb}{\ensuremath{\overline{x}}}
\newcommand{\rex}{\ensuremath{\re (x)}}
\newcommand{\imx}{\ensuremath{\im (x)}}
\newcommand{\rexb}{\ensuremath{\re (\overline{x})}}
\newcommand{\imxb}{\ensuremath{\im (\overline{x})}}
\newcommand{\rey}{\ensuremath{\re (y)}}
\newcommand{\imxm}{\ensuremath{\im (x_{-})}}
\newcommand{\imxp}{\ensuremath{\im (x_{+})}}
\newcommand{\rexm}{\ensuremath{\re (x_{-})}}
\newcommand{\rexp}{\ensuremath{\re (x_{+})}}
%%%%%%%%%%%%%%%%%%%%%%%%%%%%%%%%%%%%%%%%%%%%%%%%%%%%%%%%%%%%%%
%\newcommand{\Akli}{\ensuremath{{\cal A}_{l,I}^{3\pi}}} 
%\newcommand{\Aakli}{\ensuremath{{\bar{\cal A}}_{l,I}^{3\pi}}}
\newcommand{\aklil}{\ensuremath{{\cal A}_{\mathrm L}^{l,I}}}
\newcommand{\aklis}{\ensuremath{{\cal A}_{\mathrm S}^{l,I}}} 
\newcommand{\akli}{\ensuremath{a_{l,I}}}             
\newcommand{\askli}{\ensuremath{a^{*}_{l,I}}}        
\newcommand{\deltai}{\ensuremath{\delta_{I}}}        
\newcommand{\edelti}{\ensuremath{\e^{\i\delta_{I}}}} 
\newcommand{\asxy}{\ensuremath{{\cal A}^{+-0}_\mathrm{S}(X,Y)}}   % KS ampl.
\newcommand{\asxym}{\ensuremath{{\cal A}^{+-0}_\mathrm{S}(-X,Y)}} % KS ampl.-X
\newcommand{\alxy}{\ensuremath{{\cal A}^{+-0}_\mathrm{L}(X,Y)}}   % KL ampl.
\newcommand{\alxym}{\ensuremath{{\cal A}^{+-0}_\mathrm{L}(-X,Y)}} % KL ampl.-X
\newcommand{\asxyo}{\ensuremath{{\cal A}^{000}_\mathrm{S}(X,Y)}}  % KS ampl.
\newcommand{\alxyo}{\ensuremath{{\cal A}^{000}_\mathrm{L}(X,Y)}}  % KL ampl.
\newcommand{\asxya}{\ensuremath{{\cal A}^{3\pi}_\mathrm{S}(X,Y)}} % KS ampl.
\newcommand{\alxya}{\ensuremath{{\cal A}^{3\pi}_\mathrm{L}(X,Y)}} % KL ampl.
\newcommand{\al}{\ensuremath{{\cal A}^{+-0}(X,Y)}}                % KL ampl.
\newcommand{\all}{\ensuremath{{\cal A}}}
%%%%%%%%%%%%%%%%%%%%%%%%%%%%%%%%%%%%%%%%%%%%%%%%%%%%%%%%%%%%%
\newcommand{\alcpc}{\ensuremath{{\cal A}_\mathrm{L}^{3\pi(\mathrm{CP=-1})}
                                (X,Y)}} % CP-cons. KL ampl.
\newcommand{\alcpv}{\ensuremath{{\cal A}_\mathrm{L}^{3\pi(\mathrm{CP=+1})}
                                (X,Y)}} % CP-cons. KL ampl.
\newcommand{\alcpcs}{\ensuremath{{\cal A}_\mathrm{L}^{*\ 3\pi(\mathrm{CP=-1})}
                                (X,Y)}} % CP-cons. KL ampl. *
\newcommand{\ascpc}{\ensuremath{{\cal A}_\mathrm{S}^{3\pi (\mathrm{CP=+1})}
                                (X,Y)}} % CP-cons. KS ampl.
\newcommand{\ascpcm}{\ensuremath{{\cal A}_\mathrm{S}^{3\pi (\mathrm{CP=+1})}
                                (-X,Y)}} % CP-cons. KS ampl. -X
\newcommand{\ascpv}{\ensuremath{{\cal A}_\mathrm{S}^{3\pi(\mathrm{CP=-1})}
                                (X,Y)}} % CP-viol. KS ampl.
%%%%%%%%%%%%%%%%%%%%%%%%%%%%%%%%%%%%%%%%%%%%%%%%%%%%%%%%%%%%
\newcommand{\apm}{\ensuremath{A_{+-}(\t )}}     % Asymmetry A+-
\newcommand{\apmz}{\ensuremath{A_{+-0}(\t )}}   % Asymmetry A+-0
\newcommand{\aepmz}{\ensuremath{A_{+-0}^\mathrm{exp}(\t )}}  
                                                % Exp Asy  A+-0 
\newcommand{\apmzxp}{\ensuremath{A_{+-0}(X>0, \t )}}   % Asymmetry A+-0 X>0
\newcommand{\aepmzxp}{\ensuremath{A_{+-0}^\mathrm{exp}(X>0, \t )}}          
                                                % Exp Asy A+-0 X>0
\newcommand{\apmzxm}{\ensuremath{A_{+-0}(X<0, \t )}} % Asymmetry A+-0 X<0
\newcommand{\aepmzxm}{\ensuremath{A_{+-0}^\mathrm{exp}(X<0, \t )}}       
                                                % Exp Asy A+-0 X<0
\newcommand{\relam}{\ensuremath{\re (\lz )}}
\newcommand{\imlam}{\ensuremath{\im (\lz )}}
\newcommand{\xin}{\mbox{$\xi_\mathrm{N}$}}             % Normalization xi
\newcommand{\xinxp}{\mbox{$\xi_\mathrm{N}^{X>0}$}}     % Normalization xi X>0
\newcommand{\xinxm}{\mbox{$\xi_\mathrm{N}^{X<0}$}}     % Normalization xi X<0
\newcommand{\xixy}{\mbox{$\xi_{XY}$}}                  % xi sub XY
\newcommand{\zeb}{\ensuremath{\zeta_\mathrm{B}}}       % xioft ....> zeta
\newcommand{\n}{\ensuremath{n}}
\newcommand{\m}{\ensuremath{m}}
\newcommand{\x}{\ensuremath{X}}
\newcommand{\y}{\ensuremath{Y}}
%%%%%%%%%%%%%%%%%%%%%%%%%%%%%%%%%%%%%%%%%%%%%%%%%%%%%%%%%%%%%%
\newcommand{\PPb}{\ensuremath{\mathrm{\overline{p}p}}}
\newcommand{\Pp}{\ensuremath{\mathrm{p}}}
\newcommand{\Pb}{\ensuremath{\mathrm{\overline{p}}}}
\newcommand{\pb}{\ensuremath{\mathrm{\overline{p}}}}
\newcommand{\pp}{\ensuremath{\mathrm{p}}}
\newcommand{\ppb}{\ensuremath{\mathrm{\overline{p}p}}}
%%%%%%%%%%%%%%%%%%%%%%%%%%%%%%%%%%%%%%%%%%%%%%%%%%%%%%%%%%%%%
\newcommand{\pipa}{\ensuremath{\pi\pi }}
\newcommand{\pipb}{\ensuremath{\pi\pi\pi }}
\newcommand{\pipi}{\ensuremath{\pip\pim }}
\newcommand{\pip}{\ensuremath{\mathrm{\pi^+}}}
\newcommand{\pim}{\ensuremath{\mathrm{\pi^-}}}
\newcommand{\pipm}{\ensuremath{\mathrm{\pi^{\pm}}}}
\newcommand{\pimp}{\ensuremath{\mathrm{\pi^{\mp}}}}
\newcommand{\pin}{\ensuremath{\mathrm{\pi^{0}}}}
\newcommand{\pinn}{\ensuremath{\mathrm{\pi^0\pi^0}}}
\newcommand{\pinnn}{\ensuremath{\mathrm{\pi^0\pi^0\pi^0}}}
\newcommand{\pipmn}{\ensuremath{\mathrm{\pi^+\pi^-\pi^0}}}
%%%%%%%%%%%%%%%%%%%%%%%%%%%%%%%%%%%%%%%%%%%%%%%%%%%%%%%%%%%%
\newcommand{\semi}{\ensuremath{\ell \pi \nu }}
\newcommand{\pen}{\ensuremath{\e \pi \nu }}
\newcommand{\penp}{\ensuremath{\elp \pim \net }}
\newcommand{\penm}{\ensuremath{\elm \pip \netb }}
\newcommand{\pmn}{\ensuremath{\mu \pi \nu }}
\newcommand{\elp}{\ensuremath{\mathrm{e^+}}}
\newcommand{\elm}{\ensuremath{\mathrm{e^-}}}
\newcommand{\elpm}{\ensuremath{\mathrm{e^{\pm}}}}
\newcommand{\elmp}{\ensuremath{\mathrm{e^{\mp}}}}
\newcommand{\net}{\ensuremath{\mathrm{\nu}}}
\newcommand{\netb}{\ensuremath{\mathrm{\overline{\nu}}}}
\newcommand{\dsb}{\ensuremath{\mathbf{d\overline{s}}}}
\newcommand{\dbs}{\ensuremath{\mathbf{\overline{d}s}}}
\newcommand{\sbb}{\ensuremath{\mathbf{\overline{s}}}}
\newcommand{\sss}{\ensuremath{\mathbf{s}}}
%%%%%%%%%%%%%%%%%%%%%%%%%%%%%%%%%%%%%%%%%%%%%%%%%%%%%%%%%%%
\newcommand{\kk}{\ensuremath{\mathrm{K}}}
\newcommand{\kpi}{\ensuremath{\mathrm{K}\pi}}
\newcommand{\kpipm}{\ensuremath{\mathrm{K}^{\pm}\pi ^{\mp}}}
\newcommand{\kpimp}{\ensuremath{\mathrm{K}^{\mp}\pi ^{\pm}}}
\newcommand{\kp}{\ensuremath{\mathrm{K^{+}}}}
\newcommand{\km}{\ensuremath{\mathrm{K^{-}}}}
\newcommand{\kpm}{\ensuremath{\mathrm{K^{\pm }}}}
\newcommand{\kmp}{\ensuremath{\mathrm{K^{\mp }}}}
\newcommand{\kn}{\ensuremath{\mathrm{K^0}}}
\newcommand{\knb}{\ensuremath{\mathrm{\overline{K}}{}^0}}
\newcommand{\ks}{\ensuremath{\mathrm{K_S}}}
\newcommand{\kl}{\ensuremath{\mathrm{K_L}}}
\newcommand{\kls}{\ensuremath{\mathrm{K_{L,S}}}}
\newcommand{\knin}{\ensuremath{\mathrm{K}{}^{0\text{in}}}}
\newcommand{\kbin}{\ensuremath{\mathrm{\overline{K}}{}^{0\text{in}}}}
\newcommand{\knut}{\ensuremath{\mathrm{K}{}^{0\text{out}}}}
\newcommand{\kbut}{\ensuremath{\mathrm{\overline{K}}{}^{0\text{out}}}}
\newcommand{\ksin}{\ensuremath{\mathrm{K_S}{}^{\text{in}}}}
\newcommand{\klin}{\ensuremath{\mathrm{K_L}{}^{\text{in}}}}
\newcommand{\ksut}{\ensuremath{\mathrm{K_S}{}^{\text{out}}}}
\newcommand{\klut}{\ensuremath{\mathrm{K_L}{}^{\text{out}}}}
\newcommand{\klsin}{\ensuremath{\mathrm{K_{L,S}}{}^{\text{in}}}}
\newcommand{\klsut}{\ensuremath{\mathrm{K_{L,S}}{}^{\text{out}}}}
%%%%%%%%%%%%%%%%%%%%%%%%%%%%%%%%%%%%%%%%%%%%%%%%%%%%%%%%%
\newcommand{\gls}{\ensuremath{\mathrm{\Gamma_{L,S}}}}
\newcommand{\gl}{\ensuremath{\mathrm{\Gamma_{L}}}}
\newcommand{\gs}{\ensuremath{\mathrm{\Gamma_{S}}}}
\newcommand{\gm}{\ensuremath{\overline{\Gamma}}} % (\gl + \gs)/2
\newcommand{\gz}{\ensuremath{\gamma}}
\newcommand{\tls}{\ensuremath{\mathrm{\tau_{L,S}}}}
\newcommand{\ts}{\ensuremath{\mathrm{\tau_S}}}
\newcommand{\tl}{\ensuremath{\mathrm{\tau_L}}}
%%%%%%%%%%%%%%%%%%%%%%%%%%%%%%%%%%%%%%%%%%%%%%%%%%%%%%%%%
\newcommand{\mkz}{\ensuremath{m_{\kn }}}
\newcommand{\mkb}{\ensuremath{m_{\knb }}}
\newcommand{\mls}{\ensuremath{m_{\mathrm{L,S}}}}
\newcommand{\ml}{\ensuremath{m_{\mathrm{L}}}}
\newcommand{\ms}{\ensuremath{m_{\mathrm{S}}}}
\newcommand{\dm}{\ensuremath{\Delta m}}
\newcommand{\dg}{\ensuremath{\mathrm{\Delta\Gamma}}}
\newcommand{\lz}{\ensuremath{\mathrm{\lambda}}}
\newcommand{\lzs}{\ensuremath{\mathrm{\lambda_S}}}
\newcommand{\lzl}{\ensuremath{\mathrm{\lambda_L}}}
\newcommand{\lzls}{\ensuremath{\mathrm{\lambda_{L,S}}}}
\newcommand{\dlz}{\ensuremath{\mathrm{\Delta\lambda}}}
%%%%%%%%%%%%%%%%%%%%%%%%%%%%%%%%%%%%%%%%%%%%%%%%%%%%%%%%
\newcommand{\Rb}{\ensuremath{\overline{R}}}
\newcommand{\Rz}{\ensuremath{R}}
%%%%%%%%%%%%%%%%%%%%%%%%%%%%%%%%%%%%%%%%%%%%%%%%%%%%%%
\newcommand{\delm}{\ensuremath{\mathrm{M}_{\kn\kn }-\mathrm{M}_{\knb\knb }}}
\newcommand{\delg}{\ensuremath{\Gamma_{\kn\kn }-\Gamma_{\knb\knb }}}
\newcommand{\Lz}{\ensuremath{\Lambda }}
\newcommand{\Vl}{\ensuremath{V_L }}
\newcommand{\Vr}{\ensuremath{V_R }}
\newcommand{\PLaa}{\ensuremath{\mathrm{\Lambda_{\kn\kn }}}}
\newcommand{\PLab}{\ensuremath{\mathrm{\Lambda_{\kn\knb }}}}
\newcommand{\PLba}{\ensuremath{\mathrm{\Lambda_{\knb\kn }}}}
\newcommand{\PLbb}{\ensuremath{\mathrm{\Lambda_{\knb\knb }}}}
\newcommand{\PLij}{\ensuremath{\mathrm{\Lambda_{\alpha\alpha^{\prime}}}}}
\newcommand{\PLik}{\ensuremath{\mathrm{\Lambda_{\alpha\alpha^{\prime}}}}}
\newcommand{\Mz}{\ensuremath{\mathrm{M}}}
\newcommand{\Mzaa}{\ensuremath{\mathrm{M_{\kn\kn }}}}
\newcommand{\Mzab}{\ensuremath{\mathrm{M_{\kn\knb }}}}
\newcommand{\Mzba}{\ensuremath{\mathrm{M_{\knb\kn }}}}
\newcommand{\Mzbas}{\ensuremath{\mathrm{M^*_{\knb\kn }}}}
\newcommand{\Mzbb}{\ensuremath{\mathrm{M_{\knb\knb }}}}
\newcommand{\Mzij}{\ensuremath{\mathrm{M_{\alpha\alpha^{\prime}}}}}
\newcommand{\fm}{\ensuremath{\mathrm{\phi_{M}}}}

\newcommand{\Gz}{\ensuremath{\mathrm{\Gamma}}}
\newcommand{\Gzaa}{\ensuremath{\mathrm{\Gamma_{\kn\kn }}}}
\newcommand{\Gzab}{\ensuremath{\mathrm{\Gamma_{\kn\knb }}}}
\newcommand{\Gzba}{\ensuremath{\mathrm{\Gamma_{\knb\kn }}}}
\newcommand{\Gzbas}{\ensuremath{\mathrm{\Gamma^*_{\knb\kn }}}}
\newcommand{\Gzbb}{\ensuremath{\mathrm{\Gamma_{\knb\knb }}}}
\newcommand{\Gzij}{\ensuremath{\mathrm{\Gamma_{\alpha\alpha^{\prime}}}}}
\newcommand{\fg}{\ensuremath{\mathrm{\phi_{\Gamma}}}}

\newcommand{\Xz}{\ensuremath{\mathrm{X}}}
\newcommand{\Iz}{\ensuremath{\mathrm{I}}}
\newcommand{\CPx}{\ensuremath{{\pmb {\mathcal CP}}}}
\newcommand{\CPz}{\ensuremath{\mathrm{\cal CP}}}
\newcommand{\Cz}{\ensuremath{\mathrm{\cal C}}}
\newcommand{\Pz}{\ensuremath{\mathrm{\cal P}}}
\newcommand{\Tz}{\ensuremath{\mathrm{\cal T}}}
\newcommand{\CPTz}{\ensuremath{\mathrm{\cal CPT}}}

\newcommand{\Hst}{\ensuremath{\mathrm{{\cal H}_{st}}}}
\newcommand{\Hem}{\ensuremath{\mathrm{{\cal H}_{em}}}}
\newcommand{\Hwk}{\ensuremath{\mathrm{{\cal H}_{wk}}}}
\newcommand{\Hz}{\ensuremath{\mathrm{{\cal H}}}}
\newcommand{\Vz}{\ensuremath{\mathrm{V}}}

\newcommand{\mean}[1]{\langle #1 \rangle}
\newcommand{\AT}{\ensuremath{A_\mathrm{T}}}
\newcommand{\ATl}{\ensuremath{A_\mathrm{T}^{\ell }}}
\newcommand{\ATexp}{\ensuremath{A_\mathrm{T}^{\text{exp}}}}
\newcommand{\avATexp}{\ensuremath{\langle A_\mathrm{T}^{\text{exp}}\rangle }}
\newcommand{\ACPT}{\ensuremath{A_\mathrm{CPT}}}
\newcommand{\ACPTl}{\ensuremath{A_\mathrm{CPT}^{\ell }}}
\newcommand{\ACPTexp}{\ensuremath{A_\mathrm{CPT}^{\text{exp}}}}
\newcommand{\Ad}{\ensuremath{A_{\del }}}
\newcommand{\Adl}{\ensuremath{A_{\del}^{\ell}}}
\newcommand{\Adexp}{\ensuremath{A_{\del }^{\text{exp}}}}
\newcommand{\avAdexp}{\ensuremath{\langle A_{\del }^{\text{exp}}\rangle }}
\newcommand{\avedm}{\ensuremath{\langle \dm \rangle }}
\newcommand{\avets}{\ensuremath{\langle \ts \rangle }}
\newcommand{\Adm}{\ensuremath{A_{\dm }}}
\newcommand{\Admexp}{\ensuremath{A_{\dm }^{\text{exp}}}}
\newcommand{\Admc}{\ensuremath{A_{\dm }}}
\newcommand{\Adml}{\ensuremath{A_{\dm }^{\ell }}}
\newcommand{\Admcexp}{\ensuremath{A_{\dm }^{\text{exp}}}}
\newcommand{\Aexp}{\ensuremath{A^{\text{exp}}}}
\newcommand{\Afexp}{\ensuremath{A_f^{\text{exp}}}}
\newcommand{\Af}{\ensuremath{A_f}}
\newcommand{\ACP}{\ensuremath{A_\mathrm{CP}}}
\newcommand{\ACPf}{\ensuremath{A_\mathrm{CP}^f}}
\newcommand{\Aast}{\ensuremath{A_{+-}^*}}
\newcommand{\Apm}{\ensuremath{A_{+-}}}
\newcommand{\Apmexp}{\ensuremath{A_{+-}^{\text{exp}}}}
\newcommand{\Ann}{\ensuremath{A_{00}}}
\newcommand{\Annexp}{\ensuremath{A_{00}^{\text{exp}}}}
\newcommand{\Apmn}{\ensuremath{A_{+-0}}}
\newcommand{\Apmnexp}{\ensuremath{A_{+-0}^{\text{exp}}}}
\newcommand{\Annn}{\ensuremath{A_{000}}}
\newcommand{\Annnexp}{\ensuremath{A_{000}^{\text{exp}}}}
\newcommand{\modulus}[1]{\left| #1 \right|}
%%%%%%%%%%%%%%%%%%%%%%%%%%%%%%%%%%%%%%%%%%%%%%%%%%
\newcommand{\rrf}{\ensuremath{R_f}}
\newcommand{\rrfb}{\ensuremath{\overline{R}_f}}
\newcommand{\rrpm}{\ensuremath{R_{\pipi}}}
\newcommand{\rrpmb}{\ensuremath{\overline{R}_{\pipi}}}
\newcommand{\rr}{\ensuremath{R}}
\newcommand{\rrb}{\ensuremath{\overline{R}}}
\newcommand{\rrp}{\ensuremath{R_+}}
\newcommand{\rrm}{\ensuremath{R_-}}
\newcommand{\rrpb}{\ensuremath{\overline{R}_+}}
\newcommand{\rrmb}{\ensuremath{\overline{R}_-}}
%%%%%%%%%%%%%%%%%%%%%%%%%%%%%%%%%%%%%%%%%%%%%%%%%%%
\newcommand{\nrfb}{\ensuremath{\overline{N}_f}}
\newcommand{\nrf}{\ensuremath{N_f}}
\newcommand{\nrpb}{\ensuremath{\overline{N}_+}}
\newcommand{\nrp}{\ensuremath{N_+}}
\newcommand{\nrmb}{\ensuremath{\overline{N}_-}}
\newcommand{\nrm}{\ensuremath{N_-}}
\newcommand{\nrpm}{\ensuremath{N_{\pm}}}
\newcommand{\nrpmb}{\ensuremath{\overline{N}_{\pm}}}
%%%%%%%%%%%%%%%%%%%%%%%%%%%%%%%%%%%%%%%%%%%%%%%%%%%
\newcommand{\brr}{\ensuremath{B}}
\newcommand{\brrb}{\ensuremath{\overline{B}}}
\newcommand{\brpb}{\ensuremath{\overline{B}_+}}
\newcommand{\brp}{\ensuremath{B_+}}
\newcommand{\brmb}{\ensuremath{\overline{B}_-}}
\newcommand{\brm}{\ensuremath{B_-}}
\newcommand{\brpm}{\ensuremath{B_{\pm}}}
\newcommand{\brpmb}{\ensuremath{\overline{B}_{\pm}}}
%%%%%%%%%%%%%%%%%%%%%%%%%%%%%%%%%%%%%%%%%%%%%%%%%%
\newcommand{\nwb}{\ensuremath{\overline{N}_{w}}}
\newcommand{\nw}{\ensuremath{N_{w}}}
\newcommand{\nwpb}{\ensuremath{\overline{N}_{w+}}}
\newcommand{\nwp}{\ensuremath{N_{w+}}}
\newcommand{\nwmb}{\ensuremath{\overline{N}_{w-}}}
\newcommand{\nwm}{\ensuremath{N_{w-}}}
\newcommand{\nwpmb}{\ensuremath{\overline{N}_{w\pm }}}
\newcommand{\nwpm}{\ensuremath{N_{w\pm }}}
%%%%%%%%%%%%%%%%%%%
\newcommand{\nwrpb}{\ensuremath{\overline{N}_{+w_{r}}}}
\newcommand{\nwrp}{\ensuremath{N_{+w_{r}}}}
\newcommand{\nwrmb}{\ensuremath{\overline{N}_{-w_{r}}}}
\newcommand{\nwrm}{\ensuremath{N_{-w_{r}}}}
\newcommand{\nwrpmb}{\ensuremath{\overline{N}_{\pm w_{r}}}}
\newcommand{\nwrpm}{\ensuremath{N_{\pm w_{r}}}}
%%%%%%%%%%%%%%%%%%%
\newcommand{\nwxpb}{\ensuremath{\overline{N}_{+w_{\xi}}}}
\newcommand{\nwxp}{\ensuremath{N_{+w_{\xi}}}}
\newcommand{\nwxmb}{\ensuremath{\overline{N}_{-w_{\xi}}}}
\newcommand{\nwxm}{\ensuremath{N_{-w_{\xi}}}}
\newcommand{\nwxpmb}{\ensuremath{\overline{N}_{\pm w_{\xi}}}}
\newcommand{\nwxpm}{\ensuremath{N_{\pm w_{\xi}}}}
%%%%%%%%%%%%%%%%%%%
\newcommand{\wra}{\ensuremath{w_r}}
\newcommand{\wrb}{\ensuremath{\overline{w}_r}}
\newcommand{\wt}{\ensuremath{w}}
\newcommand{\wtb}{\ensuremath{\overline{w}}}
%%%%%%%%%%%%%%%%%%%
\newcommand{\api}{\ensuremath{(1+4\reel )\xi}}
\renewcommand{\exp}{\ensuremath{\mathrm{exp}}}
\newcommand{\expm}{\ensuremath{\mathrm{E_-}}}
\newcommand{\expp}{\ensuremath{\mathrm{E_+}}}
\newcommand{\exppm}{\ensuremath{\mathrm{E_{\pm}}}}
\newcommand{\eff}{\ensuremath{\epsilon}}
\newcommand{\DS}{\ensuremath{\Delta S}}
\newcommand{\DQ}{\ensuremath{\Delta Q}}
\newcommand{\Dt}{\ensuremath{\Delta t}}
\newcommand{\kef}{\ensuremath{\mathrm{K}_{\e 3}^0}}
\newcommand{\dd}{\ensuremath{\mathrm{d}}}
\newcommand{\bce}{\begin{center}}
\newcommand{\ece}{\end{center}}
\newcommand{\bfig}{\begin{figure}}
\newcommand{\efig}{\end{figure}}
%%%%%%%%%%%%%%%%%%%%%%%%%%%%%%%%%%%%%%%%%%%%%%%%%%%%
\newcommand{\scbs}{\ensuremath{\mathrm{S1\overline{C}S2}}}
\newcommand{\pro }{\ensuremath{{\cal P}}}
\newcommand{\Ppl}{\ensuremath{{\cal P}_+}}
\newcommand{\Pbp}{\ensuremath{\overline{\cal P}_+}}
\newcommand{\Pm}{\ensuremath{{\cal P}_-}}
\newcommand{\Pbm}{\ensuremath{\overline{\cal P}_-}}
\newcommand{\Ppm}{\ensuremath{{\cal P}_{\pm }}}
\newcommand{\Pbpm}{\ensuremath{\overline{\cal P}_{\pm }}}

\newcommand{\Np}{\ensuremath{N_+}}
\newcommand{\Nbp}{\ensuremath{\overline{N}_+}}
\newcommand{\Nm}{\ensuremath{N_-}}
\newcommand{\Nbm}{\ensuremath{\overline{N}_-}}
\newcommand{\Npm}{\ensuremath{N_{\pm}}} 
\newcommand{\Nbpm}{\ensuremath{\overline{N}_{\pm}}}
\newcommand{\Nb}{\ensuremath{{\overline{N}}}}

%% file: page_title.tex
\thispagestyle{empty}
\title{ 
INVESTIGATIONS ON \Tz\ VIOLATION AND \CPTz\ SYMMETRY \\IN THE
       NEUTRAL KAON SYSTEM \\ {\it -- a pedagogical choice --}}
\begin{center}
\vspace{6mm}
Maria Fidecaro\\
CERN, CH--1211 Gen\`eve 23, Switzerland \\[4mm]
and\\[4mm]
Hans--J\"{u}rg Gerber\\
ETH, Institute for Particle Physics, CH--8093 Z\"urich, Switzerland\\[10mm]
\end{center}
\begin{abstract}
During the recent years experiments with neutral kaons have yielded remarkably 
sensitive results which are pertinent to such fundamental phenomena as \CPTz\ 
invariance (protecting causality), time-reversal invariance violation, coherence 
of wave functions, and entanglement of kaons in pair states. We describe the 
phenomenological developments and the theoretical conclusions drawn from the 
experimental material. An outlook to future experimentation is indicated.
\end{abstract}
\begin{center}
\bigskip
March 16, 2005
\end{center} 
\vspace{10mm}
\tableofcontents
\maketitle

%% file: captions.tex
\section*{Figure captions}\label{fcap}
%%%%%%%%%%%%%%%%%%%%%%%---FIG1---\label{fig:detec}---%%%%%%%%%%%%%%%%%%%%%%%%
\subsection*{}
\begin{figure}
\vspace{5cm}
\begin{center}

\includegraphics[angle=-90,bb=90 -175 -275 469,width=10cm,clip]
                                          {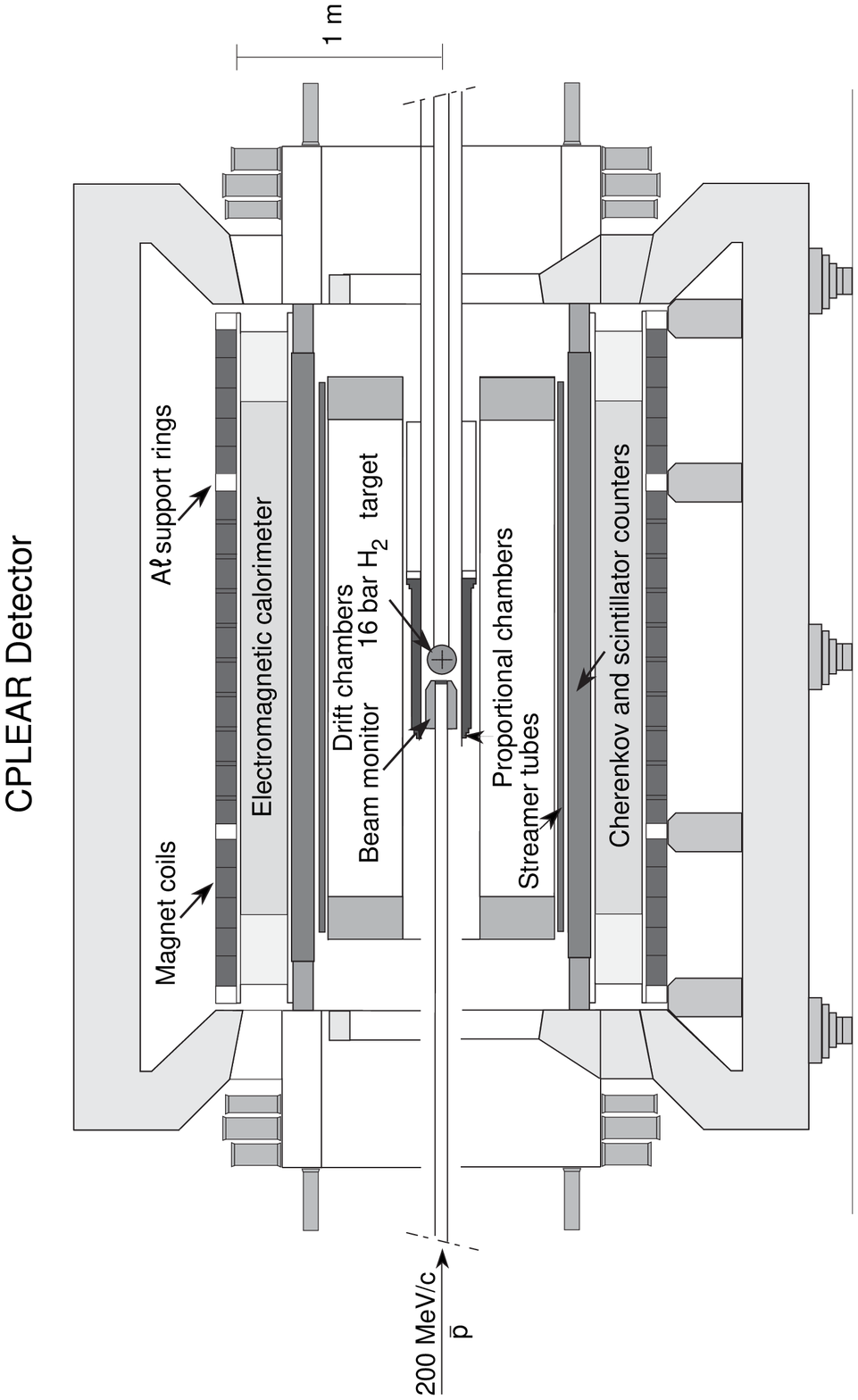}\\[5mm]
  (a) \\[15mm]
\includegraphics[bb=119 232 506 550,width=7.14cm,clip]{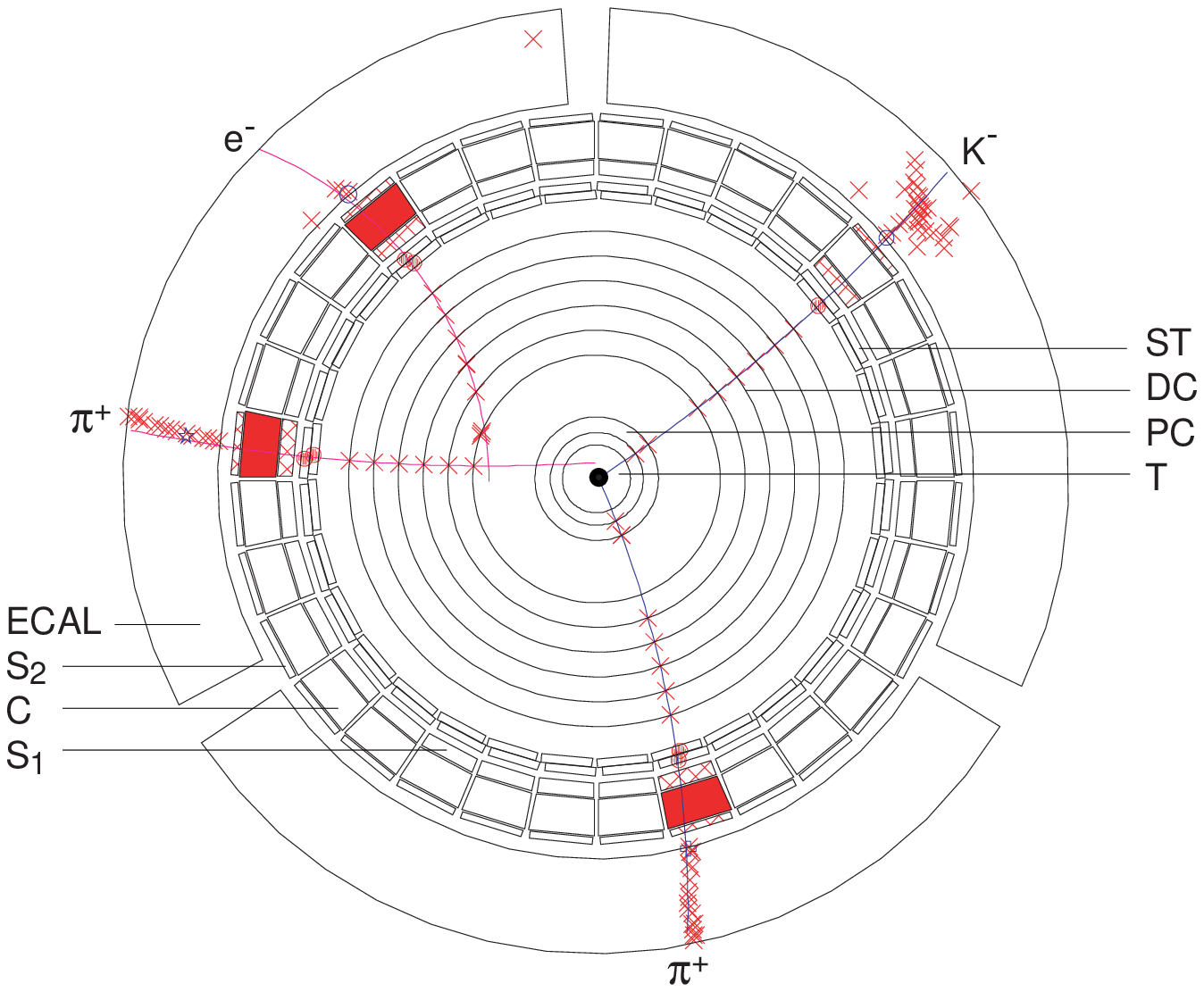}\\[5mm]
                                                      % -- evtest.eps
  (b) \\[15mm]
\caption{}
\label{fig:detec}
\end{center}
\end{figure}

   Fig.~ \ref{fig:detec} --  CPLEAR detector.\\ 
   (a) longitudinal view.\\ 
   From the left, the $200$ \mev\ \Pb\ beam delivered by LEAR  enters the 
   magnet along its axis, and through a thin scintillator (Beam monitor) 
   reaches a pressurized hydrogen gas target (T) where antiprotons stop 
   and annihilate. Cylindrical tracking detectors provide information about 
   the trajectories of charged particles in order to determine their charge 
   signs, momenta and positions: two proportional chambers (PC), six drift 
   chambers (DC) and two layers of streamer tubes (ST). The particle 
   identification detector (PID) comprising a threshold Cherenkov 
   detector (C) sandwiched between two layers of scintillators (S1 and S2)
   allows the charged-kaon identification, and also the separation of 
   electrons from pions. The outermost detector is a  lead/gas sampling 
   calorimeter (ECAL) to detect the photons from $\pi^0$ decays.\\
   (b) transverse view and display of an event.\\ 
   $\PPb$ (not shown) $\rightarrow$ \km\pip\kn\ with the neutral kaon 
   decaying to \elm\pip\netb . The view (b) is  magnified twice with respect 
   to (a) and does not show the magnet coils and outer detector components.\\ 
   From Ref. \cite{pr}.

%%%%%%%%%%%%%%%%%%%%---FIG2---\label{fig:eleff}---%%%%%%%%%%%%%%%%%%%%%%%%%%
\subsection*{}   
\begin{figure}
  \begin{center}                                                      
    \begin{tabular}{cc}
  (a) &  (b) \\                
  \includegraphics[bb=188 291 396 494,width=0.45\linewidth,clip]   
                                                 {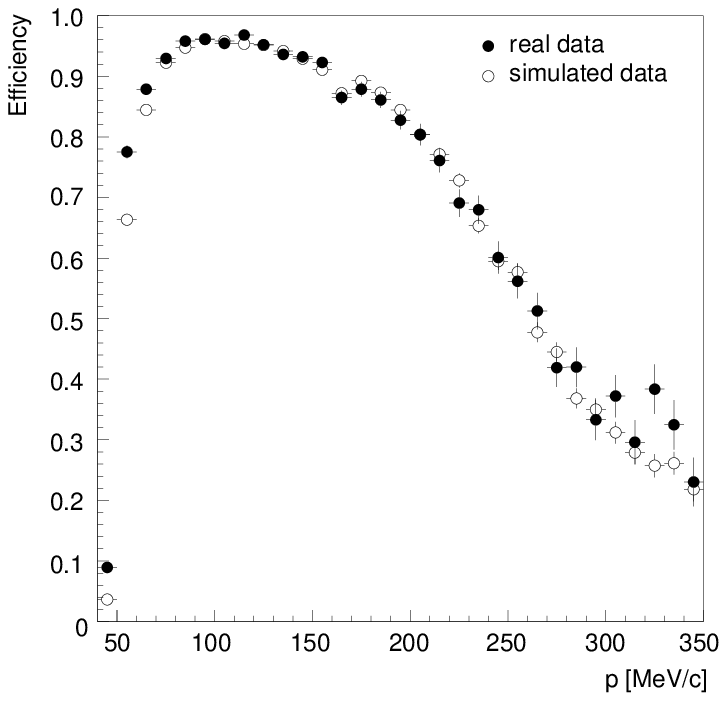} &    
  \includegraphics[bb=14 124 570 651,width=0.5\linewidth,clip]     
                                                  {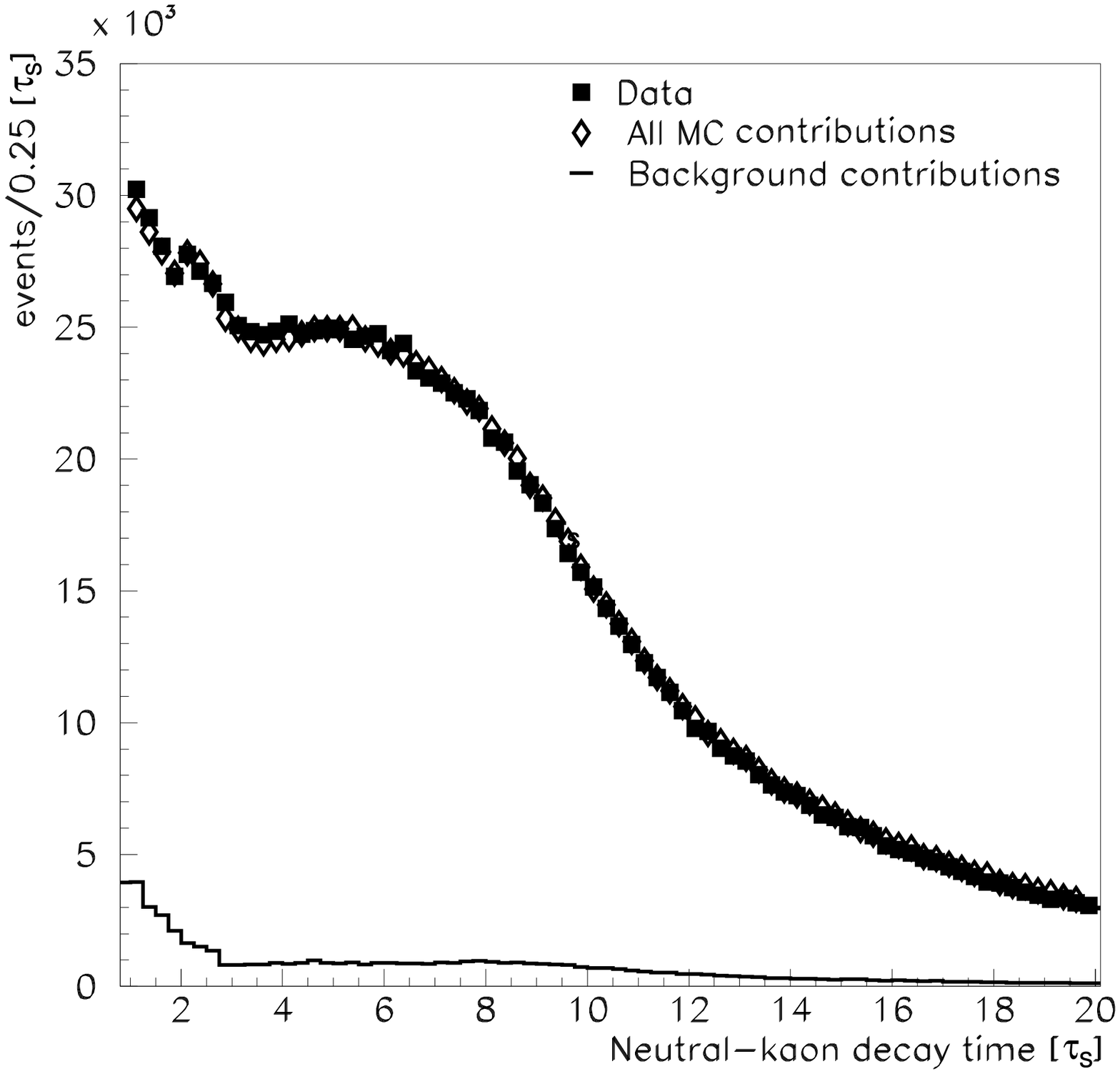}   
    \end{tabular}                                                      
  \end{center}                                                         
\caption{}
\label{fig:eleff}                                               
\end{figure}

   Fig.~ \ref{fig:eleff} -- Electron detection.\\
   (a) Electron identification efficiency as a function of 
                   momentum when $<2\%$ of pions fake electrons for real
                   $(\bullet )$ and simulated $(\circ )$ calibration    
                   data; \\                                                
   (b) Decay time distribution for real $({\tiny \blacksquare })$ 
      and simulated  
      $(\diamond )$ data. The expected background contribution is shown 
      by the solid line. \\ 
    From Ref. \cite{pr}.

%%%%%%%%%%%%%%%%%%%%---FIG3--- \label{fig:norm}---%%%%%%%%%%%%%%%%%%%%%%%%%
\subsection*{}   
\begin{figure}
\begin{center}
\includegraphics[bb=172 330 484 597,width=0.8\linewidth,clip]
                                                   {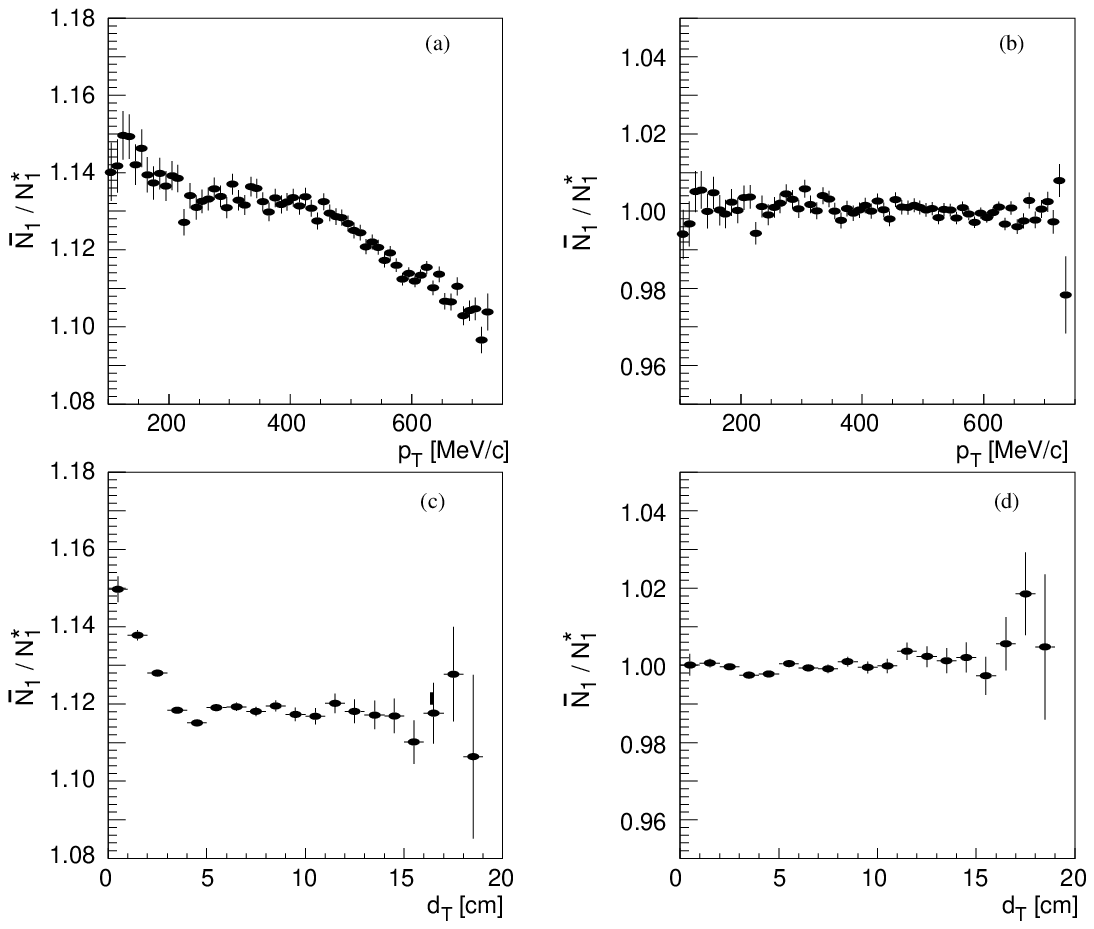}
\end{center}
\caption{}
\label{fig:norm}
\end{figure}

   Fig.~\ref{fig:norm} --  \knb\ to \kn\ normalization.\\
   The ratio between the numbers of \knb\ and \kn\ decays 
   to \pipi\ in the $(1 - 4)$ \ts\ interval, $\Nb_1 /N^*_1 $ 
   (corrected for regeneration and \CPz\  violating effects) as a 
   function of the neutral-kaon transverse momentum \pt , \\
   (a) before and\\ 
   (b) after giving each \kn\ event its normalization weight.\\
   The same ratio as a function of the decay vertex transverse separation 
   from the production vertex \dt ,\\
   (c) before and \\
   (d) after applying the normalization weights. \\
   From Ref. \cite{pipm}.

%%%%%%%%%%%%%%%%%%%%---FIG4---\label{fig:twopi_a}---%%%%%%%%%%%%%%%%%%%%%%%%%
\subsection*{}
\begin{figure}
\begin{center}                          
\includegraphics[bb=404 -245 660 9,width=0.8\linewidth,clip]
                                                {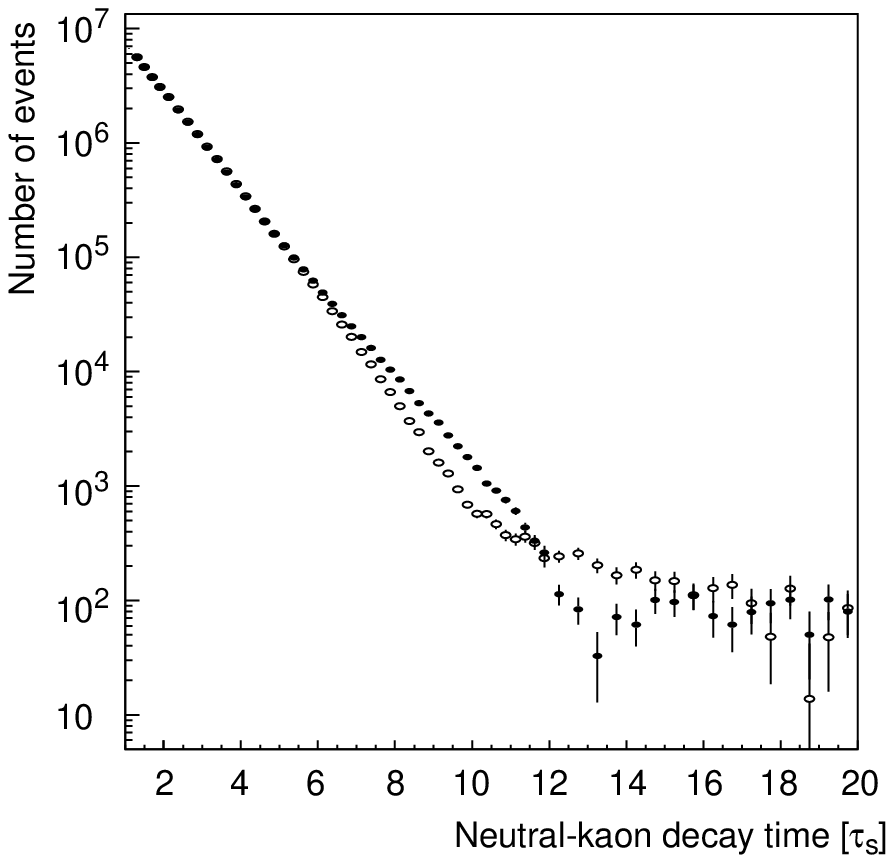}
\end{center}
\caption{}
\label{fig:twopi_a}
\end{figure}

   Fig.~\ref{fig:twopi_a} -- Neutral-kaon decay to \pipi\ : the different
                               decay rates indicate \CPz\ violation.\\ 
   The measured decay rates 
   for \kn\ ($\circ$) and \knb\ ($\bullet$) after acceptance correction
   and background subtraction.\\
   From Ref. \cite{pipm}.

%%%%%%%%%%%%%%%%%%%%---FIG5---\label{fig:twopi_b}---%%%%%%%%%%%%%%%%%%%%%%%%%
\subsection*{}
\begin{figure}
\begin{center}                          
\includegraphics[bb=0 19 232 235,width=0.8\linewidth,clip]
                                               {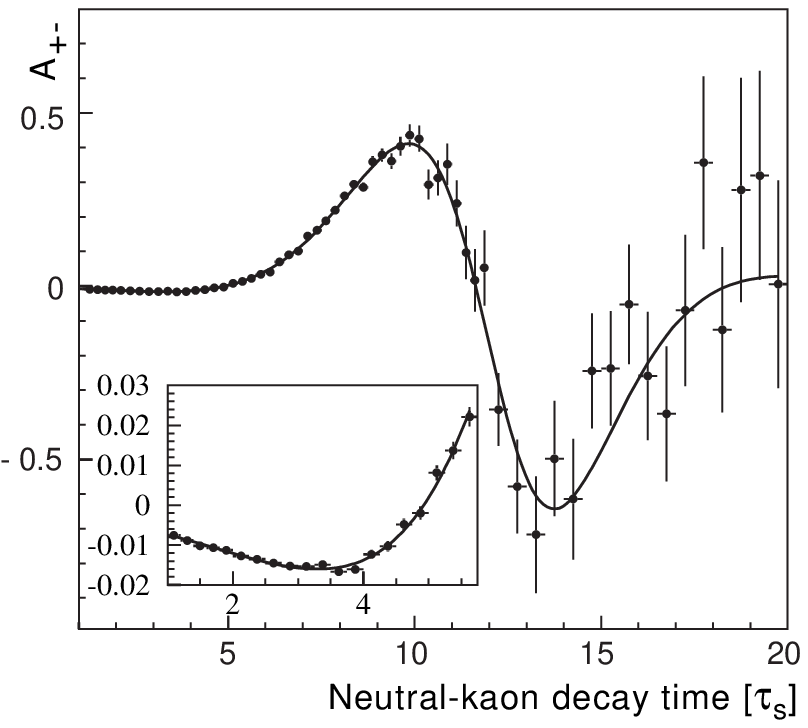}
\end{center}
\caption{}
\label{fig:twopi_b}
\end{figure}

   Fig.~\ref{fig:twopi_b} -- Neutral-kaon decay to \pipi\ : the rate
                         asymmetry \Apm ($t$) demonstrates \CPz\ violation. \\
   The measured asymmetry \Apmexp ($t$), Eq. (\ref{eq:asym_1c}), reduces to 
   \Apm ($t$) when the background is subtracted from the measured rates. The 
   dots are the data points. The curve is the result of the fit making use of 
   the rates (\ref{eq:2.65}). \\
   From Ref. \cite{pr}.

%%%%%%%%%%%%%%%%%%%%---FIG6---\label{fig:ad_mod}---%%%%%%%%%%%%%%%%%%%%%%%%%%%
\subsection*{}   
\begin{figure}
  \begin{center}
   \includegraphics[bb=0 0 368 231,width=0.8\linewidth,clip]
                                                 {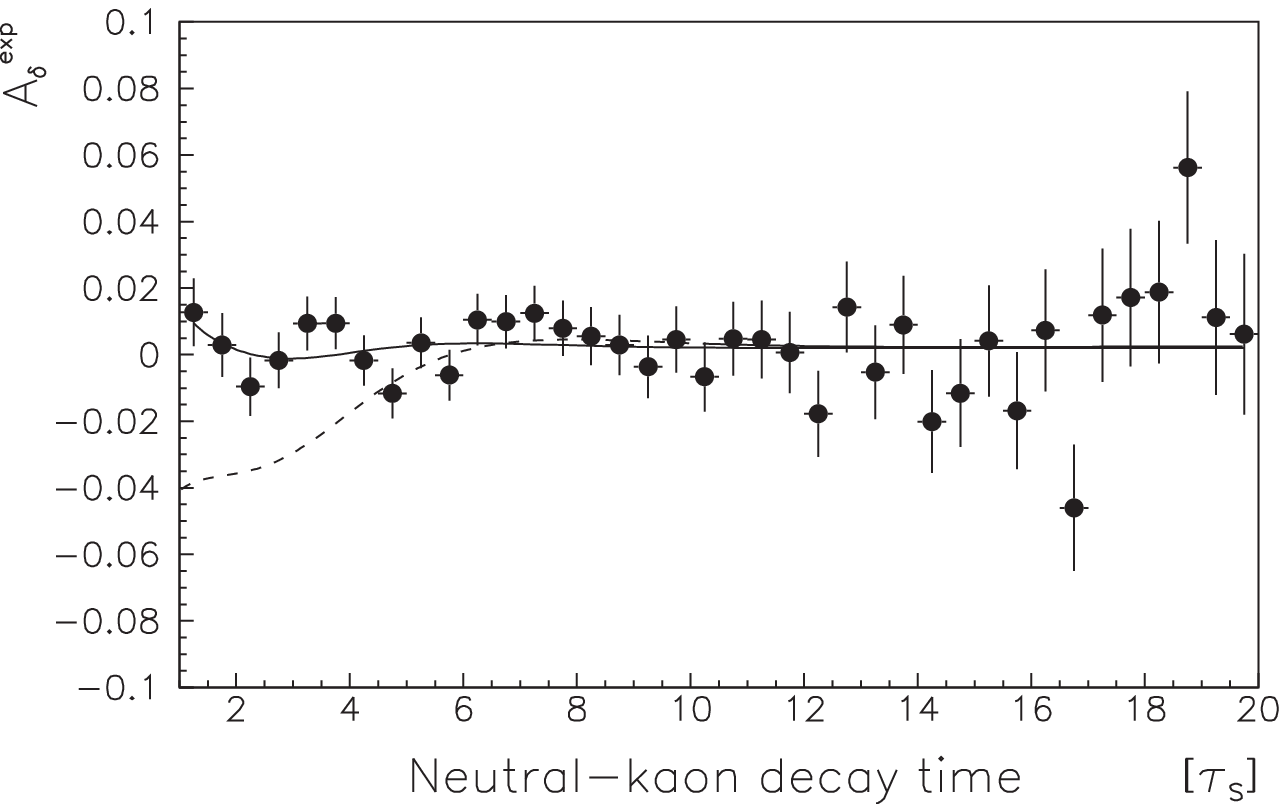}
  \end{center} 
\caption{}
\label{fig:ad_mod}
\end{figure}

   Fig.~\ref{fig:ad_mod} -- Experimental confirmation of \CPTz\ invariance 
   in the $time \ evolution$ of neutral kaons. \\ 
   The present result is the determinant input to~a measurement of the
   decay-width and mass differences between the neutral kaon
   and its antiparticle, the value of the latter being %$\Mzaa - \Mzbb =  
   $(- 1.7 \pm 2.0)\times 10^{-18}\; \geve ~.$ See Eqs. (\ref{dGdM}).\\ 
   Values for times $t\ \widetilde{>}\ 5\tau_S\ $ depend entirely on 
   a hypothetical \CPTz\ violation in the time evolution, independently 
   of further hypothetical violations of \CPTz\ invariance in the $decay$, 
   or of violations of the $\DS = \DQ$ rule.\\
   The points are the measured values of \Adexp .
   The solid line represents the fitted curve (\ref{adexp2}). The dashes 
   indicate a hypothetical violation of the $\DS = \DQ$ rule with the final 
   CPLEAR values \cite{phen2}, exagerated by one standard deviation for 
   $\imxp\ (\ra \ -4.6 \times 10^{-3})$, and for $\rexm\ 
   (\ra \ 2.5 \times 10^{-3})$.\\
   The validity of the $\DS = \DQ$ rule is confirmed.\\
   Data from Ref. \cite{pen3}

%%%%%%%%%%%%%%%%%%%%---FIG7---\label{fig:at_mod}---%%%%%%%%%%%%%%%%%%%%%%%%%%%%
\subsection*{}   
\begin{figure}                                              
  \begin{center}          
  \includegraphics[bb=108 325 490 545,width=0.8\linewidth,clip]
                                                 {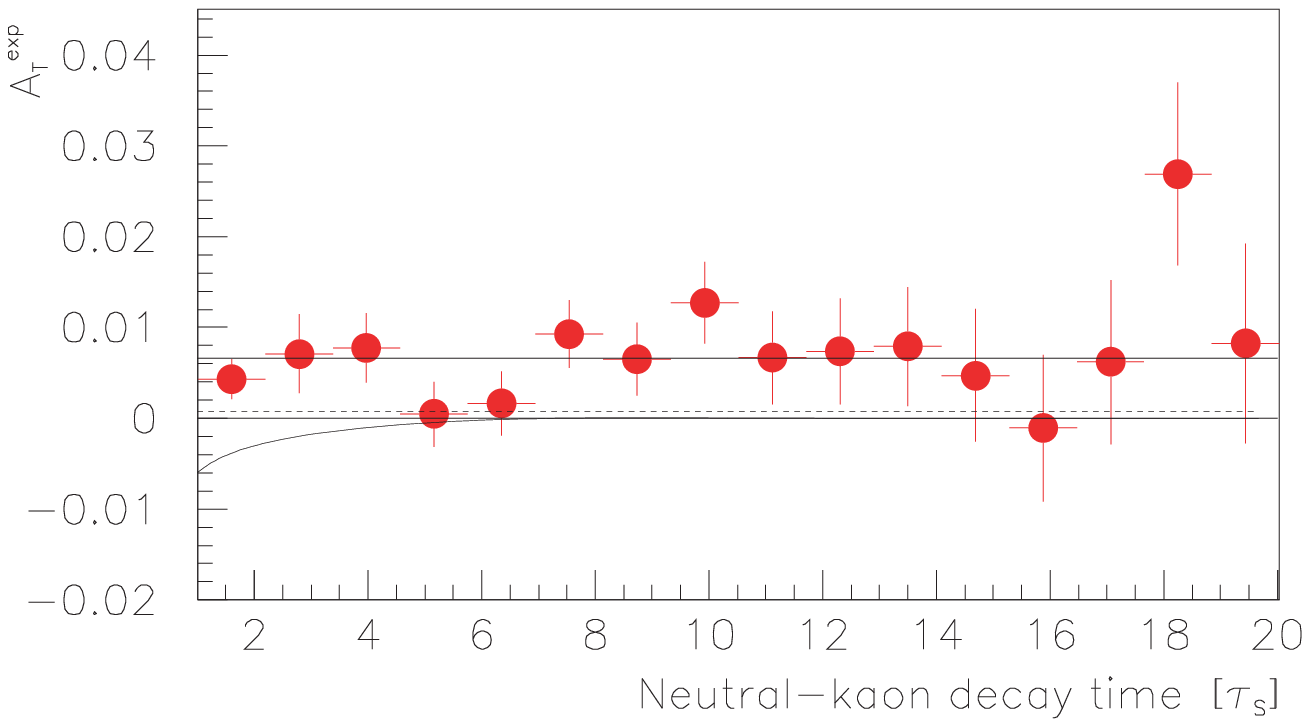} 
  \end{center}                                                    
\caption{}
\label{fig:at_mod}                                               
\end{figure}

   Fig.~\ref{fig:at_mod} -- Experimental demonstration of \Tz\ invariance 
   violation in the $time \ evolution$ of neutral kaons. \\ 
   The positive values show that a \knb\ develops into a \kn\ with higher 
   probability than does a \kn\ into a \knb . This $contradicts 
   \ \Tz ^{-1} \Hwk\ \Tz = \Hwk$ \ for neutral kaons.\\
   The points are the measured values of $\ATexp (t)$.
   The solid line represents the fitted constant \avATexp.
   All symmetry violating parameters concerning the semileptonic 
   $decay \ process$ are compatible with zero.
   The dashes indicate $- 4\re (y+x_{-}) = 0.8 \times 10^{-3}$, 
   the contribution of the constrained hypothetical \CPTz~\ invariance 
   violation in the $decay$, expressed in Eq. (\ref{atexp}).
   The time dependent curve, seen at early times, indicates a further 
   constrained hypothetical violation of the $\DS = \DQ$ rule,
   expressed by the function g (with exagerated input values). \\                                 
   Data from Ref. \cite{pen2}.
\newpage

%%%%%%%%%%%%%%%%%%%%%%%---FIG1---\label{fig:detec}---%%%%%%%%%%%%%%%%%%%%%%%%

%%%%%%%%%%%%%%%%%%%%%%%%---FIG2---\label{fig:eleff}---%%%%%%%%%%%%%%%%%%%%%%%%

%%%%%%%%%%%%%%%%%%%%%%%%---FIG3--- \label{fig:norm}---%%%%%%%%%%%%%%%%%%%%%%%%%

%%%%%%%%%%%%%%%%%%%%%%%%%---FIG4---\label{fig:twopi_a}---%%%%%%%%%%%%%%%%%%%%%%

%%%%%%%%%%%%%%%%%%%%%%%%%---FIG5---\label{fig:twopi_b}---%%%%%%%%%%%%%%%%%%%%%%

%%%%%%%%%%%%%%%%%%%%%%%%%---FIG6---\label{fig:ad_mod}---%%%%%%%%%%%%%%%%%%%%%%
                                                    
%%%%%%%%%%%%%%%%%%%%%%%%%---FIG7---\label{fig:at_mod}---%%%%%%%%%%%%%%%%%%%%%%%
 
%%%%%%%%%%%%%%%%%%%%%%%%%%%%%%%%%%%%%%%%%%%%%%%%%%%%%%%%%%%%%%%%%%%%%%%%%

%% file: new_biblio.tex
\bigskip
Figures 1, 2, 5, 6, 7 have been reprinted with the permission of Elsevier.

Figures 3 and 4 have been reprinted with the permission of Springer.